\documentclass{emulateapj}
\usepackage{apjfonts,lscape}

\begin{document}


\title{The Molecular Gas Density in Galaxy Centers and How It Connects
  to Bulges}

\shorttitle{Gas \& Bulges}
\shortauthors{Fisher et al.}
      

\author{David~B.~Fisher\altaffilmark{1}} ,
\author{Alberto~Bolatto\altaffilmark{1}},
\author{Niv~Drory\altaffilmark{2}}
\author{Francoise Combes\altaffilmark{3}}
\author{Leo~Blitz \altaffilmark{4}}
\author{Tony~Wong \altaffilmark{5}}
\altaffiltext{1}{Laboratory of Millimeter Astronomy, University of
  Maryland, College Park, MD 29742}
\email{dbfisher@astro.umd.edu}
\altaffiltext{2}{Instituto de Astronom\'ia, Universidad Nacional
  Aut\'onoma de M\'exico, A.P.\ 70-264, 04510 M\'exico, D.F.,
  M\'exico}
\altaffiltext{3}{Astronome \'a l'Observatoire de Paris, 61 Av. de l'Observatoire
F-75 014 Paris, France}
\altaffiltext{4}{University of California Berkeley, Astronomy Department,
Berkeley, CA USA}
\altaffiltext{5}{ University of Illinois at Urbana-Champaign,
  Department of Astronomy, Urbana, IL 61801
USA }

\slugcomment{Accepted for publication in ApJ}


\begin{abstract}
  In this paper we present gas density, star formation rate, stellar
  masses, and bulge disk decompositions for a sample of 60
  galaxies. Our sample is the combined sample of BIMA SONG, CARMA
  STING, and PdBI NUGA surveys. We study the effect of using
  CO-to-H$_2$ conversion factors that depend on the CO surface
  brightness, and also that of correcting star formation rates for
  diffuse emission from old stellar populations. We estimate that star
  formation rates in bulges are typically lower by 20\% when
  correcting for diffuse emission.  Using the surface brightness
  dependent conversion factor, we find that over half of the galaxies
  in our sample have $\Sigma_{mol}>100$~M$_{\odot}$~pc$^{-2}$. Though
  our sample is not complete in any sense, our results are enough to
  rule out the assumption that bulges are uniformly gas poor
  systems. We find a trend between gas density of bulges and bulge
  S\'ersic index; bulges with lower S\'ersic index have higher gas
  density. Those bulges with low S\'ersic index (pseudobulges) have
  gas fractions that are similar to that of disks. Conversely the
  typical molecular gas fraction in classical bulges is more similar
  to that of an elliptical galaxy. We also find that there is a strong
  correlation between bulges with the highest gas surface density and
  the galaxy being barred. However, we also find that classical bulges
  with low gas surface density can be barred as well. Our results
  suggest that understanding the connection between the central
  surface density of gas in disk galaxies and the presence of bars
  should also take into account the total gas content of the
  galaxy. Finally, we show that when using the corrected star
  formation rates and gas densities, the correlation between star
  formation rate surface density and gas surface density of bulges is
  similar to that of disks. This implies that at the scale of the
  bulges the timescale for converting gas into stars is comparable to
  those results found in disks.

\end{abstract}



\section{Introduction}\label{sec:intro}

The stellar populations and gas densities in the centers of nearby
bulge-disk galaxies is not uniformly old and non-star forming.  Results
from millimeter wave interferometers show that the centers of many
bulge-disk galaxies are gas rich (see for example
\citealp{regan2001bima,helfer2003}). Similarly, results from Spitzer
and GALEX show that star formation rate densities of many bulges are
high and star formation is significant compared to the bulge stellar
mass \citep{fisher2006,fdf2009}.  Also, results from stellar
populations show that many bulges are young and growing
\citep{peletier1996,ganda2007,macarthur2009}. We now understand that
there is a great variety in the radial distribution of molecular gas
in bulge-disk galaxies.

In the main bodies of disks, total gas density profiles are very well
described by exponential decay
\citep{young1995,schruba2011,bigiel2012}.  However, for radii smaller
than $\sim$25\% of the optical radius there is a very large amount of
scatter from galaxy-to-galaxy.  \cite{regan2001bima} shows that the
gas density in galaxy centers frequently breaks from the exponential
disk, and often increases in density similar to star light. These
small regions in the centers of disks are the same location as bulges
in the stellar light profiles.  Other than the special case of
outflows, the gas in the centers of bulge-disk galaxies is likely in a
thin layer that is only a few hundred parsecs in thickness
\citep[e.g.][]{garcia1999}. However, typical bulge-disk decompositions
do not consider the thickness of the bulge-component, only the surface
density of stars. Therefore, the gas in the centers of disks that is
higher in surface density than the exponential profile is likely
making stars that are, from the point-of-view of bulge-disk
decompositions, in the ``bulge.''  However, in this case the ``bulge''
is not a 3-dimensional spheroid at the center of a galaxy disk, rather
in this paper we adopt the definition of ``bulge'' to be the centrally
located high surface-density component of a galaxy surface brightness
profile, in excess over the inward extrapolation of an exponential
disk. This definition of bulge is clearly observationally motivated.


Bulges are heterogeneous in more ways than just the star formation.
Once thought to be uniformly similar to elliptical galaxies,
observations of bulge dynamics, stellar populations structure and
morphology, now show that there are at least two categories of bulges
(respectively \citealp{k93,peletier1996,carollo97,fisherdrory2010},
also see \citealp{kk04} for a review). This dichotomy of bulge
properties can be summarized as follows: some bulges have properties
similar to disk galaxies, and other bulges have properties similar to
elliptical galaxies. Those bulges with properties similar to disks are
called ``pseudobulges'' and those with properties similar to
elliptical galaxies are referred to as ``classical bulges''. Note that
as used in this paper the terms pseudobulge and classical bulge are
observationally based, and do not make any {\em a priori} assumption
about the formation mechanism. Also note, that in this paper the word
``bulge'' refers to the bulge in surface brightness profiles of
non-edge on galaxies.

This dichotomy is not a sequence in bulge-to-total ratio; at a
constant bulge-to-total ratio bulges separate into two classes,
even at small bulge-to-total ratios \citep{droryfisher2007}.  Also, it
appears to be consistent within different means of identifying
pseudobulges and classical bulges. For example, bulges with higher
3.6-8.0~$\mu$m colors, indicative of higher specific star formation
rates, have lower S\'ersic index \citep{fisherdrory2010}. Also bulges
with low velocity dispersion are more likely to be identified as
pseudobulges \citep{fabricius2012}. In this paper, we wish to
directly compare molecular gas density to pseudobulge diagnostics,
like S\'ersic index.

A number of simulations show that barred gravitational potentials lead
to central concentrations of gas in simulated disk galaxies
\citep[e.g.][]{simkin1980,combesgerin1985,heller2007b}. In principle,
the bar torques the gas, this torque facilitates an exchange of
angular momentum \citep{garcia2005}. Inside of corotation a common,
but not universal, reponse to bar torques is a angular monentum loss,
and thus the gas falls to smaller radius.  \cite{athan92} shows with
hydrodynamical models that indeed bars are capable of driving
significant masses of gas; this occurs due to shocks on the leading
edges of bars that efficiently drive inward flows of
gas. \cite{regan1997} show that the motion of gas in bars is
consistent with that of shocks predicted in the simulations of
\cite{athan92}, and the net effect is inflow in the central parts of a
barred disk galaxy.  Detailed studies of individual galaxies have
shown that torques introduced by bars in disk galaxies are likely
responsible for inward flowing gas, however the flow of gas in
galaxies is often complicated by the details of the disk
\cite{garcia2005,hunt2008,haan2009}. These studies show that bars can
drive gaseous inflows; typical rates of inflow are of order
0.1-10~$M_{\odot}~yr^{-1}$.  Several studies find a high-scatter
correlation between the presence of a bar and high densities of gas in
the central kiloparsec of disks
\citep{sakamoto1999,jogee2005,sheth2005}. Similarly, barred disks are
more likely to have bluer (younger) bulges \citep{gadotti2001}, and
are star forming \cite{fisher2006,wang2012,oh2012}.

The scatter in the bar-central gas density correlation is large enough
to suggest that simply dividing galaxies into barred vs. unbarred is an
oversimplification.  A number of factors could be adding to the
dispersion between the correlation of bars and central gas density.
Simulations show that bars can dissolve
\citep[e.g.][]{friedli1999,shen2004}.  Also, spirals can similarly
torque gas \citep{zhang99}.  Accretion or weak interactions with
neighboring galaxies could drive central concentrations of gas
\citep{combes1994,espada2010}. Indeed, \cite{bournaud2002} show that
bar driven ``secular'' evolution and evolution driven through
accretion/merging is not mutually exclusive .


Thus far, arguments that high gas densities in bulges are increasing
the stellar bulge-to-total ratio of galaxies \citep[e.g.][]{fdf2009}
are built on the assumption that star formation in bulges is the same
as in spiral arms of disks. \cite{bolatto2008} shows that typical
giant molecular clouds in nearby galaxies are of the order
$\sim$50-300~M$_{\odot}$~pc$^{-2}$. Gas densities observed in the
centers of disks can be an order of magnitude higher
\cite[e.g.][]{jogee2005}. The increased pressure from a higher density
of stars could significantly affect the formation of stars \citep[for
a discussion see][]{blitz2006}. Furthermore, emission from a nearby
AGN could ionize gas. It is therefore not clear if the formation of
stars out of gas has similar efficiency in bulges and disks over the
full range of bulge properties. In this paper we will investigate
modifications to both the CO-to-molecular gas conversion factor and
also to the calculation of star formation rates suited to the
environment of bulges, we will also investigate the star formation law
of bulges to determine if star formation in bulges is, at least to low
order, similar to what is observed in disks.


In this paper we will
implement new methods to determine both star formation rates and gas
masses in ways that are more appropriate for bulges. We will then
determine what structural properties of the bulge and galaxy are
associated with high gas density. 
\citep{schruba2011,bigiel2012}  

\section{Sample}

The properties of galaxies in our sample are listed in Table 1. Our
sample is the superset of 3 surveys which map the CO(1-0) in nearby
bulge-disk galaxies. Those surveys are the SONG \citep[described
in~][]{helfer2003} conducted on the Berkeley-Illinois-Maryland
Association (BIMA) interferometer \citep{welch1996}, the NUGA
(described in \citealp{nuga}; PIs: S. Garc\'ia-Burillo and F. Combes)
survey conducted on the Plateau de Bure (PdBI) interferometer, and the
STING (described in \citealp{rahman2012}; PI:A. Bolatto).

In addition to the molecular gas mass, for each galaxy we also
determine the star formation rate, stellar mass and carry out
bulge-disk decompositions to the near infrared stellar light profile
(these methods are described below). Therefore we require that all
galaxies have MIPS data from {\em Spitzer Space Telescope}.  Also the
galaxy must have near IR data with high enough spatial resolution to
accurately decompose the stellar light profile \citep[for a detailed
discussion see][]{fisherdrory2008}. Two galaxies are thus excluded due
to lack of ancillary data. NGC~3718 does not have MIPS data, and
NGC~6574 did not have adequate data for bulge-disk
decomposition. Also, we are interested in the properties of disks, so
we omit those galaxies that are experiencing major mergers from the
sample: NGC~5953, NGC~1961, NGC~4490.  The AGN in NGC~1068 makes both
decomposition and measurement of the star formation rate unreliable;
it is thus omitted. Finally, due to Galactic extinction we could not
reliably estimate the mass-to-light ratio of IC~342 \& NGC~1569
(E(B-V) = 0.558 \& 0.700~mag respectively, \citealp{schlegel}); they
are thus removed from the sample. The resulting combined sample is 60
galaxies.
\begin{figure}[t]
\begin{center}
\includegraphics[width=0.49\textwidth]{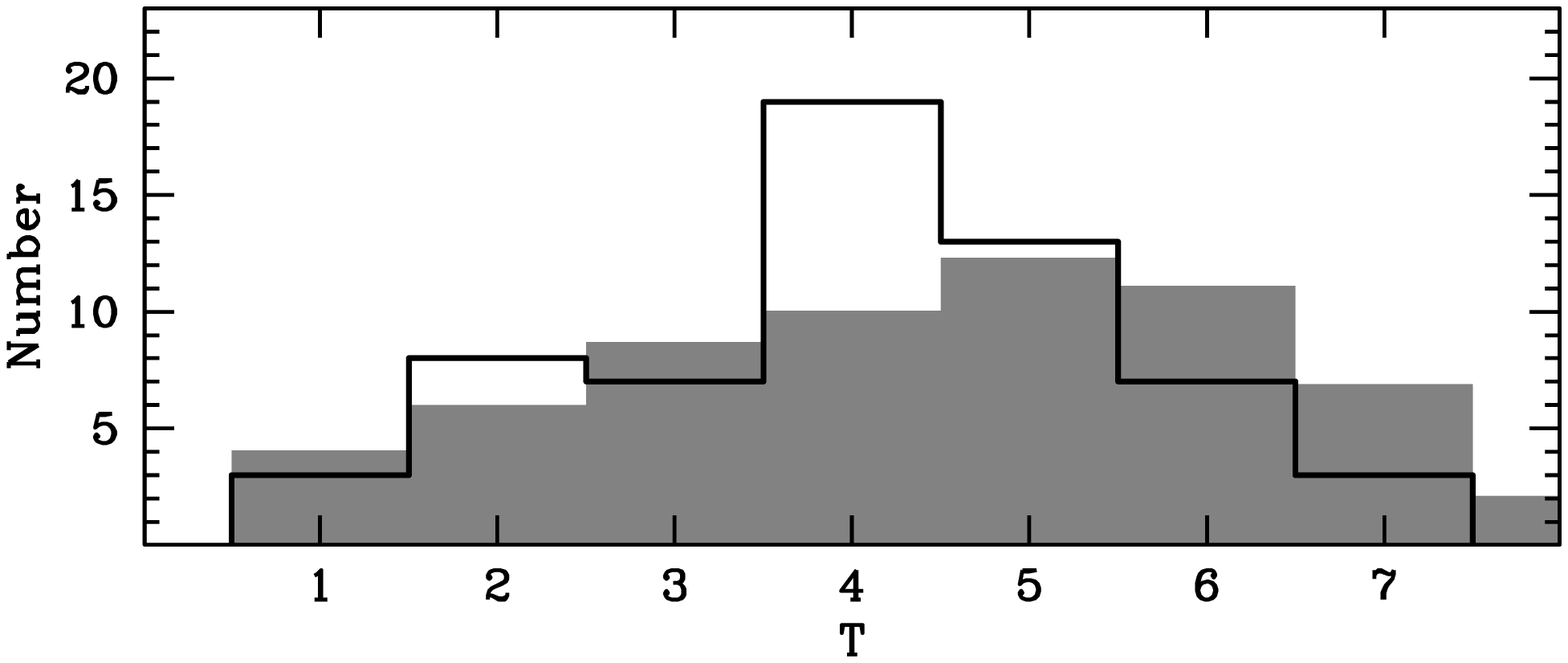}
\includegraphics[width=0.49\textwidth]{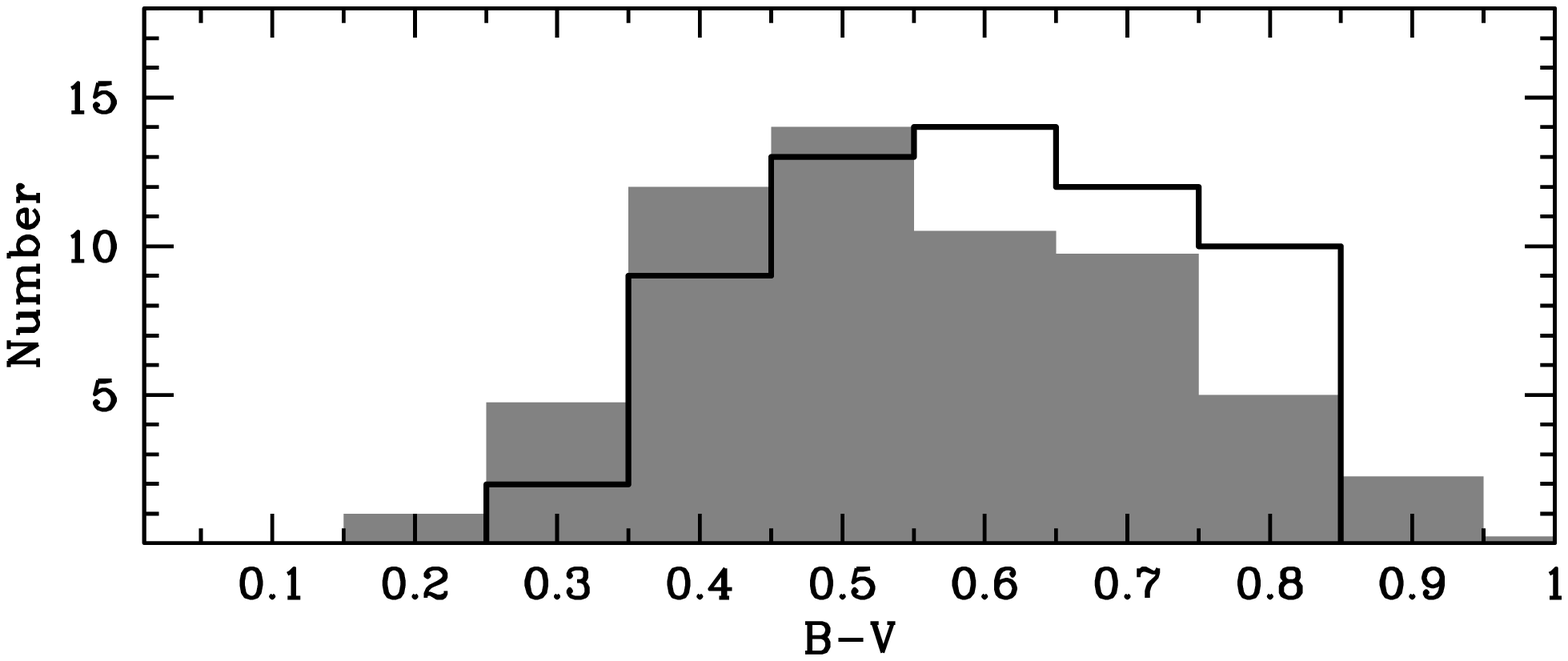}
\includegraphics[width=0.49\textwidth]{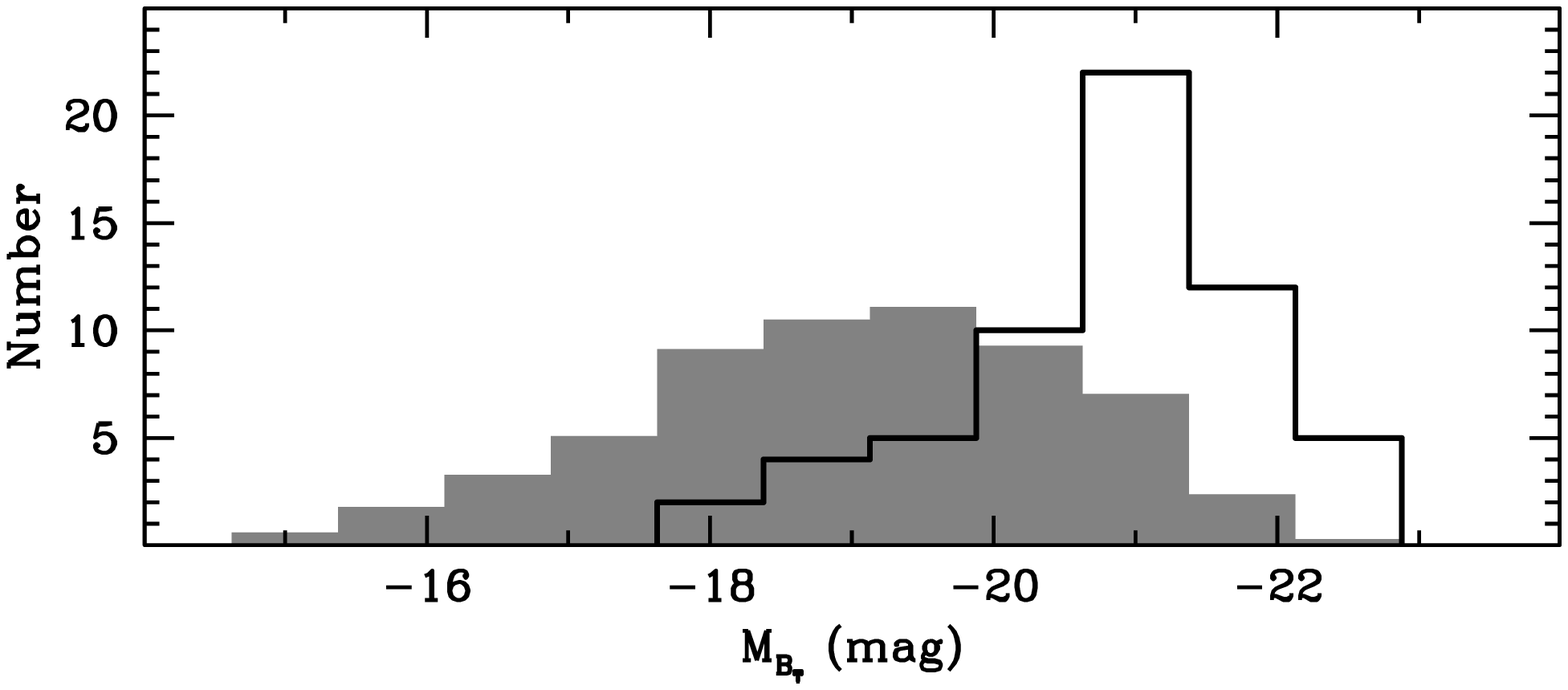}
\includegraphics[width=0.49\textwidth]{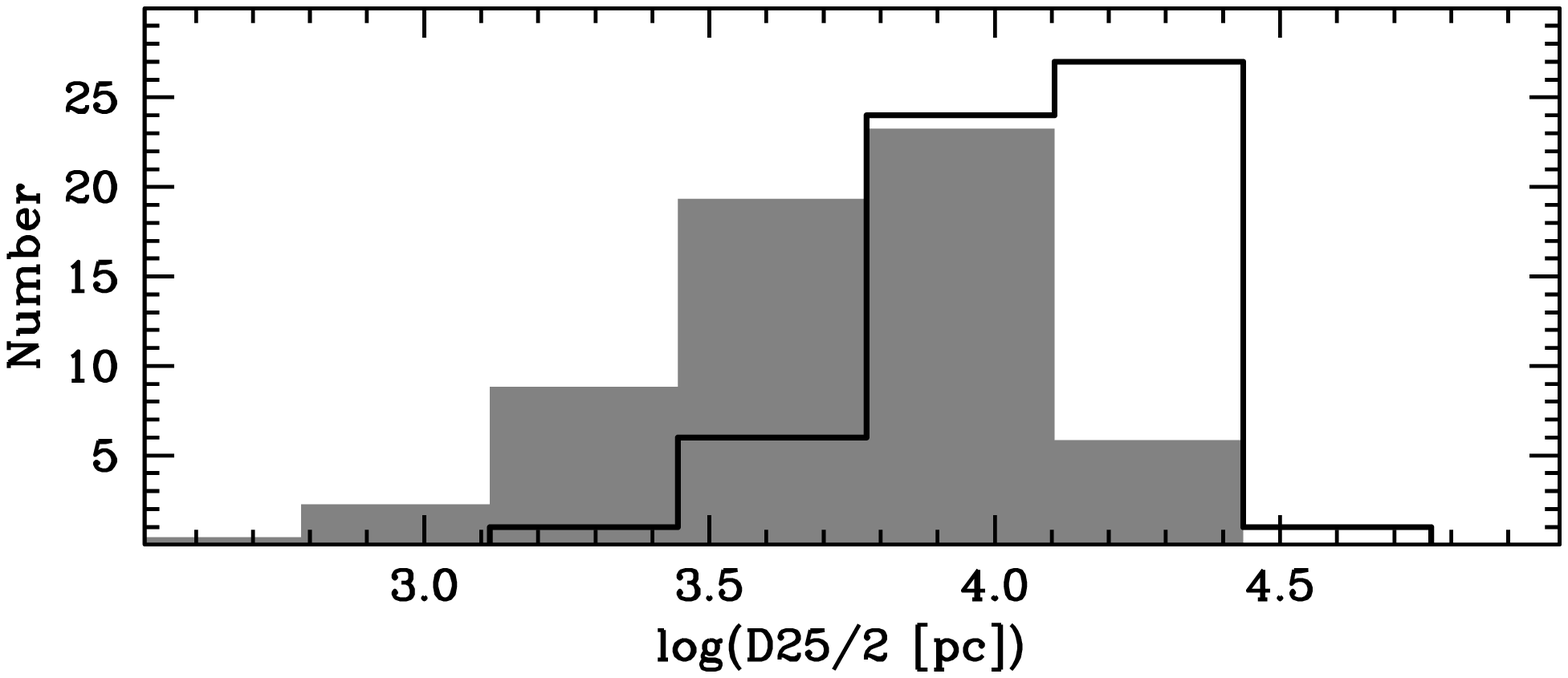}
\end{center} 
\caption{ The distribution of galaxy properties of our sample. From
  top to bottom we show the distribution of Hubble type, $B-V$ color,
  absolute B mag, and optical radius. These data are taken from
  Hyper-LEDA. The solid lines indicate the distribution for our sample
  galaxies, and the shaded histogram shows the distribution of the RC3
  galaxies covering the same range in Hubble type, scaled to match our
  sample size. \label{fig:sample}}
\end{figure}

We combine the multiple surveys simply to increase the number of
galaxies, and thus obtain better statistics. This is especially useful
for separation into sub-samples for comparison (e.g.~barred vs
unbarred or pseudo- vs classical bulges). However, the trade off is
that our sample is not based on a uniform set of selection criteria.
In Fig.~\ref{fig:sample} we show the basic properties of our
sample. For comparison we also show the distribution of basic
properties of galaxies from the RC3 \citep{rc3} that cover the same
range in Hubble type. We have scaled the numbers of the RC3 galaxies
to match our sample. In the top panel we show the distribution of
morphological T-types of our sample. The sample includes spiral
galaxies from Sa to Sd, with a preference toward intermediate
types. 
The sample will thus favor medium size bulges \citep{simien1986},
however, it should adequately cover parameter space. In the second
panel we show the distribution $B-V$ optical color, and our sample
adequately covers the range of $B-V$ in typical spiral galaxies. In
the bottom two panels we show absolute galaxy magnitude and galaxy
size. It is clear that our sample preferentially includes brighter,
larger galaxies. Therefore, the results we find are most applicable to
massive spiral galaxies.

\section{Methods}

In this paper we wish to compare gas masses of bulges, to bulge-disk
decompositions, stellar masses, and star formation rates.  Therefore,
we require data from a wide variety of sources and must employ vastly
different techniques, across a wide range in wavelengths. In this
section we will describe each technique employed.

We take distances from the NASA Extragalactic Database (NED). We use
the NED averaged redshift independent distance. In the event that an
extreme outlier exists from the average, we remove the outlier and
re-average.

\subsection{Near-IR Surface Photometry \& Bulge-Disk Decompositions} 
For each galaxy, we first carry out bulge-disk decomposition to the
starlight. This allows us to measure the radius inside which the
bulge dominates the surface brightness profile. Also we will use the
decomposition parameters to study how gas density scales with those
properties of the starlight. Our method is described in detail in
several publications, including
\cite{fisherdrory2008,fisherdrory2010,kfcb}.

A key part of our decomposition method is combining surface photometry
from multiple data sources to construct a single 2-D surface
brightness distribution. This technique reduces systematic errors in a
particular data set (e.g. sky subtraction and point-spread function),
as well as increasing dynamic fitting range, in radius, by combining
high spatial resolution (typically HST) data with wide field
ground-based data.  This method is specifically optimized to reduce
the uncertainty in bulge paramaters. Maximizing the dynamic fitting
range in log(radius), not necessarily adding azimuthal information, is
critical to accurately fit the curvature in the surface brightness
profile, which is represented by the S\'ersic index, for a particular
bulge.

Our primary source of data for measuring the bulge-disk decomposition
is the 2MASS H-band maps. Although 2MASS $K_{s}$ band is slightly less
sensitive to dust emission and variations in mass-to-light ratio, the
$H$-band data goes much deeper in surface brightness. The improved
sensitivity to the faint isophotes gives a better constraint to the
disk parameters in the bulge-disk decompositions.
When available we use the higher quality images from
the 2MASS Large Galaxy Atlas \citep{jarrett2003}. Otherwise $H$-band
images are taken from the 2MASS archive\footnote{2MASS archive is
  available at http://irsa.ipac.caltech.edu}. We also frequently add
data, when available, from the OSU Bright Galaxy Survey
\citep{eskridge2002}.  

For all ground-based images we calculate sky values by fitting a plane
to the areas of the image not affected by galaxy light or bright
stars. We then subtract the surface representing the sky from the
image. As we state above, comparison of multiple data sources
facilitates identification of inaccurate sky-subtraction. When a
galaxy has only one source of wide-field $H$-band data we then carry
out the surface photometry using data available from Spitzer/IRAC at
3.6~$\mu$m. Uncertainties in sky subtraction of $H$-band images are
typically $\lesssim$0.1~mag.

Fine spatial resolution is crucial for accurate bulge-disk
decomposition using the S\'ersic function \citep[for a detailed
discussion
see][]{fisherdrory2008,fisherdrory2010}. \cite{fisherdrory2010} finds
that in most cases resolution of $\sim$100~pc is sufficient for robust
decompositions.  Since 2MASS has a typical resolution of $\sim$2'' we
must include higher resolution data in decompositions of galaxies at
greater distance than 10~Mpc.  For nearby galaxies including high
resolution data significantly reduces the uncertainty in the fit. The
most common source of high resolution data is publicly available
images in the HST archive\footnote{Hubble Legacy Survey can be found
  at http://hla.stsci.edu/}. HST data is always measured with the
F160W filter (H-band). 

To calculate the 2-D surface brightness distribution, we use the isophote
fitting routine of \cite{bender1987}. First, interfering foreground
objects are identified in each image and masked via automatic routines
in Source Extractor \citep{bertin1996}. We then run multiple
iterations of manual masking.  Isophotes are sampled by 256 points
equally spaced in an angle $\theta$ relating to polar angle by $\tan
\theta = a/b\,\tan \phi$, where $\phi$ is the polar angle and $b / a$
is the axial ratio.  An ellipse is then fitted to each isophote by
least squares. The software determines six parameters for each
ellipse: relative surface brightness, center position, major and minor
axis lengths, and position angle along the major axis. We then combine
the surface photometry of different data sources by averaging the
profiles. 
\begin{figure*}[t]
\begin{center}
\includegraphics[width=0.89\textwidth]{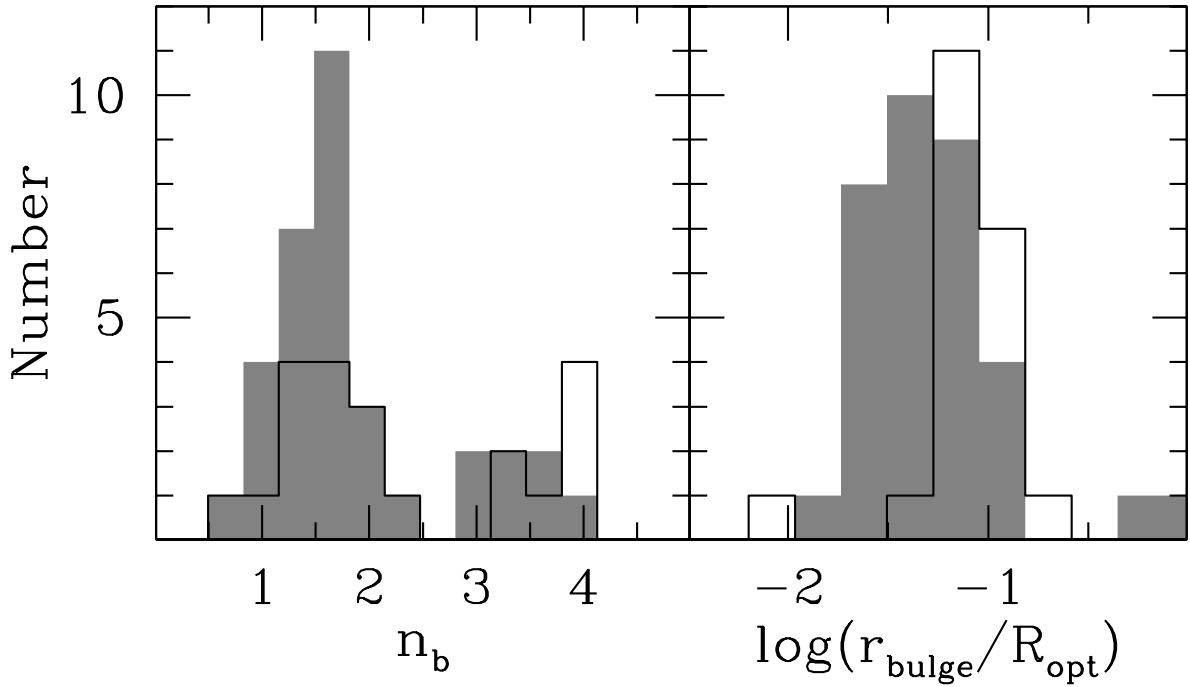}
\end{center} 
\caption{ The distribution of barred and unbarred bulge
  parameters. The left panel shows bulge S\'ersic index, and the right
  panel shows the size of bulges normalized by the optical radius of
  the galaxy. Shaded histograms represent barred disks and open
  histogram represents unbarred. The distribution of both S\'ersic
  index and bulge size for barred and unbarred galaxies overlap
  significantly.  \label{fig:bar_check}}
\end{figure*}

All profiles are scaled to match the zero-point of 2MASS
data. For example, when we include Spitzer~3.6~$\mu$m data we
calculate the H-3.6 color over a region that is trustworthy (not likely
affected by either point-spread-function or sky subtraction), and then
shift the surface brightness of the 3.6~$\mu$m to match the 2MASS
image. This is carried out for all data sources. 

We determine bulge and disk parameters by fitting each
surface brightness profile with a one-dimensional S\'ersic function
plus an exponential outer disk,
\begin{equation}
I(r)=I_0\exp\left[-(r/r_0)^{1/n_b} \right ] + I_d\exp\left[-(r/h)
\right ]\, ,
\label{eq:decomp}
\end{equation}
where $I_0$ and $r_0$ represent the central surface brightness and
scale length of the bulge, $I_d$ and $h$ represent the central surface
brightness and scale length of the outer disk, and $n_b$ represents
the bulge S\'ersic index. The half-light radius, $r_e$, of the bulge is
obtained by converting $r_0$,
\begin{equation}
r_e=(b_n)^nr_0,
\end{equation}
where the value of $b_n$ is a proportionality constant defined such
that $\Gamma (2n)=2\gamma(2n,b_n)$. $\Gamma$ and $\gamma$ are the
complete and incomplete gamma functions, respectively. We use the
approximation $b_n\approx2.17n_b-0.355$. We restrict our range in
possible S\'ersic indices to $n_b>0.33$ to ensure that the
approximation of $b_n$ is accurate \citep{caon94,macarthur2003}.  A
more precise expansion is given in \citealp{macarthur2003}.  (Also,
see \citealp{graham2005} for a review of the properties of the
S\'ersic function.) 

Because the S\'ersic parameters are only determined to fit the major
axis profile, they do not include information about the azimuthal
shape of the bulge or disk. We therefore adjust all luminosities by
the average ellipticity in the region in which that parameter
dominates the light.

Despite its successes, the S\'ersic bulge plus outer exponential disk
model of bulge-disk galaxies does not account for many features of
galaxy surface brightness profiles. Disks of intermediate type
galaxies commonly have features such as bars, rings, and lenses (see
\citealp{k82} for a description of how these manifest in surface
brightness profiles), similar features are now well known to exist in
the centers of galaxies (e.g. nuclear rings and nuclear star
clusters). In general there are two methods for dealing with
perturbations to a fit. In some cases authors attempt to fit the extra
components, other authors mask those components. We choose the
later. Both have advantages and disadvantages. Fitting extra
components (such as bars and nuclear star clusters) introduces
degeneracies between the two components, where as masking data can
introduce selection biases.  Detailed testing and description of our
method to deal with these structures can be found in
\cite{fisherdrory2008}.  The appendix of \cite{fisherdrory2008}
investigates the stability of S\'ersic parameters to the masking of
bars and nuclear clusters. In cases in which the bulge is well
resolved these methods recovers bulge and disk parameters that are
equivalent \citep[also][]{macarthur2003}.  In the subsequent
paragraphs we describe how we deal with these structure:

{\em Bars:} We mask the isophotes that are affected by bars
and rings, and they are not included in the fit. This is a subjective
procedure, as it requires selectively removing data from a galaxy's
profile, and undoubtedly has an effect on the resulting parameters.
\cite{fisherdrory2008} investigate the affects of this procedure in
detail on surface brightness profiles; see the appendix of that paper.
Typically, the structure of the ellipticity profile aids in
identification of bars, as described in \cite{marinova2007}. In a
galaxy in which we identify a bar (either using the ellipticity
profile or near-IR imaging) we identify isophotes associated with the
bar, then remove those isophotes, and re-fit Eq.~\ref{eq:decomp}. We
continue this process, iteratively, until the systematic deviations in
the residual profile converge on a robust solution.
\cite{fisherdrory2008} find that the largest effect of masking larger
ranges is to increase the uncertainty. The size of the bulge remains
roughly constant; there is a tendenancy for the S\'ersic index to
decrease very slightly ($\Delta n \sim 0.1$). We stress that the
reason we can do this is that our technique relies upon resolving the
bulge well, either from the galaxy being very close or from having HST
data.

Bulges in unbarred galaxies are larger than those in barred
galaxies.This is not likeley an artifact of our method to remove bars
from fits because the same result can be found in the samples of other
works that carry out bulge-bar-disk decomposition. This is true both
when we compare physical size, and the size of bulges that has been
normalized by the size of the galaxy. We will discuss later (in \S
3.6) that the small size difference is not what is responsible for the
differences in bulge CO surface brightness between either barred or
unbarred galaxies or classical or pseudobulges.

In our sample we find that the median size of bulges in unbarred
galaxies are 1.5$\times$ the size of bulges in bared galaxies. To
check that this is not a bias introduced by our method, we compile
data from both \cite{laurikainen2004} and \cite{gadotti2009} to find a
similar result; both of these papers use 2-D decompositions, that
include parameterizations for the bulge, bar, and disk.
\cite{gadotti2009} finds, similar to us that the median bulge
half-light radius for classical bulges ($n>2$) is $\sim1.5\times$ the
median r$_e$ of pseudobulges ($n<2$). We reconfirm this with data from
\cite{laurikainen2004} who also carries out bulge-bar-disk
decomposition to 181 bright spiral galaxies in the OSUBGS
\citep{eskridge2002}. It is important to take into account the size of
the galaxy; bigger bulges may simply reside in bigger halos. So we
recheck this result, this time we normalize all samples by the optical
radius of the galaxy ($R_{opt} = D25/2\cong 4\times h$). We find the
same result. The median bulge in an unbarred disk is $\sim 2\times$
the size of that in a barred disk (when normalized by the disk
scale-length).

In Fig.~\ref{fig:bar_check} we show the distribution of S\'ersic
indices and bulge sizes for our sample galaxies. The size of the bulge
$r_{bulge}$ is the radius at which the S\'ersic function of the bulge,
and the exponential function of the disk have equal surface
brightness.  We do not see a significant offset in either $n_b$ or
bulge size.  The distribution of S\'ersic indices in this sample
appears to be very similar to that of
\cite{fisherdrory2008,fisherdrory2010}; the bimodality in bulge
S\'ersic index is clearly visible for barred and unbarred galaxies. In
bulge size there is a slight preference for unbarred bulges to be
larger fraction of the optical radius. But as we state above, this is
likely an unavoidable physical phenomenon.  We will discuss below that
this difference does not bias our CO surface brightness measurements.
In our sample, bulges in unbarred galaxies are typically 1.5$\times$
the sizes of bulges in barred galaxies, when normalized either by disk
scale-length or $R_{opt}$.  

A few pseudobulges have very small S\'ersic indices ($n<1$). They are
NGC~0925, 2403, 2903. \cite{gadotti2008bars} shows that bars can have
very low S\'ersic index. It is therefore possible that one could
misidentify a bar as a pseudobulge, with very low S\'ersic index.  In
NGC~0925 and NGC~2903 we detect very clear, prominent large-scale
bars in the stellar light distribution. They are masked from the
fit. Therefore, in these galaxies the bulge is not a misidentified
large-scale bar. In NGC~2403 we do not identify a bar. The galaxy is
flocculent, and morphologically similar to M~33. The bulge in NGC~2403
is small. It has $B/T~0.05$ and the radial size of the bulge of is
roughly 4\% of the disk scale length.  Also, the isophotes of NGC~2403
do not become flatter (that is more bar like) in the bulge region.  We
feel this small size and isophote shape likely indicate that this is a
bulge and not a bar.
 
{\em Nuclear Point Sources:} We do not include nuclear star clusters
in the fit. Similar to bars they are masked. We do not attempt to
model the PSF of instrument in the surface brightness profile. We
simply do not attempt to fit isophotes that are smaller than the beam
size of the instrument, and if a nuclear point source is identifiable
we do not fit that region either. \cite{fisherdrory2008} describe in
detail the uncertainty introduced in this method. Similar to removing
a bar, removing the isophotes of a nuclear point source is an
iterative process in which we try to minimize systematic deviations in
 the residual profile. 

{\em Rings, Warps and Tidal Features:} In our sample 14 galaxies show
signs of interaction. In some cases the interaction is significant
enough to manifest as perturbations in the surface brightness profile
at large radius. Also galaxies can contain large scale rings that are
not exponential in a surface brightness profile.  Our decomposition
method assumes that the surface brightness profile becomes exponential
at large radius. Violations to this assumption are well documented
\citep[eg.][]{vanderkruit1981,k82,erwin2005}.  A few examples in which
this occurs in our sample are NGC1637, NGC5371,NGC3953. In cases where
the galaxy clearly does not agree with this assumption we truncate the
fit at a radius shorter than the disturbing feature. As with bars and
nuclear point sources the exact radius is chosen through iteratively
refitting.

\subsection{Pseudobulge Identification}

Identifying bulges as pseudobulges or classical bulges remains a topic
of ongoing research. A common method is to base the bulge
classification on the morphology of the broadband optical emission at
high spatial resolution \citep[for detailed descriptions
see][]{kk04,fisherdrory2008}. In this method, it seems quite plausible
that the spiral will be more easily identified in a bulge with more
dust. In this paper, we are specifically studying the distribution of
gas, therefore identifying bulges based on morphology may 
introduce a circular selection criteria.

\cite{fisherdrory2008,fisherdrory2010} show that the S\'ersic index,
from Eq.~\ref{eq:decomp}, fitted to near infrared surface photometry
offers a means of identifying pseudobulges and classical bulges that
is less affected by the dust and gas content of the bulge.  A typical
disk has a roughly exponential decline, and therefore has $n=1$ where
as a typical elliptical galaxy has larger S\'ersic indices
$n=2-10$. The bulges in \cite{fisherdrory2008} show a bimodal
distribution of S\'ersic indices, and the minimum of the distribution
is at $n=2.1$.  They also show that 90\% of bulges with disk like
morphology have $n<2$. Also \cite{fisherdrory2010} show that bulges
with $n<2$ are not on the same scaling relations as elliptical
galaxies (such as size-luminosity) but those with larger S\'ersic
index are on those correlations.  \citep{fabricius2012} show that
using the S\'ersic index to identify bulge, agrees with dynamical
methods to identify pseudobulges. Though more work on the field of
pseudobulge identification is certainly needed it appears that the
S\'ersic index is able to distinguish physically different
objects. Lower S\'ersic index bulges are, on average, lower velocity
dispersion systems and establish different scaling relations between
size, luminosity and surface density.

\subsection{Stellar Masses}

We estimate the mass-to-light ratio separately for bulges and disk to
account for variation in the stellar populations and extinction within
the galaxy. We use multicolor imaging to estimate the mass-to-light
ratio in the bulge separately from the disk.  We use two separate
methods to determine the mass-to-light ratio, depending upon available
data. All of our sample galaxies have data available in 2MASS, and 42
galaxies are covered in the {\em Sloan Digital Sky Survey} (SDSS) data
release 8. The 42 galaxies with available SDSS data allow us to carry
out direct SED fits to determine the mass-to-light ratio.  Note that
all magnitudes are corrected for Galactic extinction using values from
\cite{schlegel}, and extinction curves from \cite{cardelli1989}.

\begin{figure}[t]
\begin{center}
\includegraphics[width=0.49\textwidth]{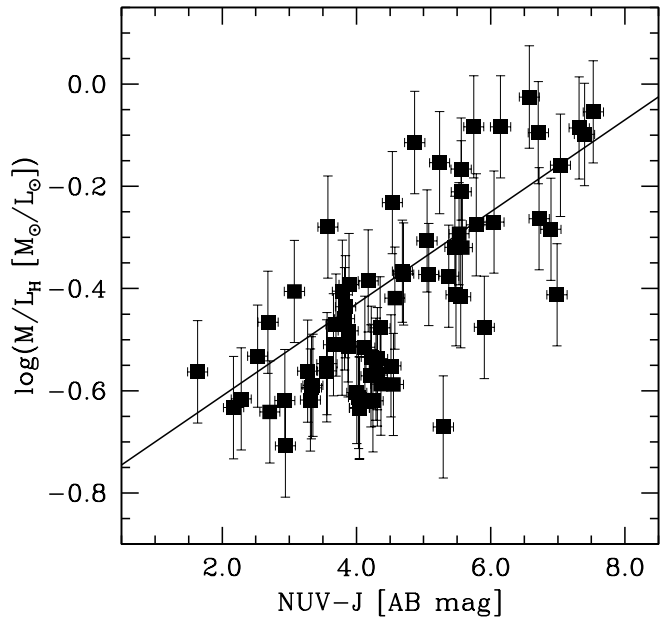}
\end{center}
\caption{The correlation between NUV-J color and mass-to-light ratio
  for galaxies in our sample. We use the NUV-J color to determine
  mass-to-light ratio for those galaxies that do not have SDSS
  data. \label{fig:m2l}}
\end{figure}
For the galaxies covered in SDSS we measure the stellar mass directly
from fitting \cite{bruzualcharlot} models to the multicolor
(ugriz-JHK) photometry. The models include composite stellar
populations of varying star formation history, age, metallicity,
burst fractions, and dust content, and assume a Salpeter initial mass
function. The uncertainty in the fit to these models is typically
$\sim$0.1~dex.  For a detailed description of the model see
\cite{drory2004}.

Not all galaxies in our sample are covered in SDSS. We therefore must
choose a means of estimating the mass-to-light ratio that is available
on all galaxies. Sources of data that are widely available on nearby
galaxies include near-IR colors (J-K), a combination of GALEX and
2MASS data (NUV-J), and RC3 (B-V)$_T$ total galaxy colors. To
determine which metric is best, we measure the correlation for those
galaxies with mass-to-light ratios determined with SDSS data. We find
that $NUV-J$ has the smallest scatter and the strongest correlation
coefficient. This correlation is shown in Fig.~\ref{fig:m2l}.  We find
a relationship such that,
\begin{equation}
log(M/L) = 0.10\times(NUV-J)-0.89.
\end{equation}
The relationship between $NUV-J$ and $M/L$ is shown in
Fig.~\ref{fig:m2l}. A Pearson's correlation coefficient for this
relationship is r=0.74.  From our SDSS sample, we determine that the
scatter in mass-to-light ratio using the NUV-J color is 0.1~dex.

 For the $H$-band flux used to measure the bulge mass, we use
the total flux in the same region that the FUV and 24~$\mu$m are
measured. To be clear this is not the flux from a S\'ersic fit. Using
the flux from a S\'ersic function returns masses that are
systematically large by roughly 20\%. 

For each galaxy we note which method was used to determine the stellar
mass in Table 2 \& 3. In our sample 42 galaxies have mass-to-light
ratios determined with full optical-near IR SED, and 18 are determined
only with the single color. The uncertainty in the final stellar mass
is determined by adding the uncertainty from the bulge-disk
decomposition with the uncertainty in the mass-to-light ratio. We use
the typical uncertainty of 0.1~dex for galaxies with the full SED fit and 0.15~dex
for the NUV-J determined stellar masses.

\subsection{Global Star Formation Rates} 
We calculate the star formation rate (SFR) by combining $GALEX$ $FUV$
and $Spitzer$~24~$\mu$m fluxes. The reason for combining $FUV$ and
24~$\mu$m data to trace the SFR is fairly straightforward. The
emission of young O and B stars heavily dominates the wavelength range
covered by the $FUV$ filter. However, this very blue light is heavily
affected by extinction. The emission at 24~$\mu$m around a star
forming region, is mainly from hot dust grains. For a more detailed
discussion of these processes we refer the reader to
\cite{calzetti1995,kennicutt98,buat2002,calzetti2007,kennicutt2009}. Indeed,
both $FUV$ and 24~$\mu$m luminosity have been shown to strongly
correlate with SFR independently
\citep[respectively][]{salim2007,calzetti2007}.  The method of
combining UV and IR data has been used by many authors \citep[for a
detailed discussion see][]{leroy2012}. A very practical reason for
this method is the large number of galaxies with available data in
Spitzer and GALEX archives, and these sources are well matched in
resolution to typical interferometric maps of CO(1-0).

When available we use the high quality maps from the SINGS survey
\citep{sings} otherwise we download the post-basic calibrated data
from the Spitzer Archive\footnote{available at
  http://sha.ipac.caltech.edu}. Similarly, when available we use $FUV$
data from the $GALEX$ Nearby Galaxies Survey \citep{galexngs},
otherwise we take the deepest available $FUV$ image of each
galaxy. Images are then background subtracted, in a similar way to the
sky subtraction described in the photometry subsection. That is we fit
a surface to parts of the image that are not affected by galaxy
emission. We measure fluxes in $FUV$ and $24$~$\mu$m bands for the
region where the bulge dominates the starlight, and also the total
galaxy. We estimate disk flux as the total minus the bulges,
$F_{disk}=F_{total}-F_{bulge}$.
\begin{figure}[t]
\begin{center}
\includegraphics[width=0.49\textwidth]{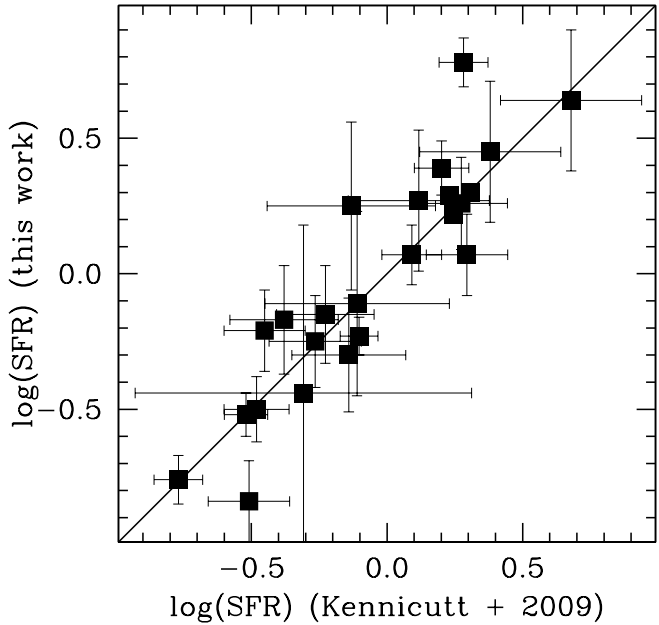}
\end{center}
\caption{The correlation between our FUV+24~$\mu$m based star formation
  rates and those from \cite{kennicutt2009} based on
  H$\alpha$+24~$\mu$m emission.  \label{fig:sfr}}
\end{figure}

The stated assumptions that (1) the star formation rate can be
broken down into two components of emission, unobscurred and obscured
star formation, and (2) these components can be
traced, respectively, by the FUV and 24~$\mu$m luminosities, is
quantified with the following equation,
\begin{equation}
 SFR = a\times L(FUV)+b\times L(24). \label{eq:sfr}
\end{equation}
Where we define the luminosities as $L(FUV)\equiv\nu_{fuv}L_{fuv}$ and
similarly for IR, $L(24)\equiv\nu_{24~\mu m}L_{24~\mu m}$. The $a$
coefficient relating the FUV emission to the unobscurred star
formation is taken from \cite{salim2007}; they find $SFR(UV) =
3.44\times10^{-44} L(FUV)$, where SFR is in $M_{\odot}~yr^{-1}$ and
L(FUV) is in erg~s$^{-1}$. To determine the coefficient relating
obscured star formation to the L(24) we solve $b = (SFR-a\times
L(FUV))/L(24))$. For this we take the data from \cite{kennicutt2009}
to use as a control set of star formation rates, we find $<b> =
1.0\pm0.6\times 10^{-44}$. In our sample 3 galaxies did not have FUV
data, for these galaxies, we use the 24~$\mu$m emission alone from
\cite{calzetti2007}. Fig.~\ref{fig:sfr} we compare our SFR to those
from \cite{kennicutt2009}, for the 24 galaxies which exist in both
samples. The solid line indicates the line of equality (y=x), the two
are strongly correlated with Pearson's correlation coefficient of
r=0.9. The scatter in the correlation is $\sim$0.2~dex. Therefore, our
uncertainty for star formation rates will be determined by adding
0.2~dex with measurement uncertainties of 24~$\mu$m and FUV flux. In
the case of bulge star formation rates, a significant source of
uncertainty comes from the relative beam size of GALEX and MIPS to the
size of bulges. We estimate this simply by determining the flux in
regions $\pm$1 beam sizes, the uncertainty from beam sampling is
typically a few percent of the total flux. The beam size of Spitzer
24~$\mu$m and GALEX FUV are both 6\" roughly. Two galaxies, NGC~4273
\& 4654, have bulges that are smaller than the beam size of these
instruments. In these galaxies we take all fluxes to be the size of
the MIPS beam.

\subsection{Estimating Star Formation Rates of Bulges} 
The luminosity we measure in Spitzer and GALEX images can be thought
of as a superposition of emission from both young stars and old
non-star forming populations, as follows,
\begin{equation} 
L_{24} = L_{24}(old) + L_{24}(young) \label{eq:24}
\end{equation}
\begin{equation}
L_{FUV} = L_{FUV}(old) + L_{FUV}(young) \label{eq:fuv}
\end{equation}

The old stars emit some light in the UV, hence $L_{FUV}(old)$. Also
the old stellar populations could heat dust causing it to reradiate,
hence $L_{24}(old)$. This is some times referred to as ``diffuse
emission'' or ``cirrus emission.''  The determination of the star
formation rate in Eq.~\ref{eq:sfr} relies upon the assumption that the
FUV and 24~$\mu$m flux is coming from, or at least strongly dominated
by, young stellar populations. At the least star formation rate
indicators that are calibrated to unobscured emission for HII regions
using Pa$\alpha$ lines, \citep[e.g.][]{kennicutt2009} implicitly
account for deviations from this assumption. Furthermore, the total
flux from a galaxy in FUV and 24~$\mu$m is likely to be heavily
dominated by emission from star forming regions.  Therefore, global
star formation rates, likewise include a correction for this.
However, measuring the star formation rate for substructure within
galaxies may result in specious estimations of the star formation
rate. The problem of diffuse emission affecting star formation rate
indicators is discussed in numerous papers, for recent discussion see
\cite{leroy2012,kennicuttevans2012arxiv}.

This phenomenon is likely stronger in bulges because they are
systematically higher in stellar surface density than the outer
disk.
In the most extreme case some bulges in the nearby Universe, for
example M~31, are very old \citep{saglia2010} and may be completely
empty of young star forming regions \citep{groves2012}. Although this
is clearly not the case for all bulges.  Those containing nuclear
rings, like NGC~3351, show clear signs of active star formation that
is easily identifiable in NICMOS Pa$\alpha$
maps.  

\begin{figure}[t]
\begin{center}
\includegraphics[width=0.49\textwidth]{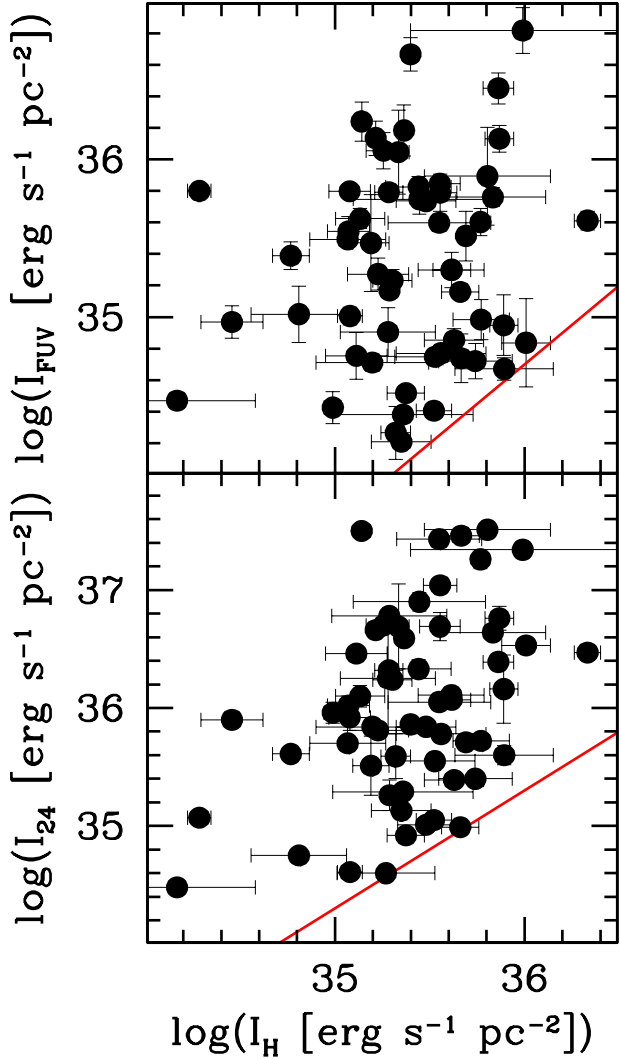}
\end{center}
\caption{ Comparison of surface brightness of bulges in H-band to the
  surface brightness at FUV (top panel) and 24~$\mu$m (bottom
  panel). The redline represents the adopted corrections. The units
  in all axes are erg~s$^{-1}$~pc$^{-2}$.  \label{fig:huv24}}
\end{figure} 

\cite{leroy2012} estimate the diffuse radiation field through models
of dust emission. Such an approach is grounded in physically based
arguments, but requires a significant amount of ancillary data that is
not always available for large samples of galaxies, or at adequate
spatial resolution to study bulges.  In this paper, we adopt an {\em
  ad hoc} approach to estimating the contribution of old stars in
bulges.  Our method for estimating the flux from old stars is based on
the assumption that at least some of the bulges in our sample are
dominated by emission from old stellar populations, $L_{24}(old) >>>
L_{24}(young)$ and $L_{FUV}(old) >>> L_{FUV}(young)$.  In these bulges
we are assuming that the 24~$\mu$m emission is due to dust grain
heating from evolved stellar populations.  In this limit we then make
the approximation that $L_{24} \sim L_{24}(old) \propto L_{H}$ (and
similarly for FUV), where $L_{H}$ is the luminosity of the same region
in $H$-band.  The bulges with the lowest values of $L_{24}/L_{H}$ and
$L_{FUV}/L_{H}$ are then used to determine the correction for the
24~$\mu$m and FUV fluxes.  
\begin{figure}[t]
\begin{center}
\includegraphics[width=0.49\textwidth]{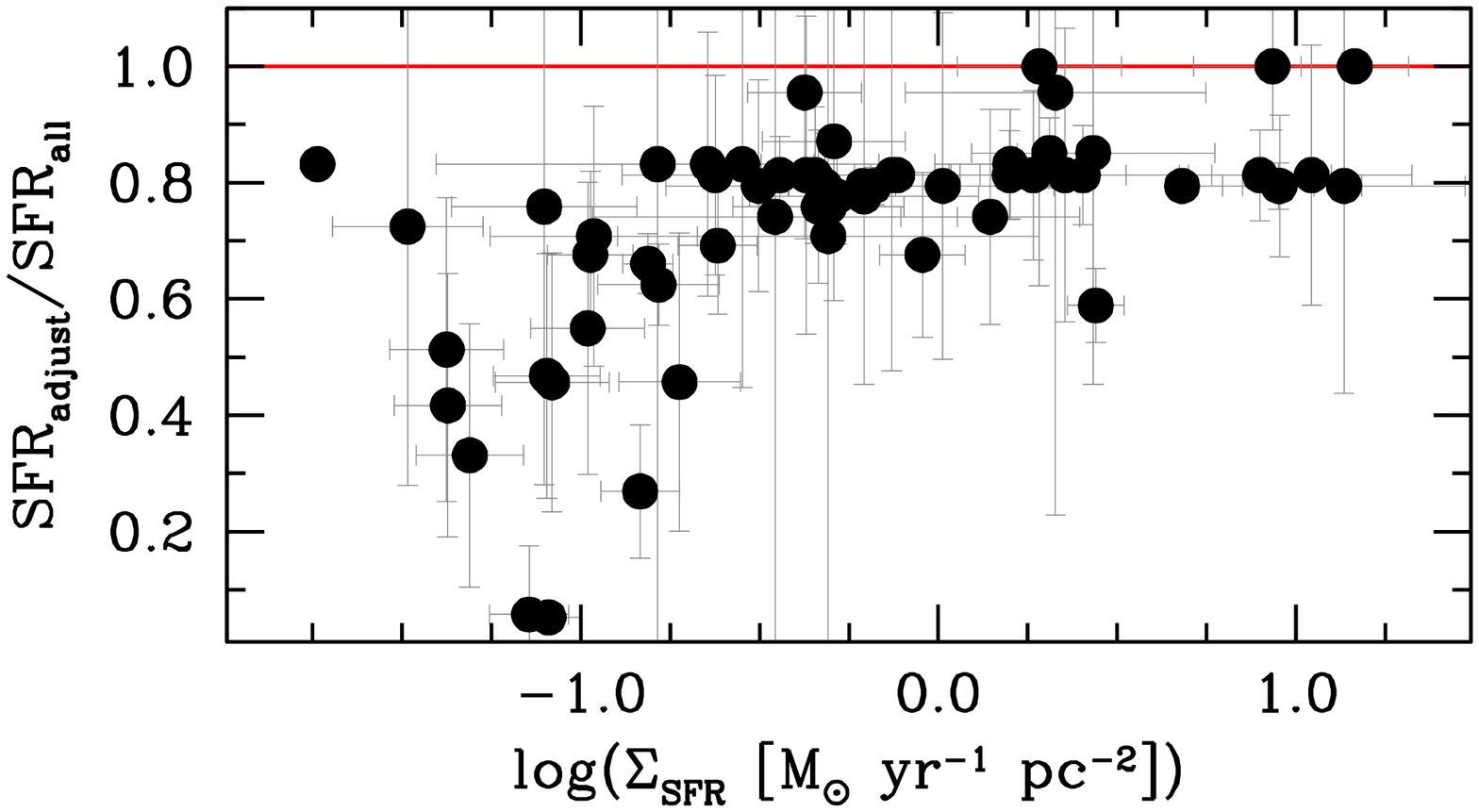}
\includegraphics[width=0.49\textwidth]{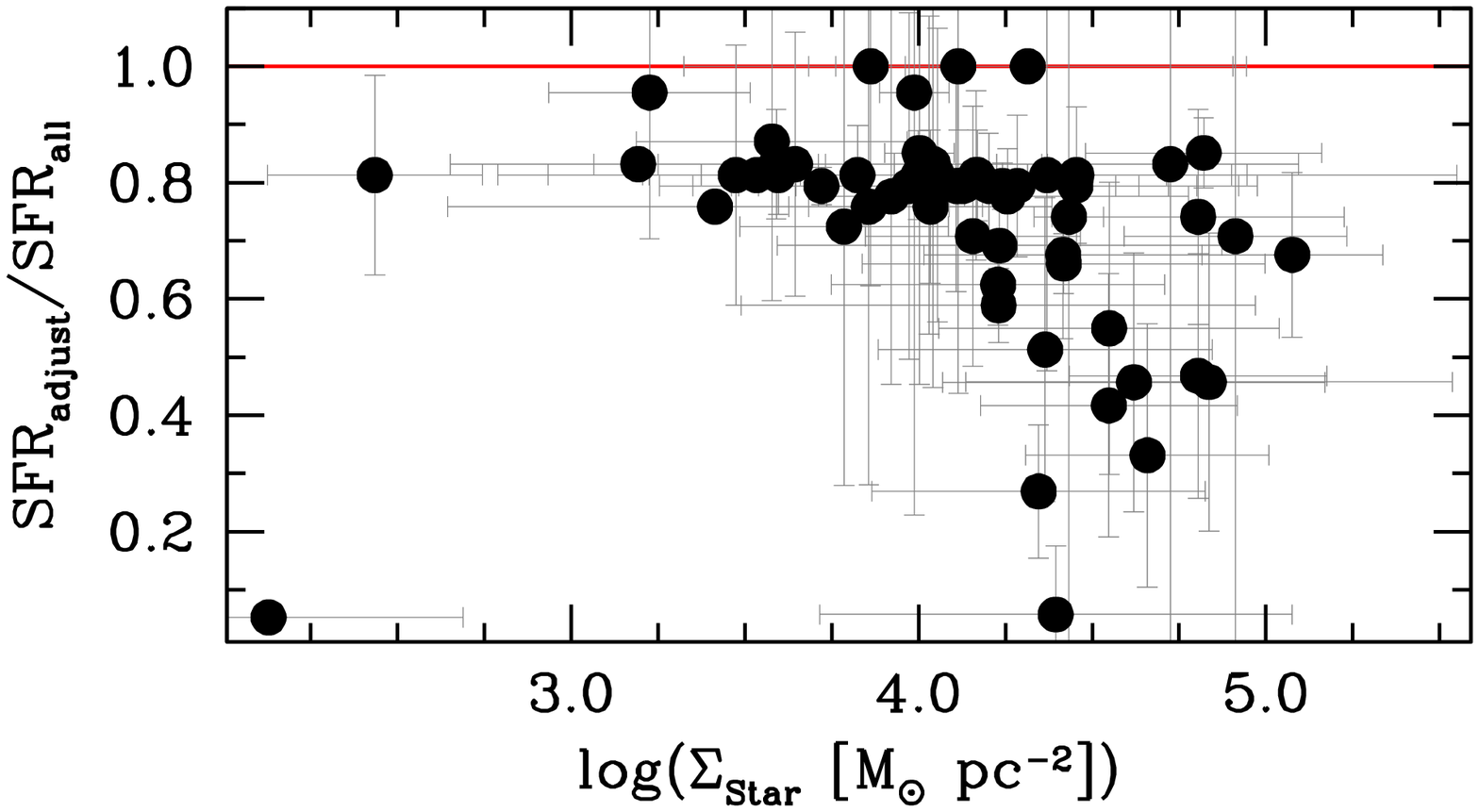}
\end{center}
\caption{Ratio of star formation rates that are corrected for emission
  from old stars to the same quantity that has not been corrected is
  plotted against the surface density of star formation, measured with
  uncorrected emission.  The bottom panel shows the same ratio plotted
  against stellar surface density. The red line indicates a ratio of
  unity.  \label{fig:sfrcomp}}
\end{figure} 

In Fig.~\ref{fig:huv24} we plot the surface brightness of bulges in
FUV ($I_{FUV}$, top) and 24~$\mu$m ($I_{24}$, bottom) against the
surface brightness in $H$-band, $I_{H}$. These figures are intended to
illustrate the magnitude of the correction in each band.  For the
$H$-band flux we use the total flux in the same region that the FUV
and 24~$\mu$m are measured. To be clear this is not the flux from a
S\'ersic fit.  We find that the 24~$\mu$m luminosity is never lower
than 20\% of $L_H$, and the FUV luminosity is never less than 5\%
(units are erg~s$^{-1}$). Red lines indicate these ratios in
Fig.~\ref{fig:huv24}. Average flux ratios for bulges in our sample are
$<L_{24}/L_H> \sim 4$ and $<L_{FUV}/L_H> \sim 0.8$.  We then invert
Eqs.~\ref{eq:24}~\&~\ref{eq:fuv}, and substitute in the estimates of
the old stellar populations
\begin{equation} 
L_{24}(young) = L_{24} - 0.2\times L_H \label{eq:24c}
\end{equation}
\begin{equation}
L_{FUV}(young) = L_{FUV}-0.05\times L_H  \label{eq:fuvc}.
\end{equation}

We now calculate the SFR of a bulge the same as total galaxies,
however, we now use corrected fluxes from Eqs.~\ref{eq:24c} \&
\ref{eq:fuvc}.  The corrections in Eqs.~\ref{eq:24c} \& \ref{eq:fuvc}
are intended to be taken as rough estimates. A better approach would
include detailed modeling of both the optical SED (preferably from IFU
data) and the IR SED including Spitzer \& Herschel data; such analysis
is beyond the scope of this work. We will investigate the effect of
using corrected and uncorrected fluxes on the star formation law later
in this work.

In Fig.~\ref{fig:sfrcomp} we compare the star formation rates
determined from the total flux at FUV \& 24~$\mu$m to those corrected
emission from old stars.  In the top panel, we plot the ratio of star
formation rates against the star formation rate density from the
uncorrected flux. We find that in bulges the median ratio of corrected
to uncorrected star formation rate is 80\%. At low star formation rate
surface densities the correction becomes much larger. There appears to
be a break near $\Sigma_{SFR} \sim 0.25~M_{\odot}~yr^{-1}~kpc^{-2}$,
below which the $FUV$ and $24$~$\mu$m flux is more heavily affected by
old stars.  In the bottom panel, we plot the ratio of star formation
rates against the stellar surface density of bulges. As one would
expect, the emission from old stars becomes a larger fraction at
higher stellar mass densities. However, it is notable that having a
large stellar mass surface density does not necessarily mean that the
flux in Eqs.~\ref{eq:24}~\&~\ref{eq:fuv} is dominated by old stars.
There are several bulges with high stellar surface densities that are
still dominated by young stars.  We remind the reader, that the
correction of almost 100\% of the flux in two galaxies is set by
definition.

The galaxies in which the correction to the SFR is more than 50\% are
NGC~2841, 3521, 7217, 4725, 3031, 3953, 3992, 3486. 6 of 8 are
classical bulges (larger S\'ersic index). NGC3953 is a pseudobulge,
but the galaxy is very gas poor, in fact the CO(1-0) bulge flux is not
measurable, that is it is an upperlimit. NGC~3486 is a very low
surface brightness galaxy, which is also relatively gas poor. These
galaxies preferentially occupy the low star formation - low gas
density region of the correlation between gas and star formation rate
density.  Using the corrected star formation rates these galaxies are
not outliers in the $\Sigma_{SFR}-\Sigma_{mol}$ relationship, and have
a similar average depletion time ($\Sigma_{mol}/\Sigma_{SFR}$) as the
rest of the sample. However using uncorrected star formation rates,
they have depletion times that are significantly shorter.

\begin{figure}[t]
\begin{center}
\includegraphics[width=0.49\textwidth]{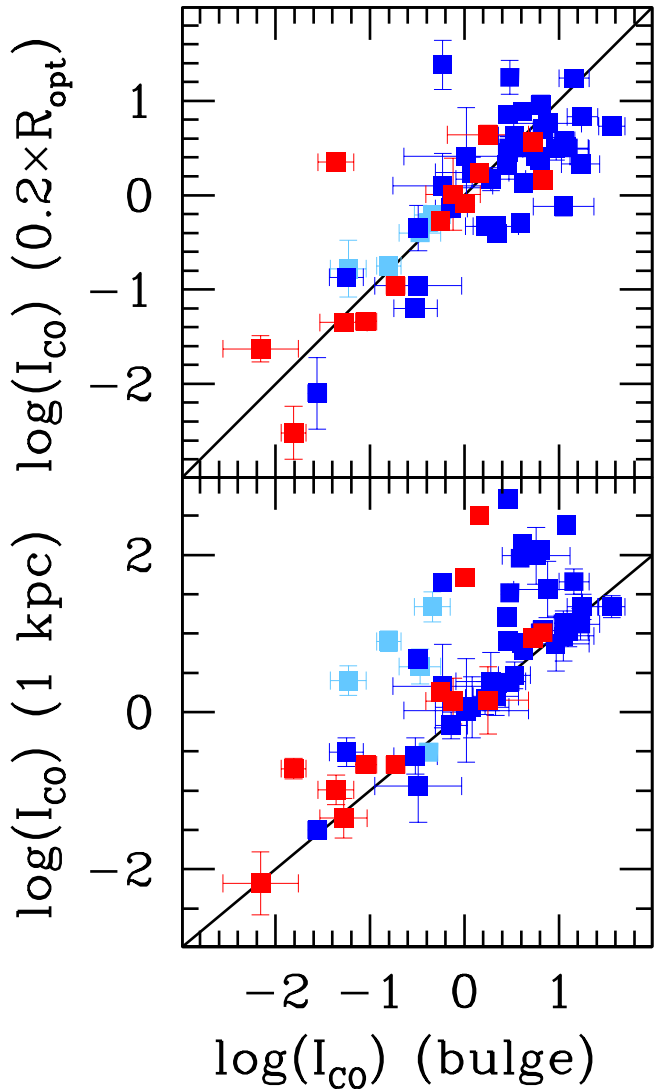}
\end{center}
\caption{The comparison of CO surface brightness measured in the
  central 1~kpc to that measured in the region where the bulge
  dominates the light. The units of all axes are
  log(Jy~km~s$^{-1}$~arcsec$^{-2}$) \label{fig:kpccomp}}
\end{figure}

\subsection{Molecular Gas Masses} 
\begin{figure*}[t]
\begin{center}
\includegraphics[width=0.99\textwidth]{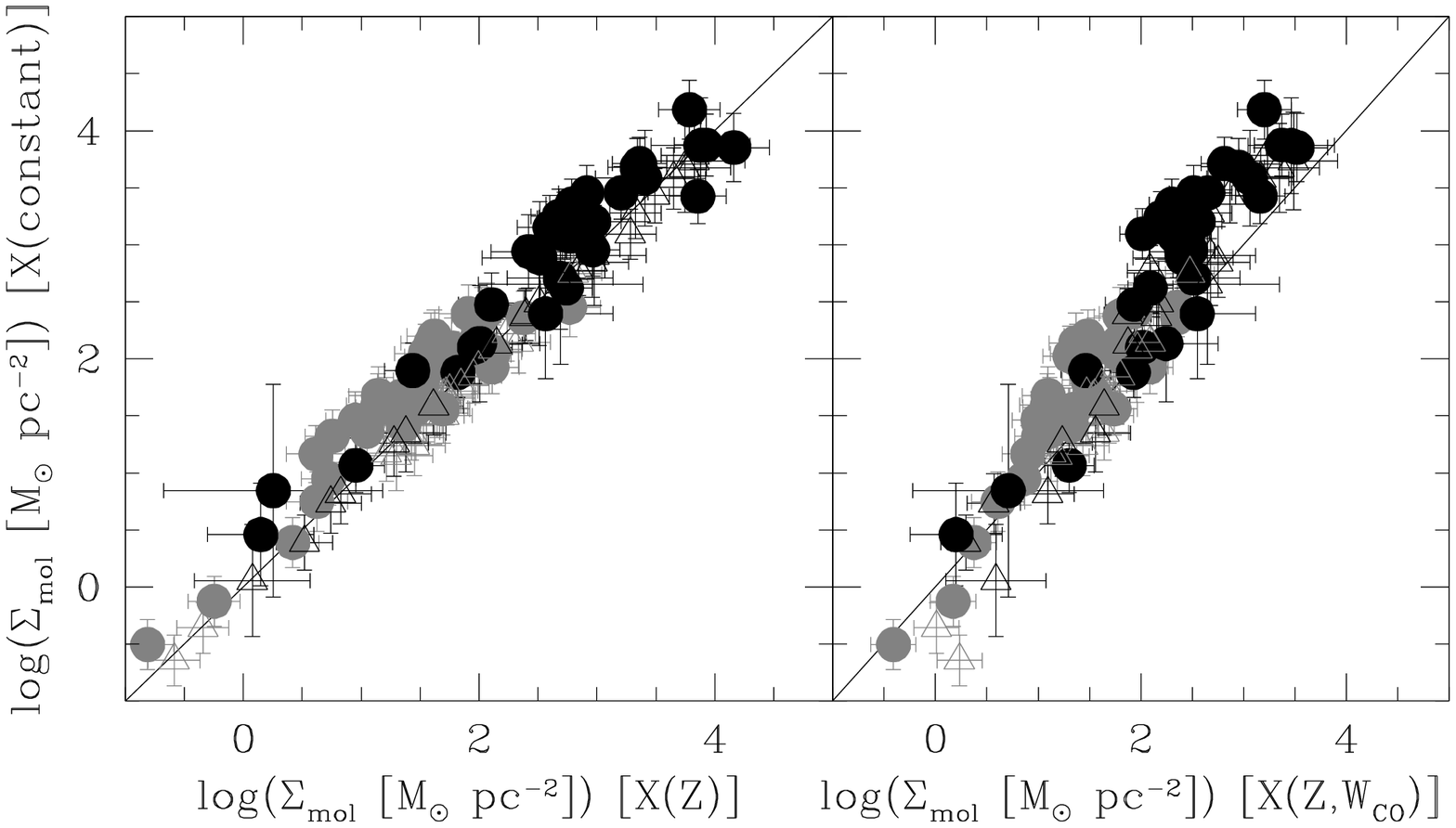}
\end{center}
\caption{We compare gas densities derived from a constant $X_{CO}$ to
  the metallicity dependent conversion factor of CO(1-0) to molecular
  gas (left), and also the conversion factor which depends on both
  metallicity and CO brightness temperature. Black symbols represent
  the gas densities of bulges, grey symbols represent the gas density
  of disks. Closed symbols represent those galaxies with published
  metallicities, open symbols represent those in which the metallicity
  is derived from the mass-metallicity relationship.  \label{fig:xco}}
\end{figure*}
As we state above our sample is the combination of 3 recent
interferometric surveys of CO(1-0) emission from nearby
galaxies. They are the BIMA SONG, the CARMA STING and the PdBI
NUGA. Here we will briefly describe observations of each survey, and
then how the CO(1-0) fluxes are converted to molecular hydrogen gas
masses.

{\bf BIMA SONG:} For a detailed description of SONG data and data
reduction see \cite{helfer2003}. The SONG maps CO(1-0) emission in 44
nearby spiral galaxies using the BIMA interferometer and the NRAO~12~m
single dish telescope at Kitt Peak. The aim of SONG was to map CO(1-0)
on a representative set of galaxies. The selection criteria included
$V_{Hel}<2000~km~s^{-1}$, declination higher than $-20^{o}$, not too
highly inclined ($i<70^{o}$) and brighter than $B_T<11$. Typical beam
sizes of SONG maps were 6'' with robust fields of view 3' across. The
data were taken in 10~km~s$^{-1}$ channels, with typical noise of
$\sim$58~mJy~beam$^{-1}$. The SONG maps are publicly available in the
NASA/IPAC Extragalactic Database.

{\bf CARMA STING} The CARMA STING maps CO(1-0) emission from 23 nearby
spiral galaxies. The CARMA STING survey aims to map the CO(1-0)
emission out to one-quarter to one-half of their optical radii.  The
sample is composed of northern ($\delta > -20^{o}$), moderately
inclined (i~<~75~deg) galaxies from the IRAS Revised Bright Galaxy
Sample within 45~Mpc \citep{sanders2003}. The galaxies were selected
to uniformly sample 10 mass bins distributed between M$_{star}$=$10^9$
and 3$\times10^{11}$M$_{\odot}$. The STING maps are robust over a
diameter of 2', and have typical beam sizes of 3-5 arcseconds. STING
maps employ a similar velocity masking and weighting techniques as
discussed in \cite{helfer2003}. The STING data are publicly available
on a dedicated website
\footnote{http://www.astro.umd.edu/rareas/lma/research/alberto\_bolatto/galaxies.html}. The
STING maps have not been merged with single dish measurements,
therefore we refrain from measuring flux scales larger than 1
arcminute. The typical sensitivity of STING maps is 22~mJy~beam$^{-1}$
in 10~km~s$^{-1}$ channels. STING galaxies were consciously chosen to
not have CO(1-0) maps, and therefore represents a significant increase
in sample size.

{\bf PdBI NUGA} The NUGA survey is described in \cite{nuga}. The
survey consists of CO(1-0) maps of 12 nearby galaxies within a
distance of 40~Mpc. The NUGA survey specifically targets galaxies with
low-luminosity AGNs. All NUGA observations are made on the PdBI. NUGA
maps have the highest resolution in the sample (0.5-1''). The sample
overlaps significantly with SONG \& STING, the finer resolution helps
in studying the radial distribution of $^{12}$CO. The NUGA maps used
in this paper have not been merged with IRAM~30~m data, therefore we
refrain from measuring flux at scales greater than 45''. NUGA maps
typically have sensitivities 1-5~mJy beam$^{-1}$.

For each galaxy we simply integrate the CO(1-0) emission inside the
region where the bulge dominates the light. For those galaxies without
bulges, we integrate to 20\% of the disk scale-length, which gives a
comparable size of bulges, especially pseudobulges
\citep{courteau1996,fisherdrory2008}. In one galaxy, NGC~2403, the
bulge diameter is smaller than the beam size of the CO map, in this
galaxy we use the resolution of the 6\" as the bulge radius in all
calculations, so that it is resolved in CO, 24~$\mu$m and
FUV. Uncertainty in CO(1-0) flux is determined from a combination of
the sensitivity of the map over the area being measured and the error
introduced in the beam size of the image. We are able to check that
our method of measuring the flux returns sensible results by measuring
the flux in 1~kpc diameter apertures for the SONG sample, as this has
been published in \cite{sheth2005}. Differences between our fluxes and
those of \cite{sheth2005} is typically smaller than the uncertainty in
the measurements. We divide the mass in the bulge region by the area
of the bulge to compute the gas surface density.

The observed difference between the size of classical bulges and
pseudobulges is not enough to account for the difference in we measure
CO surface brightness. In this paper we are specifically interested in
the amount of gas that is inside the radius where bulges dominate the
light.  We therefore measure the gas mass in the same region in which
the surface brightness profile of stars shows a bulges in the
bulge-disk decomposition. Classical bulges are, on average, larger
than pseudobulges \citep[see discussion above,
also][]{gadotti2009,fisherdrory2008,fisherdrory2010}, therefore it is
possible that classical bulge surface density may be lower simply
because we are averaging over larger regions.  In
Fig.~\ref{fig:kpccomp} we make a straightforward comparison of the CO
surface brightness inside of bulges to that of the central kiloparsec
(bottom panel), and to that measured inside $0.2~R_{opt}$ (top panel).
The solid line represents the line of equality.  We find that there
are not significant differences between the surface density of bulges
and those of other commonly used, or physically motivated regions
methods of defining the galaxy center.

A common way to measure the central gas density is to choose a fixed
size in parsecs in the center of galaxy
\citep[e.g.][]{sheth2005,komugi2008}. This has the advantage of being
more homogeneous, however is less physically motivated.  Recently,
multiple studies \citep{schruba2011,bigiel2012} show that the
transition between the exponential profile and the inner profile
occurs roughly near 20-30\% of the optical radius. (R$_{opt}$ is the
radius at which the $B$ band surface brightness profile reaches
25~mag~arcsec$^{-2}$.)  Bulges are typically smaller than
$0.2~R_{opt}$, the median bulge radius, in our sample, is
$0.06~R_{opt}$. Only two bulges are larger than
$0.2~R_{opt}$. Therefore, our measurements are inside the range were
the assumption of exponential decay of gas surface brightness profiles
is not valid.

In Fig.~\ref{fig:kpccomp} a few bulges have higher CO surface
brightness in the central kiloparsec than the entire bulge, but these
are pseudobulges. Note that for simplicity we do not include bulges
with upperlimits in this figure.  Despite being larger on average,
classical bulges (red points) do not show a systematic trend, they do
not deviate more than pseudobulges (blue) or the centers of bulgeless
galaxies (light blue). Based on these results we feel that the
differences we discuss later between the surface density of classical
and pseudobulges is likely real, and not an effect of the radius we
choose.

Also, the slight differences in sizes of pseudobulges and classical
bulges (described in the {\em Bulge-Disk Decomposition Section} ) are
not large enough to account for the differences in CO surface
brightness.  As we state above, classical bulges are larger by a
factor of 1.5-2$\times$ the size of pseudobulges (bulges in unbarred
disks are larger by a similar factor, so the same argument can be
applied to them).  If a galaxy had all the CO flux coming from a
radius $R<R_{pseudo}$, and we used the radius $R_{classical}\sim
2\times R_{pseudo}$ then the surface brightness would be lower by a
factor of 4$\times$.  Examination of Fig.~\ref{fig:kpccomp} is clear
that the typical difference between classical bulge and pseudobulge CO
surface brightness is much higher than this, and is typically an order
of magnitude. Based on this we feel that size difference alone is not
likely accounting for the difference between classical bulges and
pseudobulges.

Another check is simply to investigate the shape of the CO surface
brightness profiles. We plot all CO surface brightness profiles in
Fig.~\ref{fig:prof}. We will discuss the surface brightness profiles
in more detail later.  Using profiles we can ask if this hypothetical
case ever exists.  Examination of the profiles suggests that it does
not.  Pseudobulge and classical bulge CO surface brightness profiles
are different. Classical bulges frequently have holes in the center of
the CO map (and show their peak CO surface brightness in the disk);
pseudobulges are frequently peaked in the center of the galaxy. The
few classical bulges that are centrally peaked all have either a
strong bar or are interacting. This suggests a physical difference. We
therefore take both of these results in Figs.~\ref{fig:kpccomp} \&
\ref{fig:prof} to indicate the differences we observe between the
surface density of molecular gas in bulges are not simply due to
classical bulges being systematically larger.

For most of our sample we are using interferometric maps that do not
have single dish fluxes, to estimate the CO flux in the bulge. These
maps are therefore insensitive to very slowly varying
emission. However, \cite{sheth2005} shows that in BIMA SONG galaxies
the median difference for the central surface brightness measured with
and with out single dish data is roughly 5\%. Therefore, we expect
this difference to have little effect on our results.

Next we must convert the CO(1-0) emission into molecular hydrogen
masses. For review discussions of the conversion of CO(1-0)
emission to molecular
gas mass see \cite{blitz2006,kennicuttevans2012arxiv}.

Typical conversions of CO(1-0) fluxes to the mass in molecular
hydrogen assume a constant conversion factor ($X_{CO}$), typical
values are $X_{CO}=1-4\times 10^{20}$~cm$^{20}$~(K~km~s$^{-1}$)$^{-1}$
\citep{youngscoville1991}. However, a growing body of evidence
suggests that $X_{CO}$ is not completely uniform from galaxy to
galaxy, or even within in a given galaxy. \cite{solomon1997} show that
the conversion factor varies as the square root of H$_2$ density
divided by the brightness temperature $X_{CO} \propto n^{1/2}/T_b$.
Recent work has shown that X$_{CO}$ varies with metallicity
\citep{leroy2011,bolatto2011,genzel2012}. Also, naively applying
typical values of $X_{CO}$ to ultra-luminous IR galaxies returns
molecular gas masses which exceed the virial mass
\cite[e.g.]{solomon1997}.  Similar results have been found in the
centers of nearby galaxies \cite{israel2009a,israel2009b,meier2010}.
A possible explanation for this phenomenon is that high densities of
gas or stars- a common feature of both luminous IR galaxies and
galactic centers- lead to higher temperatures in molecular clouds
which then leads to a lower $X_{CO}$ \citep{maloney1988}.  For more
discussion see \cite{narayanan2011} and \cite{tacconi2008}.  It is
likely that results from Herschel Space Observatory, and especially
the KINGFISH survey \citep{kingfish}, will prove enlightening on the
radial variations of $X_{CO}$ in typical spiral galaxies. Yet at this
point the exact nature of the relationship between $X_{CO}$ and galaxy
properties is tenuously known at best.

For this work it is possible that the variety in bulge and disk
densities makes application of a single $X_{CO}$ to all galaxies a
poor assumption. We therefore investigate the use two non-constant
functions for $X_{CO}$. First we use the metallicity dependent
$X_{CO}$ factor from \cite{genzel2012}, and secondly we use the
metallicity and CO surface brightness dependent $X_{CO}$ factor from
\cite{narayanan2011}. The purpose of this is to determine if
measurements of molecular gas mass in bulges and disks are sensitive
to the choice of CO-to-H$_2$ conversion factor.  Therefore to
calculate the molecular gas
mass we use the following 3 formulae \\
Constant $X_{CO}$:
\begin{equation}\label{eq:xc}
log(M_{H_2}) =  2\times log_{10}(d) + 4.041 + log_{10}(f_{CO}),
\end{equation}
where $d$ is the distance in $Mpc$ and $f_{CO}$ is the flux density in Jy~km~s$^{-1}$.
Metallicity dependent $X_{CO}(Z)$:
\begin{equation}\label{eq:xg}
log(M_{H_2}) = 2\times log_{10}(d) - 1.14\times Z_{KK} + 13.973 + log_{10}(f_{CO})
\end{equation}
Where $Z = log(O/H) + 12$, we convert the metallicity dependance of
\cite{genzel2012} to that of \cite{kobulnicky2004} according to the
transformations in \cite{kewley2008}.  \\
Metallicity and CO surface brightness dependent $X_{CO}(Z)$:
\begin{equation}\label{eq:xn}
log(M_{H_2}) = 2\times log_{10}(d)
+log_{10} \left ( \frac{W_{CO}^{-0.32}}{Z'^{0.65}} \right ) + 
 4.424 + log_{10}(f_{CO}),
\end{equation}
where $W_{CO}$ is the CO surface brightness in K~km~s$^{-1}$ and $Z'$
is the metallicity in solar abundance. Finally, we multiply all 

\begin{figure*}[t]
\begin{center}
\includegraphics[width=0.9\textwidth]{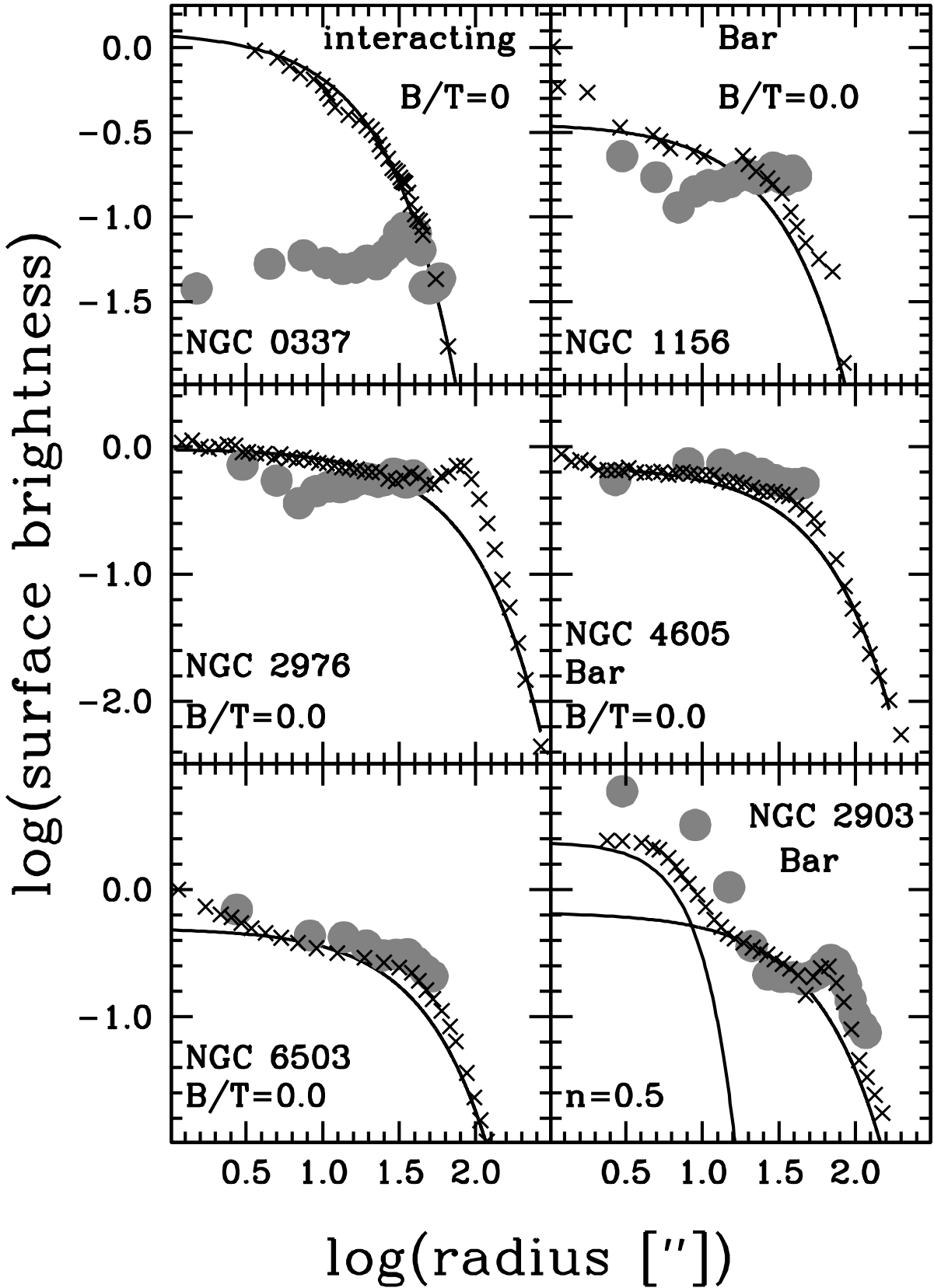}
\end{center}
\caption{The x's represent the H-band surface brightness; solid lines
  represent the bulge-disk decomposition, and filled circles represent
  CO surface brightness. The H-band
  profiles have been normalized by the surface brightness at 1''\ . The CO flux has
  been shifted to match the stars in the outer disk.  CO annuli are measured in steps
  equal to the beam size.   \label{fig:prof}}
\end{figure*}

\setcounter{figure}{9}
\begin{figure*}[t]
\begin{center}
\includegraphics[width=0.9\textwidth]{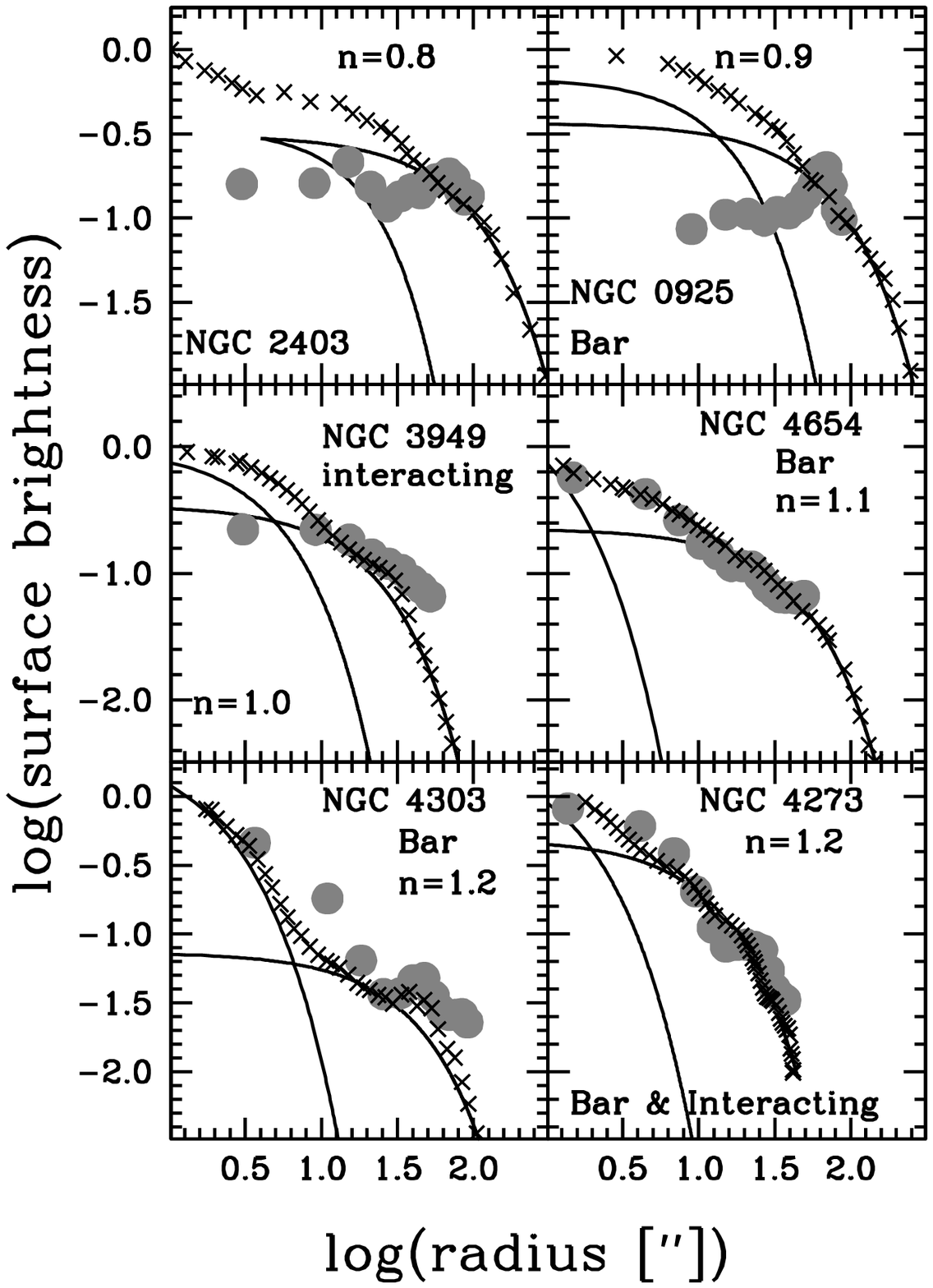}
\end{center}
\caption{The x's represent the H-band surface brightness; solid lines
  represent the bulge-disk decomposition, and filled circles represent
  CO surface brightness.  The H-band
  profiles have been normalized by the surface brightness at 1''\ . The CO flux has
  been shifted to match the stars in the outer disk. CO annuli are measured in steps
  equal to the beam size.    \label{fig:prof}}
\end{figure*}

\setcounter{figure}{9}
\begin{figure*}[t]
\begin{center}
\includegraphics[width=0.9\textwidth]{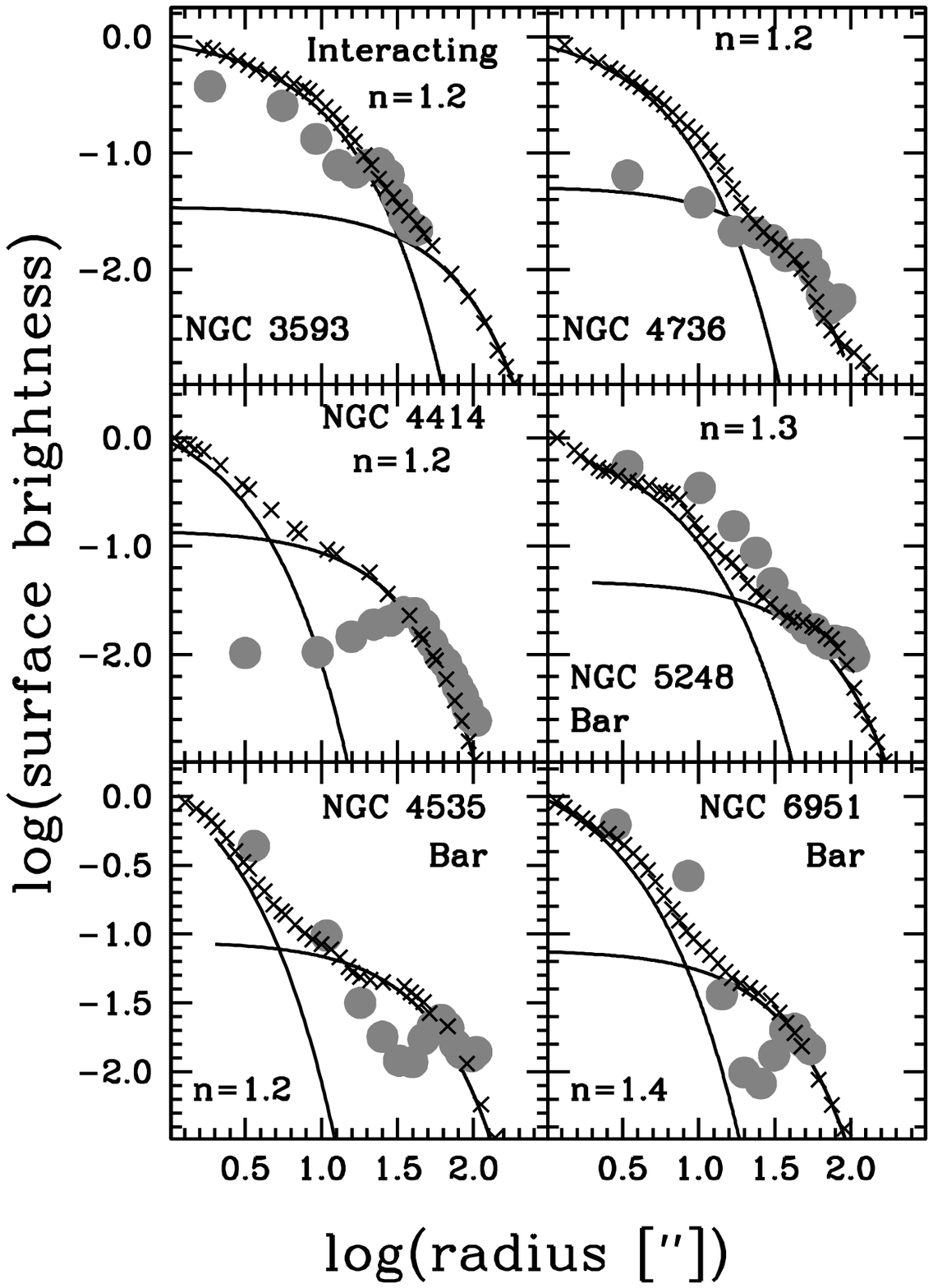}
\end{center}
\caption{ The x's represent the H-band surface brightness; solid lines
  represent the bulge-disk decomposition, and filled circles represent
  CO surface brightness. The H-band profiles have been normalized by
  the surface brightness at 1''\ . The CO flux has been shifted to
  match the stars in the outer disk. CO annuli are measured in steps
  equal to the beam size.   \label{fig:prof}}
\end{figure*}

\setcounter{figure}{9}
\begin{figure*}[t]
\begin{center}
\includegraphics[width=0.9\textwidth]{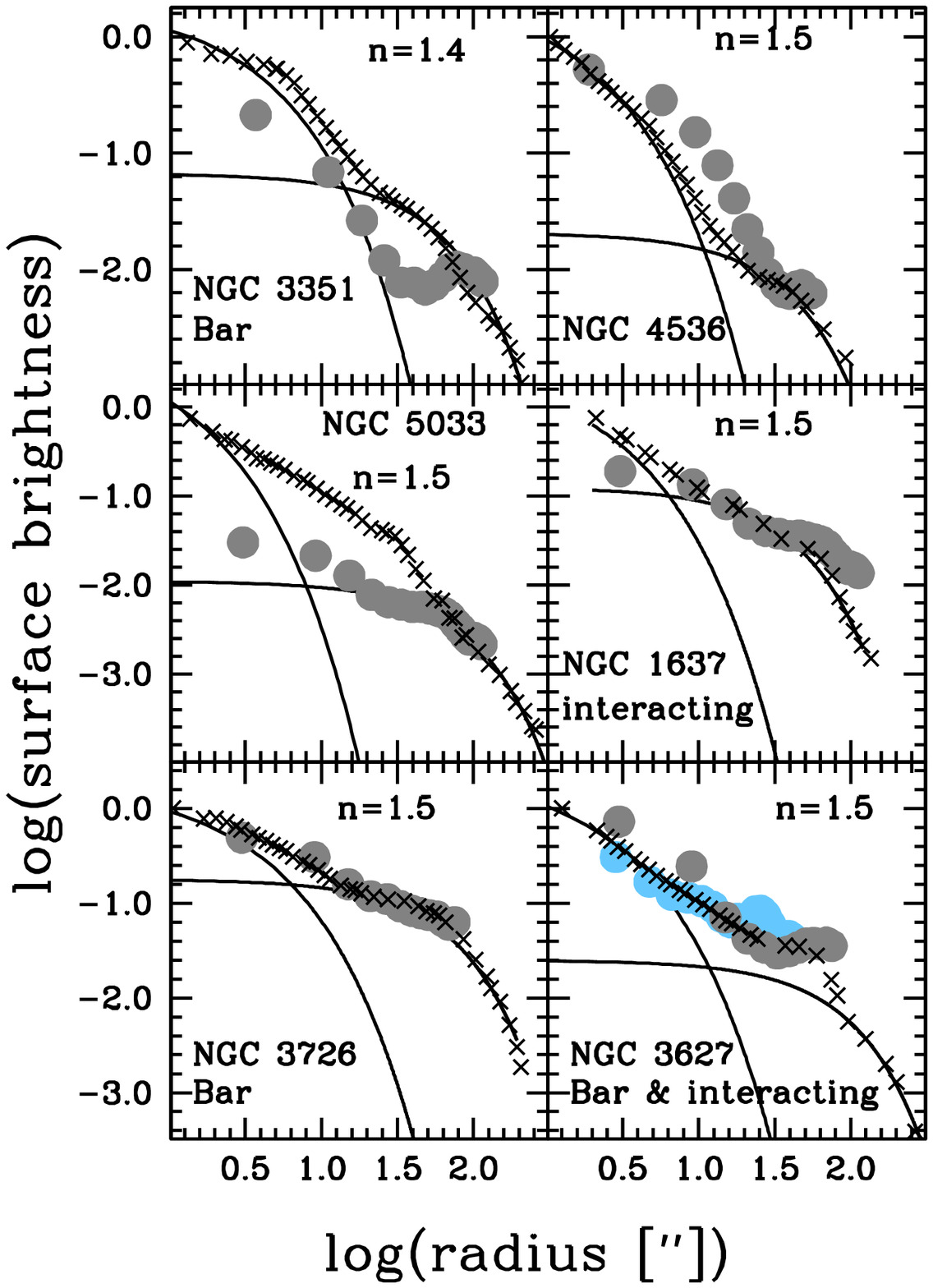}
\end{center}
\caption{The x's represent the H-band surface brightness; solid lines
  represent the bulge-disk decomposition, and filled circles represent
  CO surface brightness.  The H-band profiles have been normalized by
  the surface brightness at 1''\ . The CO flux has been shifted to
  match the stars in the outer disk. CO annuli are measured in steps
  equal to the beam size.   \label{fig:prof}}
\end{figure*}

\setcounter{figure}{9}
\begin{figure*}[t]
\begin{center}
\includegraphics[width=0.9\textwidth]{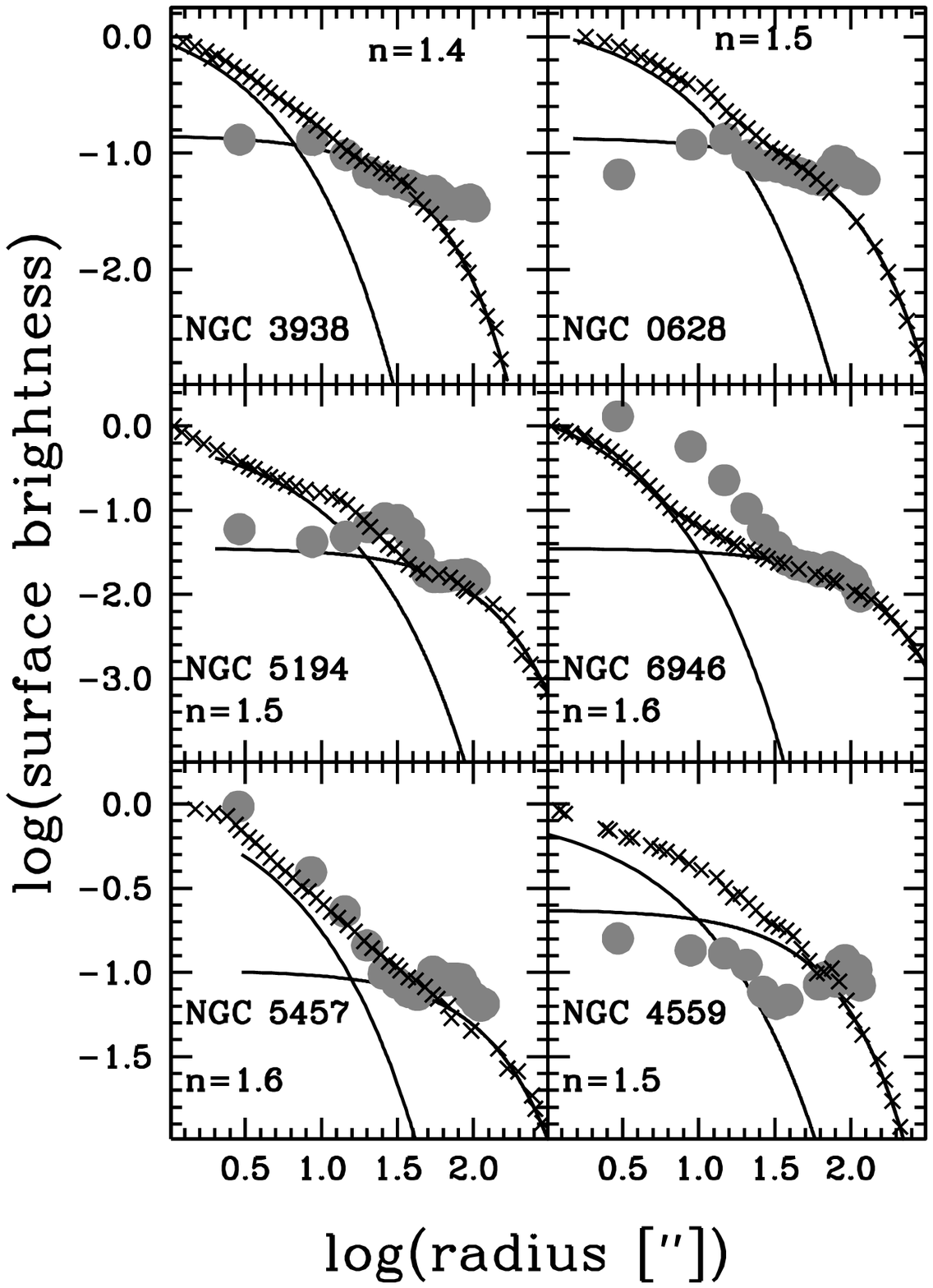}
\end{center}
\caption{The x's represent the H-band surface brightness; solid lines
  represent the bulge-disk decomposition, and filled circles represent
  CO surface brightness. The H-band
  profiles have been normalized by the surface brightness at 1''\ . The CO flux has
  been shifted to match the stars in the outer disk.  CO annuli are measured in steps
  equal to the beam size.   \label{fig:prof}}
\end{figure*}

\setcounter{figure}{9}
\begin{figure*}[t]
\begin{center}
\includegraphics[width=0.9\textwidth]{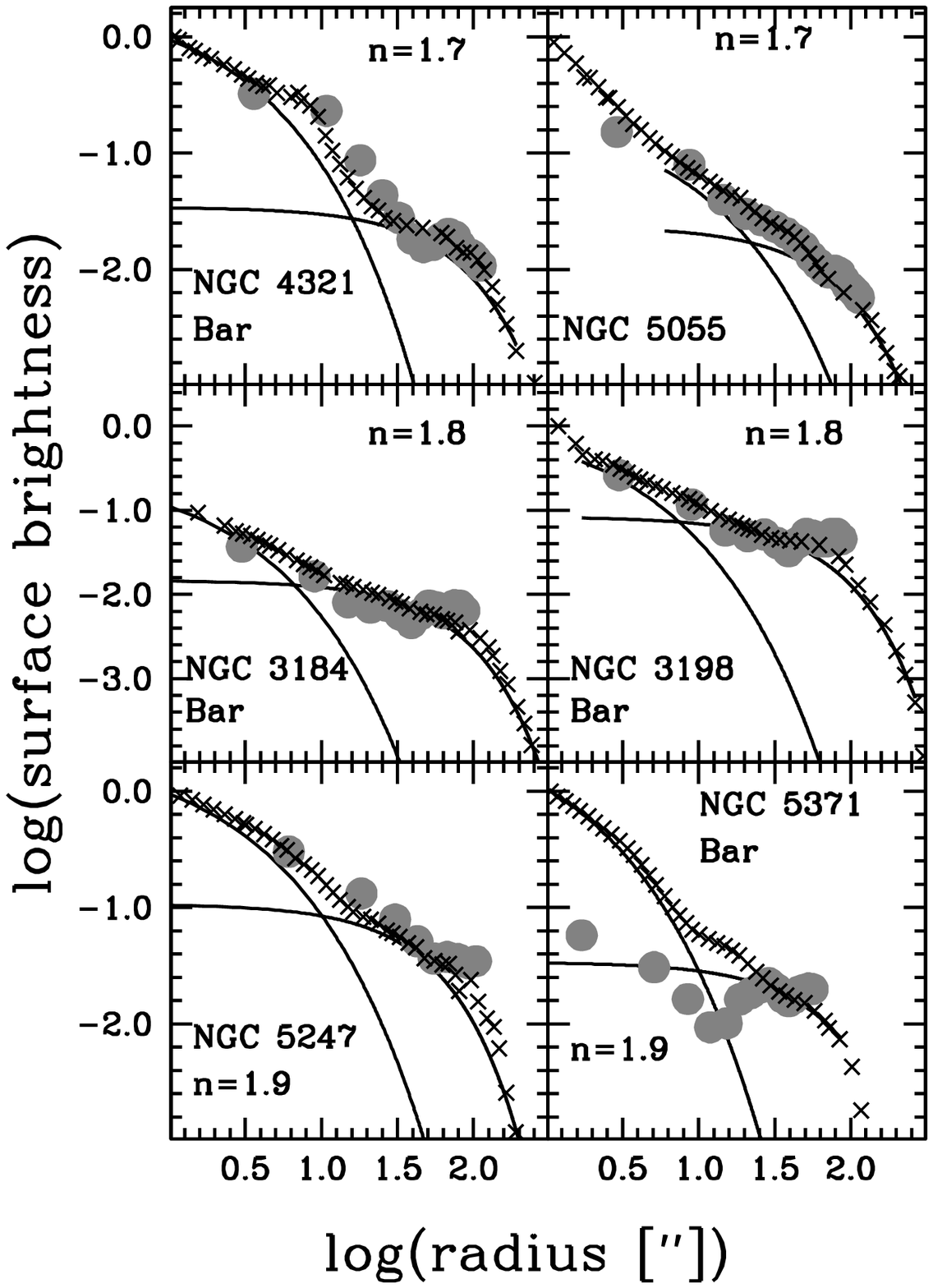}
\end{center}
\caption{The x's represent the H-band surface brightness; solid lines
  represent the bulge-disk decomposition, and filled circles represent
  CO surface brightness. The H-band
  profiles have been normalized by the surface brightness at 1''\ . The CO flux has
  been shifted to match the stars in the outer disk. CO annuli are measured in steps
  equal to the beam size.    \label{fig:prof}}
\end{figure*}

\setcounter{figure}{9}
\begin{figure*}[t]
\begin{center}
\includegraphics[width=0.9\textwidth]{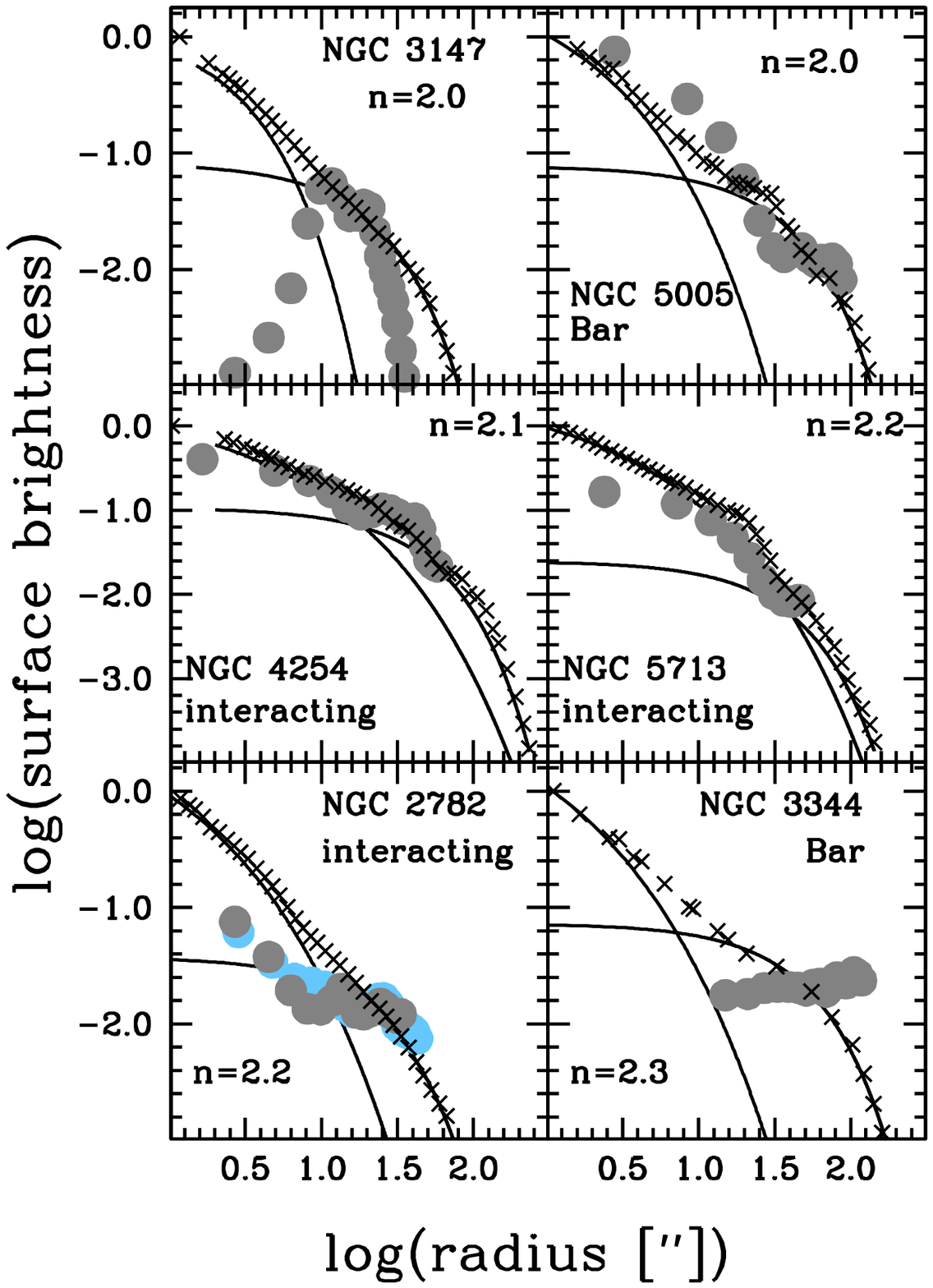}
\end{center}
\caption{The x's represent the H-band surface brightness; solid lines
  represent the bulge-disk decomposition, and filled circles represent
  CO surface brightness. The H-band
  profiles have been normalized by the surface brightness at 1''\ . The CO flux has
  been shifted to match the stars in the outer disk. CO annuli are measured in steps
  equal to the beam size.    \label{fig:prof}}
\end{figure*}

\setcounter{figure}{9}
\begin{figure*}[t]
\begin{center}
\includegraphics[width=0.9\textwidth]{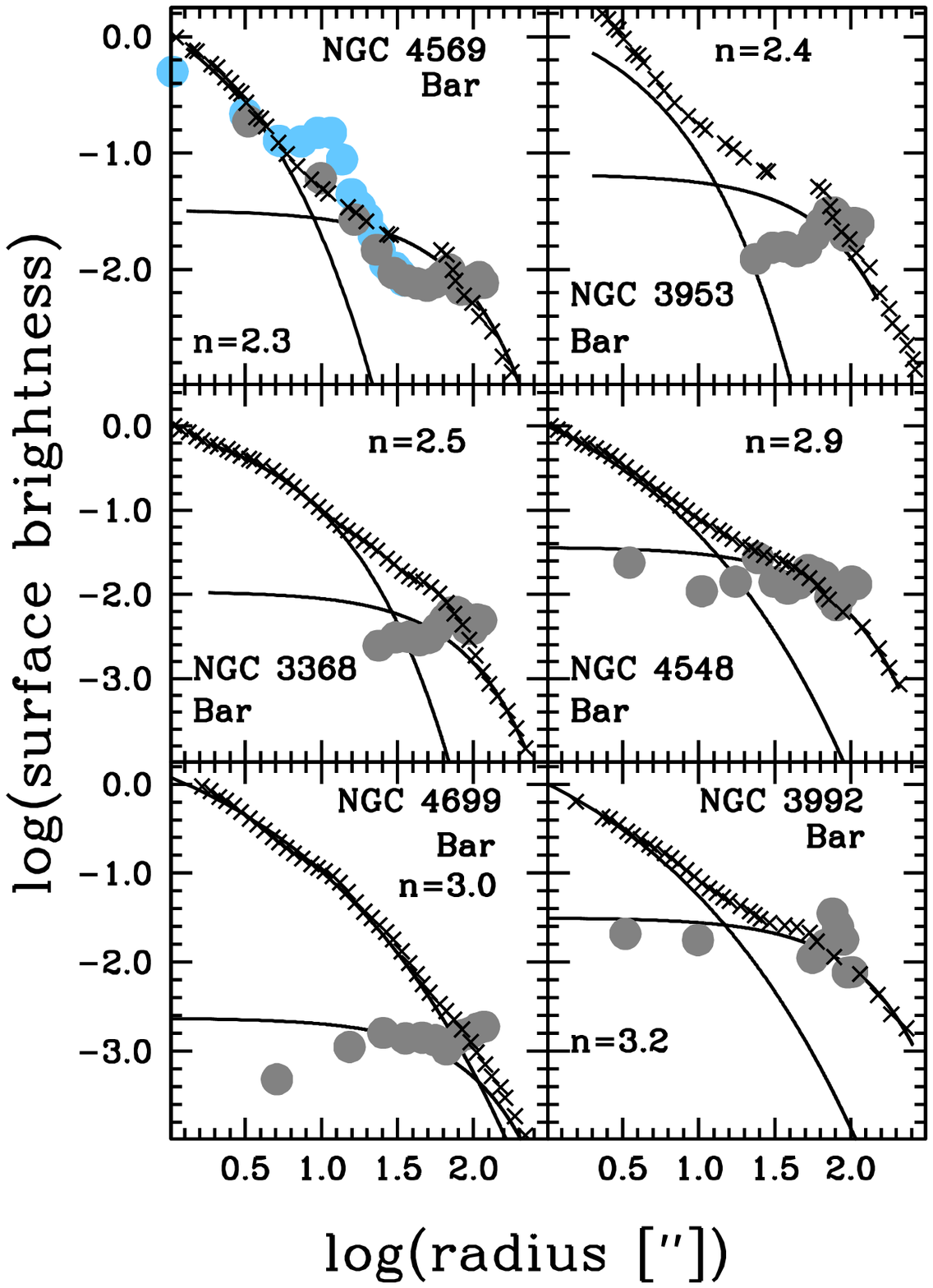}
\end{center}
\caption{The x's represent the H-band surface brightness; solid lines
  represent the bulge-disk decomposition, and filled circles represent
  CO surface brightness. The H-band
  profiles have been normalized by the surface brightness at 1''\ . The CO flux has
  been shifted to match the stars in the outer disk. CO annuli are measured in steps
  equal to the beam size.    \label{fig:prof}}
\end{figure*}

\setcounter{figure}{9}
\begin{figure*}[t]
\begin{center}
\includegraphics[width=0.9\textwidth]{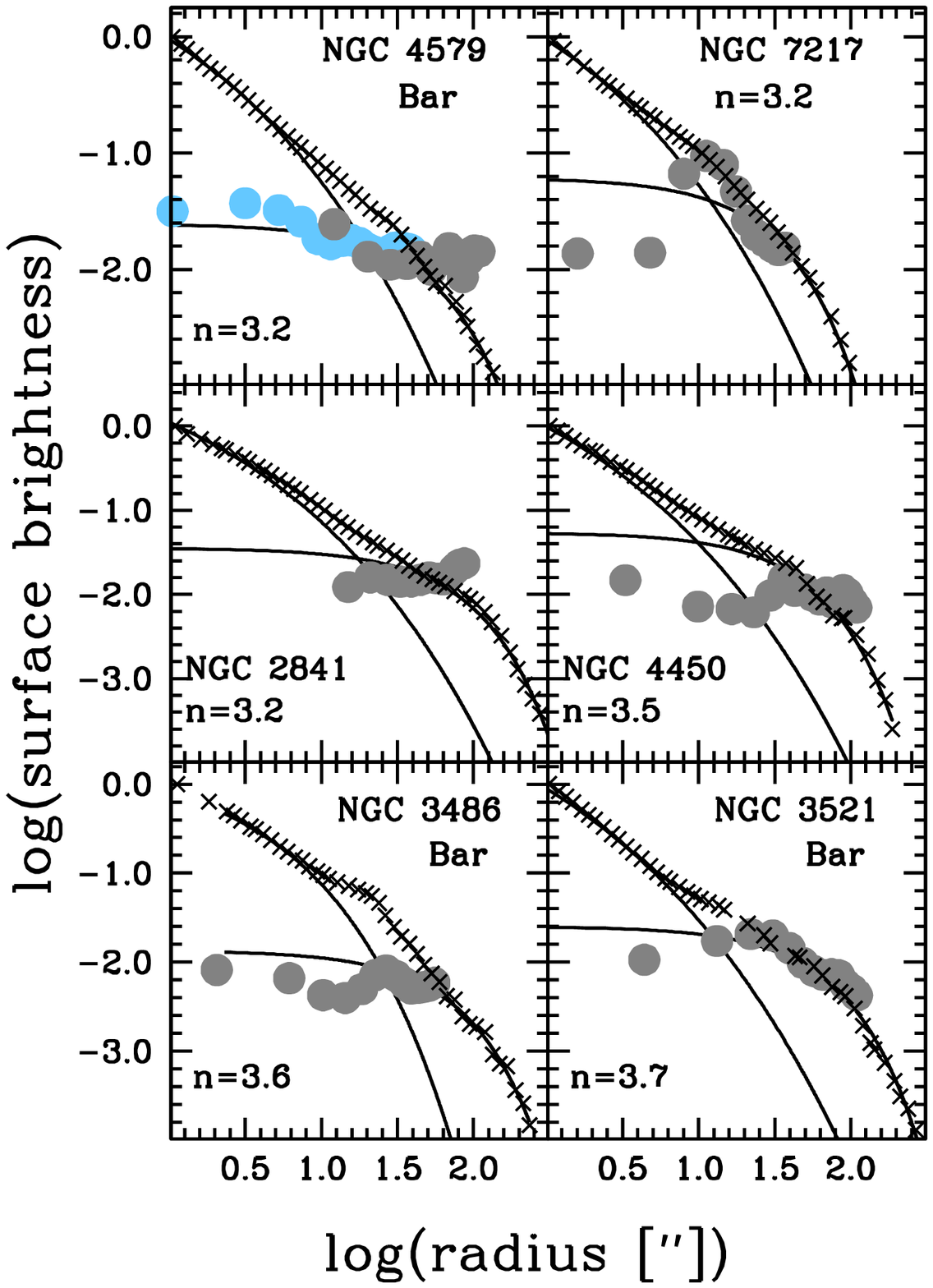}
\end{center}
\caption{The x's represent the H-band surface brightness; solid lines
  represent the bulge-disk decomposition, and filled circles represent
  CO surface brightness.  The H-band
  profiles have been normalized by the surface brightness at 1''\ . The CO flux has
  been shifted to match the stars in the outer disk. CO annuli are measured in steps
  equal to the beam size.    \label{fig:prof}}
\end{figure*}
\setcounter{figure}{9}
\begin{figure*}[t]
\begin{center}
\includegraphics[width=0.9\textwidth]{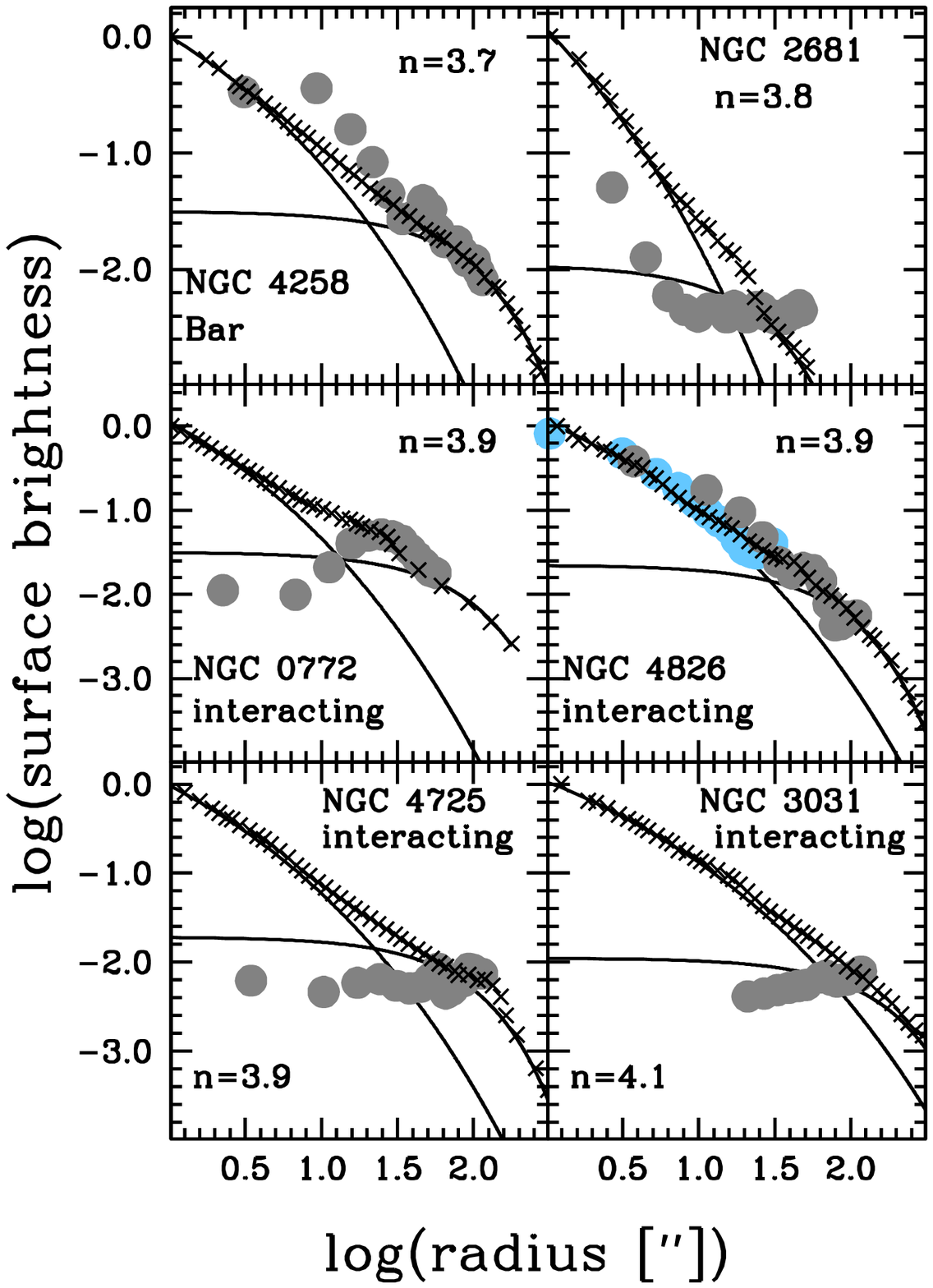}
\end{center}
\caption{The x's represent the H-band surface brightness; solid lines
  represent the bulge-disk decomposition, and filled circles represent
  CO surface brightness. The H-band profiles have been normalized by
  the surface brightness at 1''\ . The CO flux has been shifted to
  match the stars in the outer disk. CO annuli are measured in steps
  equal to the beam size.   \label{fig:prof}}
\end{figure*}
\clearpage
\noindent molecular hydrogen masses by 1.36 to account for
Helium. Therefore the final mass is $M_{mol}=1.36\times M_{H_2}$.

We take metallicities from the literature and the source is given in
Table 1. The most common source of available metallicity for galaxies
in our sample is \cite{moustakas2010}.  We use metallicities in the
\cite{kobulnicky2004} basis, and apply conversions based on
\cite{kewley2008}. Many of our galaxies do not have published
metalicities, for these we use metallicities determined from the
stellar mass-metallicity relationship in \cite{tremonti2004}.  For
those galaxies in which the metallicity is based on the mass-metallicity
relationship we assume that the uncertainty is increase in the
molecular gas mass by 0.1~dex, which is comparable to the scatter
found in \cite{tremonti2004}.

In Fig.~\ref{fig:xco} we compare the molecular gas densities
determined from the metallicity dependent conversion factor (hereafter
$X(Z)$), and the metallicity - CO surface brightness dependent
conversion factor (hereafter $X(Z,W)$). We calculate the surface
density of molecular gas in the bulge and outer disk. To calculate
molecular gas masses of disks we use total flux measurements from
FCRAO survey \citep{young1995}, and we subtract the flux from the
bulge region. The FCRAO survey uses major axis scans to obtain total
fluxes. We are therefore subtracting a flux derived from a CO map from
a flux derived from a scan, which may lead to slightly different
values. The advantage of using \citep{young1995} is that it provides a
single uniform data source for a very large number of galaxies. A
comparison of FCRAO total fluxes to those derived from fully sampled
maps has been made by \cite{helfer2003} and also \cite{leroy2009};
both find good agreement.  To calculate the surface area of the disk
we use 4 times the scale-length of the outer disk (determined from the
bulge-disk decomposition), which on average is similar to the optical
radius.

We find very little difference between $X(Z)$ and constant $X_{CO}$,
indeed the median ratio between the two is $<X(Z)/X_{CO}>=1.0$, and
the scatter is comparable to the error-bar (0.3~dex). This simply
reflects the fact that our metallicities are not very different than
clouds of typical spirals, like the Milky Way. Contrarily, the
densities based on the \cite{narayanan2011} conversion factor,
$X(Z,W)$, are systematically different than the constant $X_{CO}$. The
median molecular gas mass of bulges based on the \cite{narayanan2011}
factor is 40\% that of the constant values. The constant and
metallicity only conversion factors yield gas surface densities that
are very high, reaching $10^4~M_{\odot}~pc^{-2}$. However, the most
extreme values of the \cite{narayanan2011} conversion factor are as
much as an order-of-magnitude lower than those determined with the
constant $X_{CO}$. There is less of an effect on disk averaged
values. The median disk mass measured with $X(Z,W_{CO})$ is 75\% that
measured with a constant $X_{CO}$ and $X(Z)$.  We do not find a
systematic difference between the galaxies with measured metallicity
and those with inferred metallicity. For the remainder of the paper we
shall report all results for both $X(Z)$ and $X(Z,W)$.

\section{Results} 
\subsection{Profiles}
In Fig.~\ref{fig:prof} we show the first direct comparison of a large
number bulge-disk decompositions of stellar light to CO(1-0) surface
brightness profiles. The galaxies are arranged in order of increasing
S\'ersic index. Bulgeless galaxies are first. We label each panel with
the galaxy identifier and the S\'ersic index. If the galaxy is barred
and/or interacting we also include that information in the panel.
(Note, if a galaxy is not barred and/or not interacting, then nothing is
mentioned to this effect.). The x's represent the H-band surface
brightness profile, and the lines represent the bulge and disk that
have been fit to that profile. For ease of comparison we have
normalize all stellar surface brightness profiles (and fits) by the
surface brightness at 1~\arcsec. The large circles represent the
CO(1-0) surface brightness profile.  In a the few cases where multiple
CO data sets exists from our samples, we plot the higher resolution
data set as light blue circles.

The CO(1-0) surface brightness profiles are calculated simply by
computing the mean surface brightness in ellipses set to match the
center, axial ratio and position angle of the H-band light. The
CO(1-0) surface brightness profile is computed in increments of the
beam size for each map. Because we are principally interested in the
illustrating the relative result, we have shifted the CO profile to
match the stellar profile in the outer disk. For the interferometric
only data from NUGA \& STING we refrain from measuring the surface
brightness profiles at scales larger that 45\arcsec and 60\arcsec
respectively, as the data becomes less sensitive to structures larger
than this, due to the lack of zero-spacing data.

\begin{figure*}
\begin{center}
\includegraphics[width=0.95\textwidth]{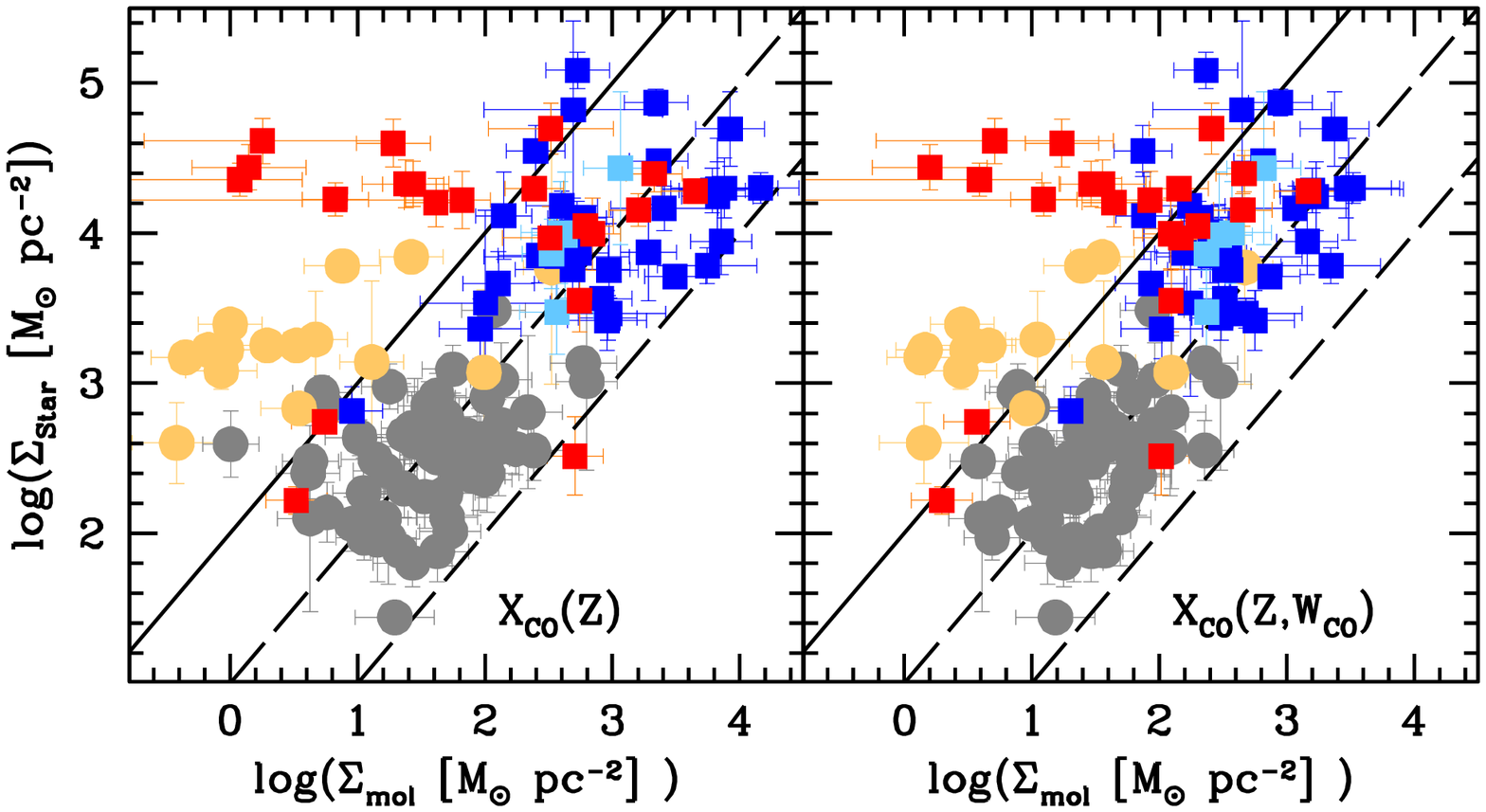}
\end{center}
\caption{ Gas surface density is compared to the surface density of
  stars for classical bulges (red squares), pseudobulges (blue
  squares), the centers of bulgeless galaxies (light blue squares),
  disks outside the radius of the bulge (grey circles), and elliptical galaxies (golden
  circles). The left panel shows gas densities calculated with X(Z),
  the right panel shows those with X(Z,W). \label{fig:gasstar}}
\end{figure*}
We classify our CO profiles into the following four
categories: (1) profiles in which the CO emission peaks at radii
larger than the bulge, (2) profiles in which the CO surface brightness
profile follows the exponential profile, (3) profiles in which the CO
emission traces the bulge or is brighter than the bulge, and (4)
profiles in which the CO emission is between (2) \& (3). We strongly
caution the reader that in a few cases this classification hinges on a
single CO(1-0) beam, and therefore should be taken with a certain
degree of skepticism. The classification is given in Table 2. 

We first note that 4 of 5 bulgeless galaxies have surface brightness
profiles that seem to follow the exponential disk. We find that 55\%
of our sample galaxies have CO(1-0) surface brightness profiles that
are at least as concentrated as the star light, and 40\% have
CO(1-0) surface brightness profiles that are as steeply concentrated
as the corresponding surface brightness profile of stars. This confirms
the initial result by \cite{regan2001bima} on a much larger sample. 

Galaxies with more concentrated CO emission are more likely to be
barred.  Roughly 70\% have bars, only half of the galaxies in our
sample with out central concentrations of gas have bars. We find no
preference for interacting galaxies to have steeper or less steep
CO(1-0) surface brightness profiles. Though this may simply be an
artifact of our sample.  The majority of galaxies with CO emission
that is more concentrated than the near-IR light have bulges with low
S\'ersic index, roughly 80\% have $n_b<2.1$. Conversely, few than half
of the galaxies with shallow central surface brightnesses of CO also
have bulges with low S\'ersic.

The general trend that emerges from examining Fig.~\ref{fig:prof},
that the concentration of CO emission is commonly as concentrated
as the star light. This connection between stellar and molecular
surface brightness is more common in barred galaxies, however many
unbarred disks also show this connection. The direct comparison of CO
flux density profiles to bulge-disk decompositions illustrates that as
a rule bulges are not gas poor. Quite the contrary, bulges are
frequently more than and order of magnitude higher in CO surface
brightness than that of the surrounding disk. 


The highest ratio of X(Z,W) in the center of the galaxy, versus that
averaged over the disk is $X_{center}/X_{disk}\leq 4$. In the
centrally concentrated galaxies, the central CO surface brightnesses
are often more than an order of magnitude greater than the disk.  This
implies that indeed at least some fraction of pseudobulges are
presently increasing the bulge-to-total ratio of their stellar mass.
It will be interesting to revisit this result in the future as methods
of converting CO(1-0) to molecular hydrogen become more robust.

\subsection{Gas Density of Bulges}

In Fig.~\ref{fig:gasstar} we show the stellar mass surface density of
pseudobulges (blue squares), classical bulges (red squares), the
centers of bulgeless disks (light blue squares), disks beyond the
radius of the bulge (grey
circles) and elliptical galaxies (gold circles) plotted against gas
mass surface density. To construct the list of elliptical galaxies we
begin with the sample of \cite{young2011}, which lists all early-type
galaxies in the ATLAS3D sample with CO(1-0) detections. We then
carry out surface photometry to the stellar light profile, if the
bulge-to-total ratio of the stars is greater than 2/3, the galaxy is
included in our figure.

In Fig.~\ref{fig:gasstar} we wish to compare the total gas density
within a given structure to the total stellar density. The boundaries
of structures are thus defined by the size in the stellar mass, as
follows: bulge sizes are the radius at which the bulge surface
brightness equals that of the disk in a bulge-disk decomposition
($R_{b=d}$), disk sizes are defined as an annulus beginning at the
bulge radius and extending to 4 times the scale-length of the disk (
in most disks this is very close to the optical radius), and the size
of early-type galaxies is twice the half-light radius of the stars.
The left panel shows the gas density calculated with X(Z) and the right
panel shows the gas density calculated with X(Z,W). The lines
represent constant proportionality $\Sigma_{mol}/\Sigma_{star}$ of
0.01,0.10,1.00 respectively.

In Fig.~\ref{fig:gasstar} it is clear that there is not a single
behavior that can describe the gas density of all bulges. Some bulges
are gas rich, and thus star forming; others have low gas density,
relative to both their own stellar density, but also the gas density
of disks.

Almost all bulges with low Sersic index have higher gas densities than
is typically found in disks. The typical gas density observed in the
pseudobulges in this sample is as high and often higher than that
typically observed in molecular clouds \cite{bolatto2008}. The average
molecular gas density for bulges with S\'ersic index, $n\leq 2$ in our
sample is $< \Sigma_{mol} > = 357$~M$_{\odot}$~pc$^{-2}$ using X(Z)
and 179~M$_{\odot}$~pc$^{-2}$ with $X(Z,W)$. In fact, 80\% of
pseudobulges ($n\leq 2$) have surface densities greater than
100~M$_{\odot}$~pc$^{-2}$.  Pseudobulges appear to extend a
relationship between surface density of gas and stars of disks, to
higher surface densities, which is to say pseudobulges have similar
gas fractions to disks. 

Conversely classical bulges seem to have no correlation between gas
density and stellar density. This behavior is similar to what is observed
in early type galaxies. \cite{crocker2011} discusses the properties of
molecular gas in early type galaxies \citep[see
also][]{young2002,young2005,combes2007}.  The high S\'ersic bulges
span a larger range of gas densities than is observed in pseudobulges.
In our sample for $n>2$ we find $< \Sigma_{mol} > =
39$~M$_{\odot}$~pc$^{-2}$ using X(Z) and 24~M$_{\odot}$~pc$^{-2}$ with
$X(Z,W)$. The choice of conversion factor has a stronger impact on the
distribution of gas densities in classical bulges. We find using
$X(Z)$ or constant $X_{CO}$, that 50\% of classical bulges have
$\Sigma_{mol} >$100~M$_{\odot}$~pc$^{-2}$, and only 10\% (2 of 19)
have that surface density when using $X(Z,W)$.


\begin{figure}
\includegraphics[width=0.49\textwidth]{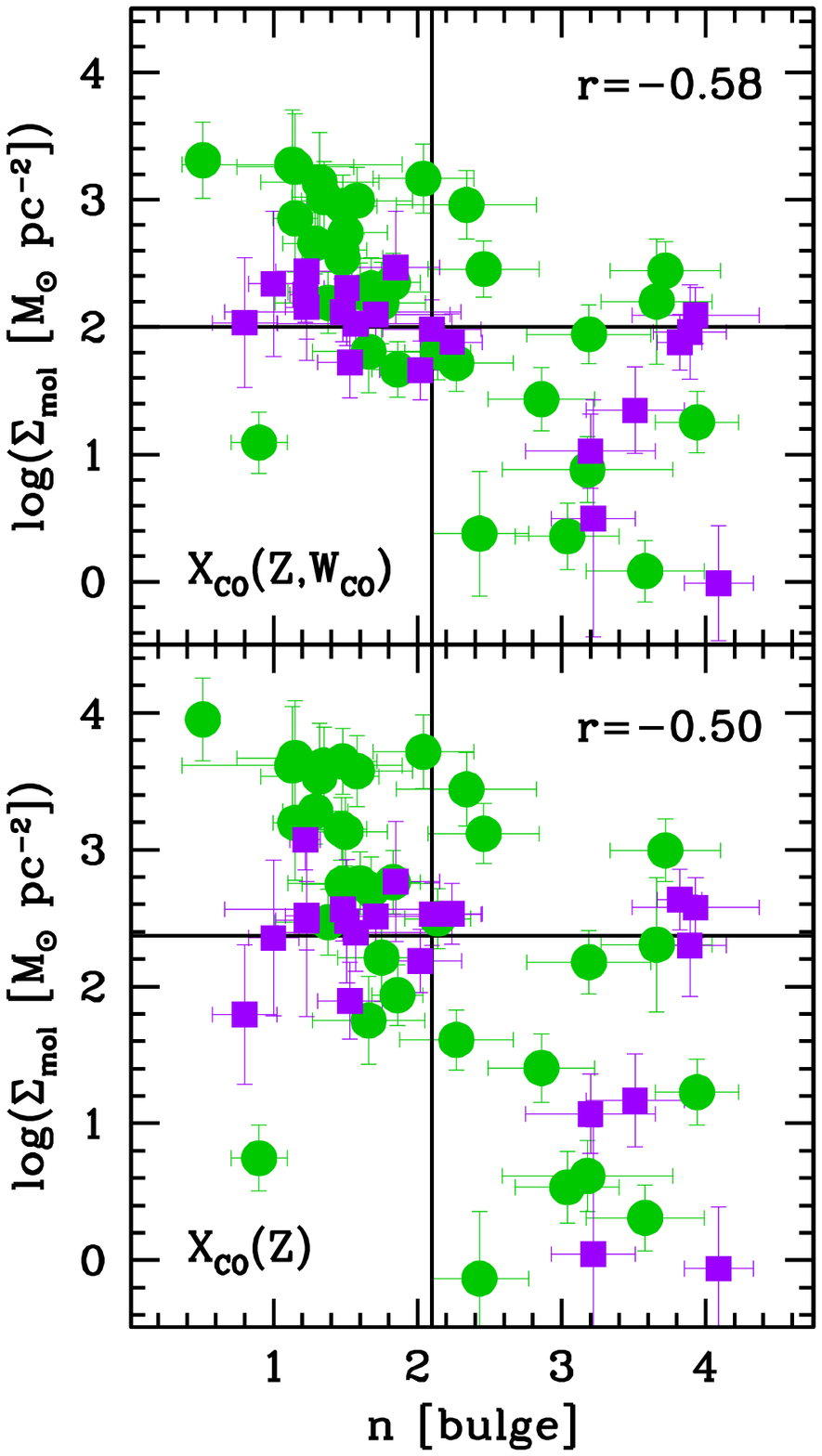}
\caption{The connection between bulge S\'ersic index and bulge gas
  surface density. Green circles represent galaxies with bars, and
  the purple squares represent bulges in unbarred disks. The vertical
  line marks the dividing line between pseudobulges and classical
  bulges and the horizontal line marks the median gas surface
  density.  \label{fig:gasnbar}}
\end{figure}
\vskip 30pt

\subsection{Correlation of Bulge Molecular Gas Density with Bars and
  S\'ersic Index}

Central concentrations of molecular gas are associated with barred
disks \citep{sakamoto1999,sheth2005,jogee2005}, and bulges with low
S\'ersic index are associated with high specific star formation rates
\citep{fisherdrory2010}. In Fig.~\ref{fig:gasnbar} we show the
dependence of gas density while simultaneous controlling for
bar/unbarred disk and the S\'ersic index. Gas densities of bulges
calculated with X(Z,W$_{CO}$) are shown in the top panel and X(Z) in
the bottom panel. We show bulges in barred galaxies as green circles
and bulges in unbarred galaxies as purple squares. In each panel we
also plot horizontal and vertical lines. The vertical line corresponds
to $n=2.1$. The horizontal line corresponds to the median
$\Sigma_{mol}$ in each panel.

For the total sample we observe a rough correlation between bulge
S\'ersic index and bulge molecular gas surface density such that
bulges with higher S\'ersic index have lower gas surface
densities. The Pearson's correlation coefficient is $r=-0.56$ with
X(Z,W$_{CO}$) and $r=-0.50$ with X(Z). There is significant scatter,
especially for intermediate S\'ersic indices. It may be more accurate
to describe the relationship between S\'ersic index and molecular gas
density of bulges as two separate distributions one scattering around
low S\'ersic index and high gas density, the other scattering
around high S\'ersic index and low gas density.

We use multiple methods to identify barred disks. First, we employ
visual inspection of $H$-band 2MASS images.  We also use the
ellipticity profile of the galaxy\citep[described
in][]{jogee2004,marinova2007}.  In this methods, bars are identified
by sharp transitions in the ellipticity profile. We compare our bar
identifications to those published in
\cite{rc3}, \cite{carnegieatlas}, \cite{laurikainen2004}, and consider
the preponderance of information when categorizing barred
galaxies. \cite{laurikainen2004} differs from \cite{rc3} and \cite{carnegieatlas}
in that the former uses bar-induced perturbation strengths to
identify bars, whereas \cite{rc3,carnegieatlas} use visual
classification.  Also, in comparing our and \cite{laurikainen2004}
classifications to the others, we take into account the known result
that bars are more easily identified in the near-IR
\citep[e.g.][]{eskridge2000,eskridge2002}. For the purposes of this
exercise we classify ovaled disks \citep[described in][]{kk04} in the
same category as bars. Also, we do not take into account the
difference between bars in early- and late-type galaxies
\citep{combes1993}.

We find that the ranking from highest to lowest average gas surface
density is as follows: the highest gas densities are in bulges with
low S\'ersic index that reside in barred disks; followed by bulges
with $n\leq 2$ in unbarred disks; then bulges with $n>2$ in barred
disks; finally the lowest gas density bulges are in those with $n>2$
in disks with no bar. The largest range in gas density exists in those
bulges with high S\'ersic index residing in barred disks. Not all
barred galaxies necessarily have high central densities of gas, and
secondly not all bulges with high S\'ersic index contain low surface
densities of gas.

It is difficult to do statistics of the gas density of classical
bulges; this work does not represent a robust sample covering the full
range of classical bulge properties.  As we discussed in the sample
selection, the criteria for the sub-samples in this paper favor IR
bright, or blue galaxies, and though adequate for studying the range
of gas densities, would be inadequate for presenting a statistical
analysis. \cite{fisherdrory2011} present a complete sample of
bulge-disk galaxies in the local Universe (assuming bulges do not
exists in galaxies with total stellar mass lower than
10$^9$~M$_{\odot}$.  They find that roughly 15\% of bulges with high
S\'ersic index also have enhanced star formation ($\Sigma_{SFR}\gtrsim
0.05$~M$_{\odot}$~yr$^{-1}$~kpc$^{-2}$. Their sample, however, only
contains 18 galaxies with classical bulges, and therefore also may
suffer from poor statistics. At this point, we cannot say how common
classical bulges with high densities of gas are, nonetheless it is
clear that several disk galaxies contain bulges in which both evidence
for a pseudobulge (high gas density) and classical bulge (high
S\'ersic index) are present and these bulges preferentially are in
barred disks.

\subsection{Bulge Molecular Gas Density of Interacting Galaxies}
\begin{figure}
\includegraphics[width=0.49\textwidth]{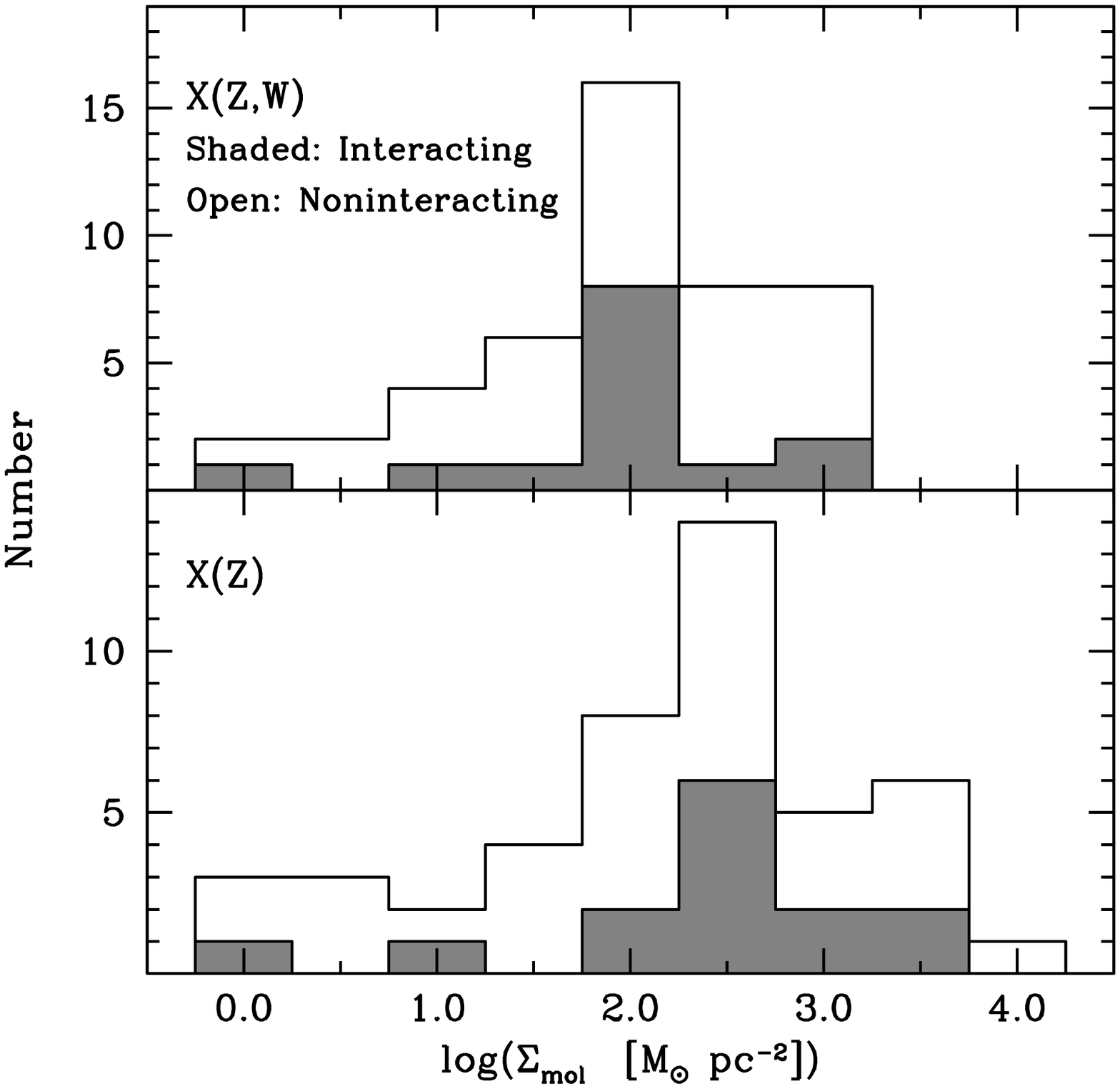}
\caption{Histogram of gas surface densities for bulges in galaxies
  that show evidence of interaction (shaded histogram) and those in
  galaxies that do not show strong evidence of an interaction. \label{fig:interact}}
\end{figure}

An alternative means of torquing gas in disks is through interactions
with other galaxies. For example, \cite{kannapan2004} show that
galaxies with nearby companions are more likely to have blue, star
forming centers. In Fig.~\ref{fig:interact} we show the distribution
of bulge molecular gas densities of interacting galaxies (shaded
region) in our sample, and those of galaxies that do not show signs of
interaction (unshaded region). We classify galaxies as interacting if
they show lopsided, or asymmetric features, or if the galaxy has a
nearby companion. Our sample selection excludes advanced major
mergers. Therefore, the asymmetries in these galaxies will be
representative of minor mergers, be they weak-distant gravitational
interactions with nearby galaxies in groups or cluster environments or
accretion of companions (e.g.~M~51). \cite{jogeeetal2009J} finds that
the low-z minor merger rate is under 10\% per gigayear. We classify 15
of 61 galaxies as interacting. Given that our sample has a slight bias
to IR bright galaxies, which are more likely to be interacting, the
frequency of interactions in our sample is in agreement with other
published minor merger rates. 

We find no significant difference between the bulge molecular gas
density of interacting galaxies in our sample and those that are not
interacting. The average molecular gas density of bulges in
interacting galaxies is $<\Sigma_{mol}>=218$~M$_{\odot}$~pc$^{-2}$
using X(Z) and $<\Sigma_{mol}>=89$~M$_{\odot}$~pc$^{-2}$ using
X(Z,W$_{CO}$). These are very similar to both the average for all
bulges in the sample, and those bulges that are not
interacting. 

We re-iterate that our sample likely under represents disk
galaxies that are low in gas surface density.  However, we can see
that the extreme high end of bulge gas surface density does not require
that a galaxy be interacting. Of the 10 bulges with the highest
molecular gas density only 3 show clear signs of
interaction. Conversely, all 10 of the bulges with the highest surface
density of molecular gas have bars, independent of whether they are
interacting or not. 

This result should not be taken to imply that these galaxies have not
experienced any minor mergers. \cite{cox2008} show that the effect on
star formation history is a strong function of the parameters of the
merger (like galaxy mass ratio). Also, \cite{bournaud2002} show that
accretion events can incite bar driven secular evolution. Therefore,
its possible that a minor merger could relax before the enhanced bulge
gas density subsides. We simply point out that high bulge gas
densities do not require that the galaxy show clear signs of ongoing
interaction.

\subsection{Total Gas Fraction} 

\begin{figure}
\includegraphics[width=0.49\textwidth]{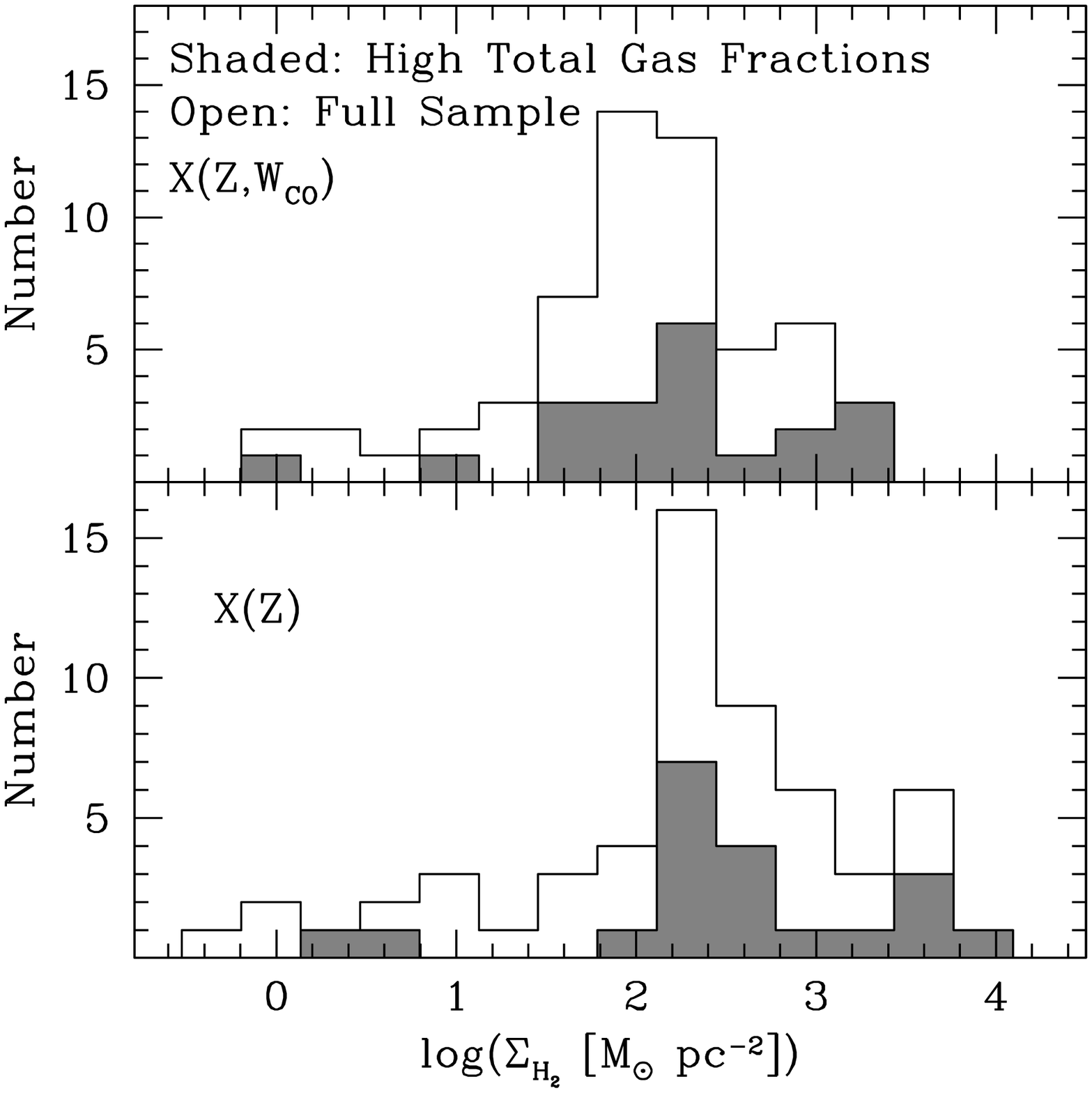}
\caption{Histogram of gas surface densities for bulges in galaxies
  with highest 1/3 of molecular gas fractions (shaded region)
  and total sample (unshaded region). \label{fig:fgas}}
\end{figure}

It seems reasonable that the total gas content of a galaxy should
affect the gas density in the center. Simply put, independent of what
process drives gas to the center of a galaxy if the galaxy has more
gas, it will more easily fill the central regions. For total gas
masses we use $CO(1-0)$ fluxes from \cite{young1995}, such that
$f_{mol} = M_{mol,tot}/M_{star,tot}$.  In Fig.~\ref{fig:fgas} we
overplot the distribution of bulge gas densities for those galaxies
with highest 1/3 of gas fractions (shaded region) over the total
distribution for all bulge gas densities for our sample (open region).
\begin{figure*}
\includegraphics[width=0.99\textwidth]{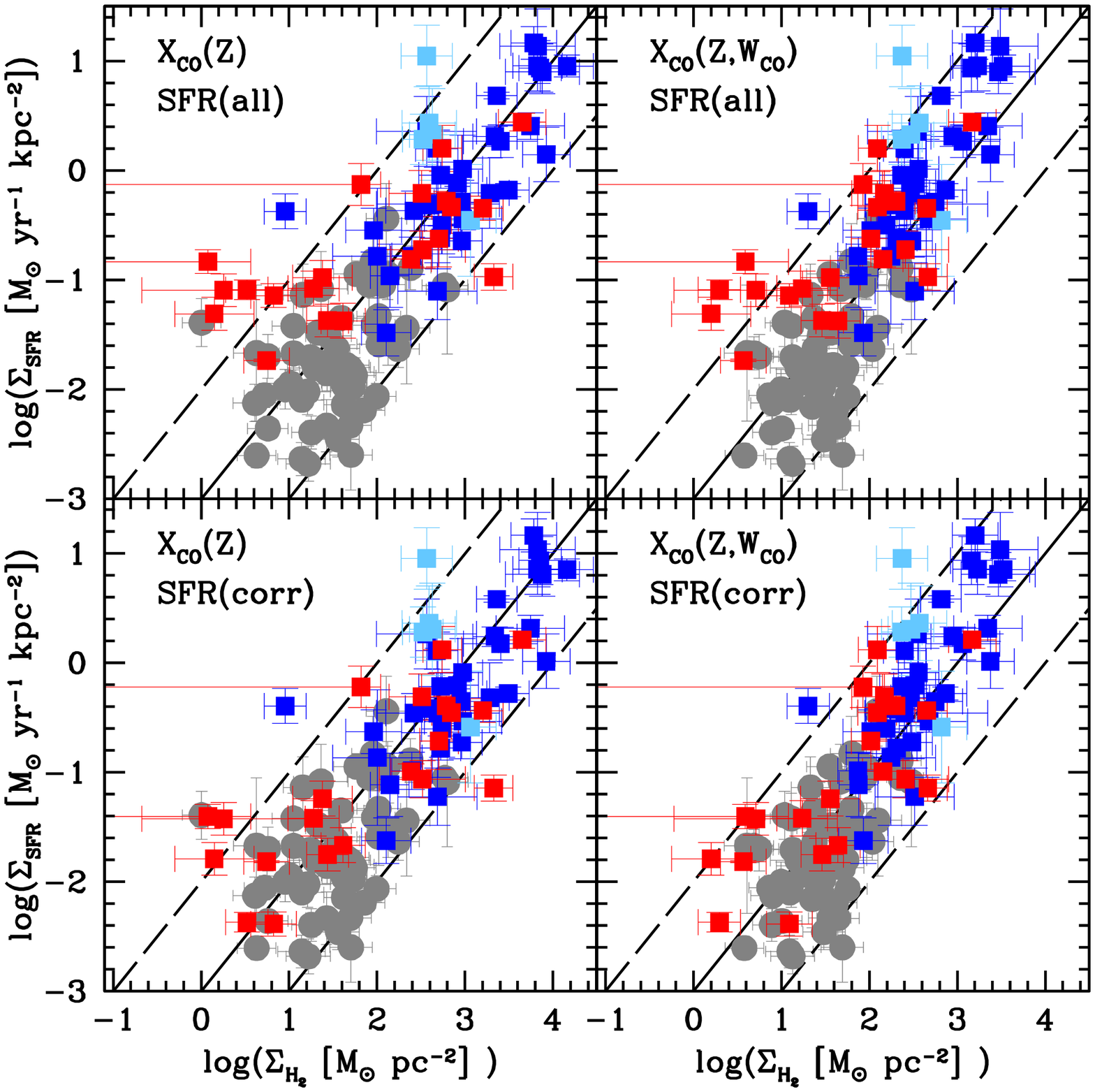}
\caption{ Gas surface density is compared to the surface density of
  star formation for classical bulges (red squares), pseudobulges (blue
  squares), the centers of bulgeless galaxies (light blue squares),
  and outer disks (grey circles). The left panel shows gas densities
  calculates with X(Z), the right panel shows those with
  X(Z,W). \label{fig:gassfr}}
\end{figure*}

The mean bulge gas density for galaxies in the highest 1/3 of gas
fractions is indeed higher than those with lower total gas
fractions. The mean bulge gas density of those galaxies in the 1/3
most gas rich galaxies in our sample is roughly 4-6$\times$, for
X(Z,W$_{CO}$) and X(Z) respectively, the mean bulge gas density for
galaxies in the 1/3 of our sample with the lowest gas fractions. 

It is clear from the histograms in Fig.~\ref{fig:fgas} that having a
higher gas fraction makes a galaxy more likely to have a higher bulge
molecular gas density. However, galaxies like NGC~4258 can have low
total gas fractions, but still have high central gas
densities. Furthermore, galaxies with very high gas fractions can also
have low central densities of molecular gas.  It thus appears that the
link between total gas fraction and bulge gas density is not very
precise, and certainly not a tight one-to-one correlation. The
correlation coefficient between $log(\Sigma_{mol}$ and $f_{mol}$ is
$r=0.38$ for both conversion factors.

\subsection{Relationship Between Star Formation and Gas in Bulges}

Finally we turn the  relationship between bulge star formation rate
surface density and gas surface density. Given that the densities of
gas in bulges are frequently much higher than what is typical of outer
disks we wish to determine if the processing of gas into stars
proceeds analogously in bulges as it does in disks.
In disks, increasing the molecular gas increases the star
formation rate density, roughly linearly
\citep{leroy2008,bigiel2008,rahman2012}. We can determine if the same
is true for bulges, and if the type of bulge has any impact on the
rate at which a bulge will consume its gas.

In Fig.~\ref{fig:gassfr} we plot the star formation rate density
against the molecular gas density for pseudobulges (dark blue
squares), classical bulges (red squares), the centers of late-type
galaxies (light blue squares) and the outer disks of the galaxies in
our sample (grey circles). In the left panels we show molecular gas
densities determined with metallicity dependent conversion factor, and
in the right panels we plot molecular gas densities with a CO-to-H$_2$
conversion factor that depends upon both gas surface brightness and
metallicity. The top panels use the total flux within the bulge to
calculate the star formation rate. The bottom panels use the 24~$\mu$m
and $FUV$ fluxes that have been corrected for diffuse emission and old
stellar populations to calculate bulge star formation rates.  The
lines indicate gas consumption times of 0.1,1, and 10~Gyr.

\cite{kennicutt98} and also \cite{kk04} show the star formation law of
nuclear rings in nearby disk galaxies. The overall nature of those
star formation laws is similar what is observed in
Fig.~\ref{fig:gassfr}. For the most part, the star formation law of
bulges appears to extend the behavior of disks to higher gas
densities. Also the way in which bulges make stars does not seem to
depend upon the type of bulge in which the gas resides. Pseudobulges,
classical bulges and the centers of late-type galaxies all seem to
reside on roughly the same correlation. The median gas depletion time,
the ratio $\Sigma_{mol}/\Sigma_{SFR}$, for bulges is $0.4 -
1.3\times10^9$~yr, depending upon the method used (uncorrected star
formation rates and the \citealp{narayanan2011} conversion factor
yield smaller depletion times). The median depletion times in disks in
our sample are similar to those in disks, differing by only 10-15\%
when using the full flux to calculate the star formation rate, and
differing by 20-30\% when using the flux that has been corrected for
diffuse emission.

The adoption of both the more complex CO-to-H$_2$ conversion factor
and diffuse emission correction to the star formation rate appears to
reduce the scatter in these plots. This is easy to see by examining
the systems with low gas and star formation rate surface
density. Moving from the top left to the lower right panels, it is
clear that the data has less scatter. We find that the scatter around
a median value of the depletion time decreases by a factor of
2.5$\times$ depending upon the $X_{CO}$ conversion factor used, where
using the conversion factor published in \cite{narayanan2011} results
in lower spread in the depletion times of bulges and disks. It is
possible that the nonlinearity in the \cite{narayanan2011} conversion
factor dampens the CO scatter, and thus the reduced scatter is by
design.  

When averaged over the whole sample the correction for diffuse
emission has little impact on the scatter about the correlation
between gas and star formation rate densities. However, as we have
shown in Fig.~\ref{fig:sfrcomp}, the diffuse emission correction
systematically affects high stellar mass density - low star forming
bulges significantly more, that is to say classical bulges. The
diffuse emission correction results in a factor of 3 reduction in the
scatter of classical bulges. Furthermore, Fig.~\ref{fig:gassfr} shows
that correcting for diffuse emission brings classical bulges into
better agreement with the correlation for galactic disks.

A power-law fit the entire sample recovers a roughly linear
relationship between $\Sigma_{mol}$ and $\Sigma_{SFR}$ that is
consistent with the results of \cite{bigiel2008}, also recently
\cite{rahman2012}. However, there appears to be a tendency for
slightly smaller gas depletion times (or more efficient star
formation) in bulges with higher surface density of star formation.
The average depletion time, and standard deviation, for the entire
sample of bulges and disks in our sample is almost exactly 1~Gyr. When
we average only the 10 bulges with the higher star formation rate
density we find that the depletion time is smaller by a factor of 2.3
using the \cite{genzel2012} conversion factor and 4.4 using the
\cite{narayanan2011} conversion factor. Therefore, this result is not
large compared to the uncertainty in the measurements of gas mass and
star formation rate, and also depends on the conversion factor used. Also, after corrections
are applied, we find little-to-no difference in the depletion times of
pseudobulges and classical bulges, and only minor differences between
bulges and disks.


\section{Summary \& Discussion}

\subsection{Summary}
In this paper we show that bulges (where bulge is defined as the high
density component of the surface brightness profile) are commonly
found to have gas densities that span ranges from 1 -
1000~M$_{\odot}$~pc$^{-2}$.  Fig.~\ref{fig:prof} shows that in the
majority of our sample, when the H-band surface brightness profile
transitions from an exponential disk to a steeper bulge, the CO(1-0)
surface brightness profile exhibits a similar transition to be more
centrally concentrated.

We find that the surface density of gas in bulges is linked to
properties of the galaxy, especially the S\'ersic index. As shown in
Fig.~\ref{fig:gasnbar}, bulges with lower S\'ersic index are much more
likely to have high surface densities of molecular gas.

Most pseudobulges and a few classical bulges have star formation rate
densities in excess of 0.1~M$_{\odot}$~yr$^{-1}$~kpc$^{-2}$.  Even
after correcting the star formation rate indicator for diffuse
emission and the CO-to-H$_2$ conversion factor for CO surface
brightness, Fig.~\ref{fig:gassfr} shows that some bulges, and a large
fraction of our sample, continue to evolve at redshift zero.  We find,
when using more nuanced metrics of the star formation rate and gas
mass, that the ratio of gas-to-star formation rate density of bulges
is similar to that of disks (Fig.~\ref{fig:gassfr}). This result
implies that the timescale for processing gas into stars in bulges is
not significantly different than that of disks, when measured on the
scale of the entire bulge. Therefore, the regulating step in low
redshift bulge growth appears to be how the bulge obtains its gas.

\begin{figure*}
\includegraphics[width=0.99\textwidth]{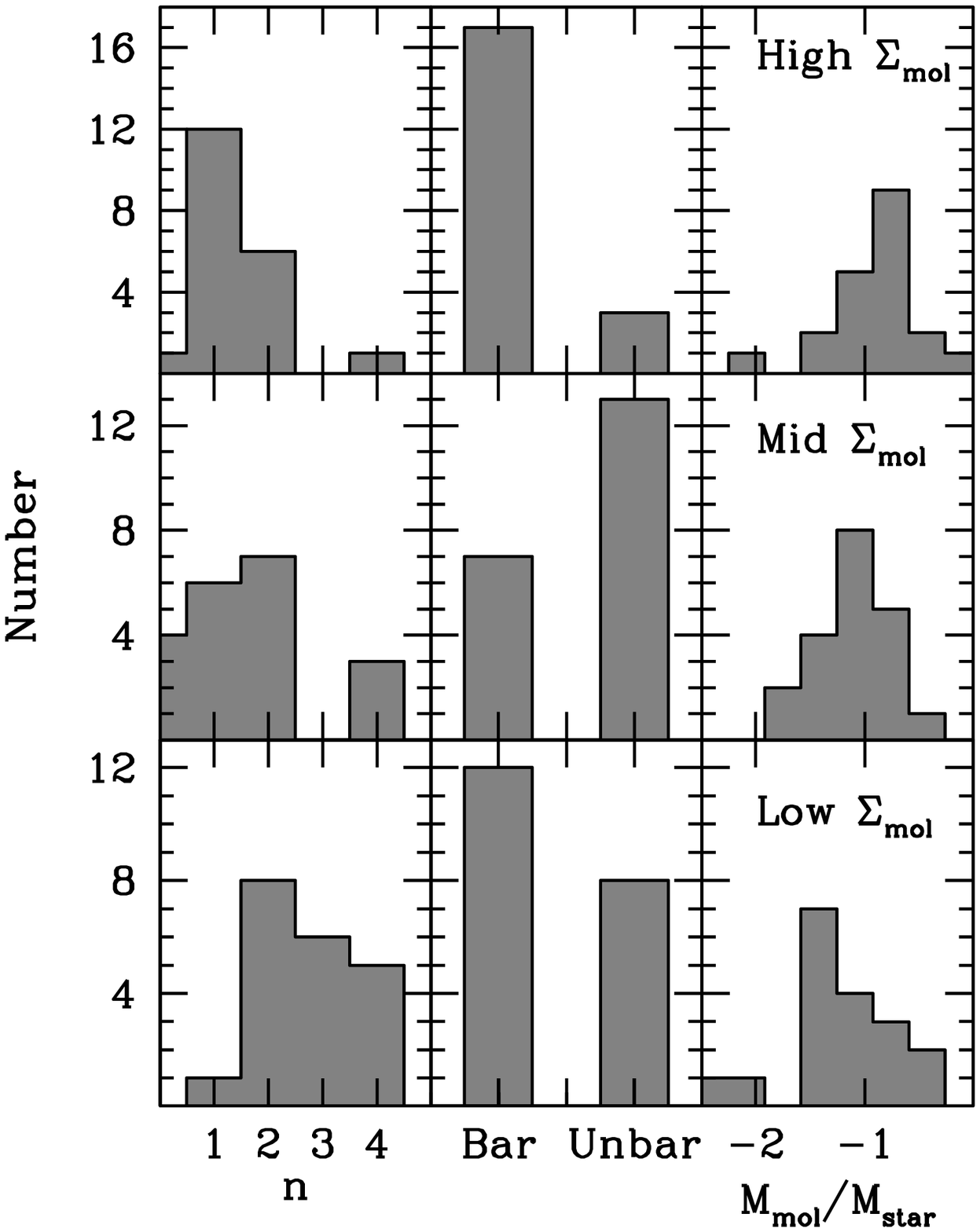}
\caption{For each set of high, middle and low gas density bulges we
  show (from left to right) the distribution of S\'ersic indices of
  the bulge, the distribution of barred and unbarred disks, and the
  distribution of total galaxy molecular gas
  fractions.  \label{fig:summary}}
\end{figure*}

In Fig.~\ref{fig:summary} we divide the sample into
3 sets of 20 galaxies ranked by gas density of the bulge. From top to
bottom, we show the 20 galaxies with highest bulge molecular gas
surface density, the 20 galaxies with bulge gas surface densities
closest to the median value, and the 20 galaxies with the lowest
surface density of gas in the bulge. For each set we show (from left
to right) the distribution of S\'ersic indices of the bulge, the
distribution of barred and unbarred disks, and the distribution of
total galaxy molecular gas fractions.  For the purposes of this
figure, we use the CO-to-molecular gas conversion factor of
\cite{narayanan2011}, and we set the S\'ersic index of those galaxies
with no bulge $n=0$. The results here change very little based on the
CO-to-H$_2$ conversion factor.

{\bf Bulges with High Molecular Gas Density:} In our sample, galaxies
with high molecular gas densities in their bulges almost always have
both low S\'ersic index ($n\lesssim 2$) and reside within barred
disks. Only 2 of our sample galaxies have high gas surface densities
in their bulges and also have high S\'ersic indices. Similarly, only 2
galaxies that do not have bars have high surface densities of gas in
their centers. There is also a tendency for bulges with high central
gas density to be in galaxies with larger total gas fraction.

{\bf Bulges with Intermediate Molecular Gas Density:} 
The main difference between galaxies in our sample with intermediate
range of bulge gas densities and those with the highest densities of
gas in their bulges is the presence of a bar. Galaxies with high
densities of gas are far more likely to have a bar than those of
intermediate gas density of the bulge. Both high and intermediate gas
density bins are predominately low S\'ersic index. Both have similar
distribution of gas fraction.

{\bf Bulges with Low Molecular Gas Density:} Contrary to naive
assumptions, in our sample the majority (12 of 20) of bulges with low
surface density of molecular gas are in barred disks. This number is
lower than the bar fraction in general. Our sample is
not sufficient to determine the frequency of gas-poor bulges in barred
galaxies. However we can show that a barred disk alone does not rule
out the possibility that a galaxy will have a gas-poor classical
bulge.  In Fig.~\ref{fig:summary} we show that along with higher
S\'ersic index bulges with low gas surface density typically are in
galaxies with lower total molecular gas fractions. A simple argument
is that if the galaxy has less gas over all even if a bar is present,
it will have less gas to send toward the center.  Also, it is possible
that the presence of a bulge with a steep density profile (as
indicated by larger S\'ersic index) would stabilize the gas in the
disk, and prevents high central densities of molecular gas. It is also
possible that the phenomenon that creates bulges with large S\'ersic
index also leaves the galaxy with a less molecular gas, and any of new
gas that enters through accretion is stabilized against inflow.

\subsection{Implications for Bulge Evolution}

The results of Fig.~\ref{fig:prof} shows, as was shown by
\cite{regan2001bima}, that clearly presence of a bulge is not mutually
exclusive with dense molecular gas. Bulges are not uniformly old,
non-starforming systems. Our sample is a much larger and we include,
for the first time, bulge-disk decompositions to star light.
Furthermore, it is clear from Fig.~\ref{fig:gasstar} that gas mass is
not a trivial fraction of the total baryonic mass of bulges. Many
bulges are building mass at low redshift. This result is similar to
that of \citep{fdf2009}, who shows that the timescales of pseudobulge
growth (based on M$_{star}/SFR$) are plausible for secular evolution,
but those timescales are orders of magnitude too long for classical
bulges.  Since bulges have similar depletion times as disks
(Fig.~\ref{fig:gassfr}), the key to understanding low redshift bulge
growth is to understand how the gas is driven into the center of a
bulge-disk galaxy.


In broad terms there are a few mechanisms that are commonly proposed
to explain the bulge-to-total ratios observed in disk galaxies. Those
mechanisms include major mergers followed by disk regrowth
\citep[e.g.][]{hammer2005}, direct accretion of satellite galaxies
\citep[e.g.][]{aguerri2001}, internal -or secular- evolution of
galactic disks \citep[e.g.][]{kk04,athan05}, and clump instabilities
in gas rich-high redshift-disk galaxies
\citep[e.g.][]{noguchi1999}. Given the dichotomous nature of bulge
properties and the connection of the bulge dichotomy to the galaxy
bimodality \citep{droryfisher2007}, it is reasonable that no one of
these theories will explain all bulge mass in all galaxies. Any
relevant theory intended to describe the formation of all intermediate
type galaxies should account for the observations that many bulges
have high densities of gas.

The idea that bulges are a result of the merging process is quite old
\citep{toomre1977}. Merging as a means to increase $B/T$ is often
proposed because galaxy-galaxy interactions occur frequently in both
the observed Universe and galaxy evolution models
\citep{whiterees1978,Cole+1994,hammer2005,Baugh2006}.  In this
scenario, the bulge we observe today is formed during the most recent
major merger event, and the remaining gas settles around the merger
remnant to form the outer disk.  
\cite{fisherdrory2011} find that high star formation rate densities in
bulges are much more common than the number of galaxies that
experience mergers in a gigayear \citep[roughly
10\%,][]{jogeeetal2009J}. Therefore, major mergers on their own
(absent subsequent accretion or secular evolution of gas) are not
likely to account for the gas in all bulges. However, if we isolate
only classical bulges the similarity of classical bulges to elliptical
galaxies is striking \citep{fisherdrory2010}. Modern simulations of
binary mergers reproduce the details of elliptical galaxies with great
detail \citep[e.g.][]{cox2006,naab2006}. In Fig.~\ref{fig:gasstar} we
show that in terms of gas density, similarities also exist between
classical bulges and elliptical galaxies. All bulges with very low
surface densities of gas are classical bulges (based on S\'ersic
index). Furthermore, as we show in Fig.~\ref{fig:summary} bulges with
the lowest gas density have lower than average total molecular gas
fractions. Both observations are consistent with the end products of
galaxy-galaxy mergers. The intense star formation during a merger
could leave a galaxy with a lower gas fraction, providing less gas to
fill the center of the galaxy. Furthermore, major mergers tend to
increase the cuspiness of the bulge
\citep{hopkins2009_sersic_merging}, which in turn is likely to
stabilize a disk. Therefore, after the star formation induced by the
merger subsides the bulge would have low gas density, as we observe in
classical bulges.

Mergers that result in bulge-disk galaxies are far more likely to
occur if the progenitor galaxies have unequal masses
\cite{bournaud2004}. Also, \cite{aguerri2001} outlines how minor
mergers can build bulges.  Indeed, in modern $\Lambda$CDM simulations
most disk galaxies acquire stellar mass through minor mergers
\citep[e.g.][]{parry2009}. Those models of galaxy evolution in which
the bulge-to-total stellar mass ratio of a galaxy is a function of
both mass-ratio and gas-fraction in the merger are well matched to
observations \citep{hopkins2009_gas}. \cite{cox2008} show that minor
mergers do not experience as dramatic increases in star formation
rate, this may leave the galaxy with enough gas to match our
observations. \cite{eliche2011} shows that minor mergers can produce
bulges with disky morphology.  However, it remains to be shown that
low mass ratio, gas-rich mergers produce bulges with the properties
quantitatively similar to observations of pseudobulges (low S\'ersic
index, scaling relations like those in \citealp{fisherdrory2010},
realistic gas and star formation rate densities). The kinematics of
merger remnants in the simulations of minor mergers in
\cite{bournaud2005} apear more like that expected for classical
bulges. Recently \cite{martig2012} show in simulations of a
large sample of disk galaxies, using a ``zoom-in resimulation
technique,'' that more violent merging history results in more
prominent bulges. Pseudobulges can be 1/3 of the galaxy mass
\citep{fisherdrory2008}, and successive accretion heats and eventually
destroys a galactic disk \citep{kaz2009,bournaud2007}.

\cite{elmegreen2008} finds that end products in clump instability
simulations tend to have high S\'ersic index. Conversely
\cite{inoue2012tmp} simulates bulges that appear similar to
pseudobulges, with low star formation rate at redshift zero. It may be
that on its own the clump instability is incapable of forming a low
redshift bulge with high densities of gas, however, at this point it
is difficult to say.

Simulations show that disk processes are quite capable of driving gas
toward the centers of galactic disks
\citep[e.g.][]{simkin1980,combesgerin1985,athan92,heller2007b}, see also
reviews \cite{kk04,athan05,combes2009}.  The connection between high
central density of gas and bars adds significant credibility to this
idea \citep[and present work]{sakamoto1999,sheth2005,jogee2005}.  Our
results show that pseudobulges are preferentially gas rich compared to
classical bulges.  \cite{kk04} argues that the similarity between
pseudobulge properties to the properties of outer disks are likely due
to pseudobulges forming secular evolution.  Indeed, in
Fig.~\ref{fig:gasstar} we show that not only are pseudobulges the mix
of stars and ISM are similar in pseudobulges and disks.  A possible
scenario is that if a gas-rich galaxy forms a bar, and there is no
stabilizing bulge with large S\'ersic index, the galaxy will then
efficiently drive gas toward the center of the disk. The gas then
forms stars in a cold-thin disk in the center of the galaxy. Since the
gas is at higher densities than the inward extrapolation of the
exponential disk (as shown in Fig.~\ref{fig:prof}), the resulting
surface density of stars is higher than the exponential
disk. Therefore, when we carry out bulge-disk decomposition then the
system is found to have a ``bulge'' in the light
profile. \cite{fabricius2012} show that bulges with low S\'ersic index
have colder stellar kinematics; also \cite{fisherdrory2008} show that
bulges with lower S\'ersic index are more likely to be flatter. In
this work we have shown that bulges with low S\'ersic index have
higher gas densities. The observations we present here, therefore, do
not contradict with the idea that the light that is frequently called
bulge light includes a significant component of cold stars that formed
though internal star formation, that frequently remains ongoing at
$z=0$.

It is likely that the total mass of a particular bulge could result
from multiple phenomena. For example, a galaxy could form a small
bulge though a very gas rich merger \citep[as described
in][]{springel2005,hopkins2009_disk_mergers} or via direct accretion,
and then secular evolution could add mass to the pre-existing
bulge. Indeed, \cite{bournaud2002} shows a simulation in which
accretion of a small-gas rich satellite instigates bar formation in a
spiral disk, and then would subsequently lead to increased gas density
in the center of a disk. \cite{fisherdrory2010} finds a sample of
bulges that are consistent with systems expectations of mixed systems.

Once the gas is in bulges, from our data, it appears to form stars
with similar efficiency as in the main body of the disk. It is
possible that multiple effects due to high pressure cancel out at
these spatial scales. For example, the intense pressure in bulges
could increase the efficiency of star formation inside of molecular
clouds in bulges, but also this intense pressure could also generate
more molecular gas in diffuse form, outside of molecular clouds
\citep{blitz2006}. Higher resolution observations than we present here
are necessary to understand the star formation in bulges at the scale
of molecular clouds. Nonetheless, at the scale of bulges (a few kpc)
our results suggest that the flow of gas into the center of the galaxy
is the main regulating factor for star formation in galaxy centers.


\acknowledgments

DBF acknowledges support from University of Maryland, the Laboratory
for Millimeter Astronomy and NSF grant AST 08-38178.  ND and DBF thank
the Max-Planck Society for support during this project. We also wish
to thank John Kormendy for helpful comments and support during the
writing of this work. We thank the anonymous referee, whose comments
significantly improved this work.  AB acknowledges partial support
form AST 08-38178, CAREER award AST 09-55836, and a Cottrell Scholar
award from the Research Corporation for Science Advancement.

This work is based on observations made with the Spitzer Space
Telescope, which is operated by the Jet Propulsion Laboratory,
California Institute of Technology under a contract with NASA. Support
for this work was provided by NASA through an award issued by
JPL/Caltech..  Some of the data presented in
this paper were obtained from the Multi-mission Archive at the Space
Telescope Science Institute (MAST). STScI is operated by the
Association of Universities for Research in Astronomy, Inc., under
NASA contract NAS5-26555. Support for MAST for non-HST data is
provided by the NASA Office of Space Science via grant NAG5-7584 and
by other grants and contracts.

Support for CARMA construction was derived from the Gordon and Betty
Moore Foundation, the Kenneth T. and Eileen L. Norris Foundation, the
James S. McDonnell Foundation, the Associates of the California
Institute of Technology, the University of Chicago, the states of
California, Illinois, and Maryland, and the National Science
Foundation. Ongoing CARMA development and operations are supported by
the National Science Foundation under a cooperative agreement, and by
the CARMA partner universities.

\bibliographystyle{apj}

\clearpage
\LongTables 
\begin{landscape}
\begin{deluxetable}{lcccccccccccc}
  \tablewidth{0pt} \tablecaption{Fluxes}
\tablehead{\colhead{Galaxy} & \colhead{Survey} & \colhead{Distance} & \colhead{12+log(O/H)} & \colhead{M$_H$} &
  \colhead{M$_H$} & \colhead{S$_{CO}$} & \colhead{S$_{CO}$\tablenotemark{(b)}} &
  \colhead{log(L$_{FUV}$)} & \colhead{log(L$_{FUV}$)} & \colhead{log(L$_{24}$)} & \colhead{log(L$_{24}$)} 
\\
\colhead{Name} & \colhead{ }  &\colhead{(Mpc)} & \colhead{\tablenotemark{(a)} } & \colhead{(mag)} &
  \colhead{(mag)} & \colhead{(Jy km~s$^{-1}$)} & \colhead{(Jy km~s$^{-1}$)} &
  \colhead{(erg~s$^{-1}$)} & \colhead{(erg~s$^{-1}$)} & \colhead{(erg~s$^{-1}$)} & \colhead{(erg~s$^{-1}$)} 
\\
\colhead{} & \colhead{ } & \colhead{} & \colhead{ } &\colhead{Bulge} &
  \colhead{Total} & \colhead{Bulge} & \colhead{Total} &
  \colhead{Bulge} & \colhead{Total} & \colhead{Bulge} & \colhead{Total} 
 }
\startdata
NGC0337 & STING & 20.3 & 9.01 (4) & -19.0 $\pm$ 0.2 & -22.2 $\pm$ 0.2 & 2.3 $\pm$ 1.2 & 55.46 $\pm$ 14.38 (9) & 41.67 $\pm$ 0.16 & 42.91 $\pm$ 0.17 & 42.07 $\pm$ 0.05 & 43.14 $\pm$ 0.05 \\
NGC0628 & SONG & 8.2 & 9.01 (4) & -20.0 $\pm$ 0.2 & -22.7 $\pm$ 0.2 & 69.0 $\pm$ 26.7 & 2160 $\pm$ 700 (14) & 41.16 $\pm$ 0.06 & 43.00 $\pm$ 0.06 & 40.76 $\pm$ 0.06 & 42.63 $\pm$ 0.06 \\
NGC0772 & STING & 27.1 & 8.63 (7) & -22.6 $\pm$ 0.6 & -25.3 $\pm$ 1.3 & 36.7 $\pm$ 24.6 & 1610 $\pm$ 310 (14) & 41.89 $\pm$ 0.16 & 43.69 $\pm$ 0.16 & 43.24 $\pm$ 0.47 & 44.29 $\pm$ 0.48 \\
NGC0925 & SONG & 8.6 & 8.64 (7) & -17.6 $\pm$ 0.2 & -21.6 $\pm$ 0.2 & 0.7 $\pm$ 0.2 & 900 $\pm$ 480 (11) & 41.80 $\pm$ 0.05 & 42.95 $\pm$ 0.05 & 41.07 $\pm$ 0.06 & 42.49 $\pm$ 0.06 \\
NGC1156 & STING & 7.8 & 9.17 (4) & -17.3 $\pm$ 0.2 & -20.1 $\pm$ 0.1 & 1.5 $\pm$ 0.7 & 24.957 $\pm$ 5.25 (10) & 41.28 $\pm$ 0.12 & 42.42 $\pm$ 0.13 & 42.54 $\pm$ 0.06 & 43.03 $\pm$ 0.06 \\
NGC1637 & STING & 9.9 & 8.59 (2) & -18.8 $\pm$ 0.9 & -21.7 $\pm$ 0.2 & 66.6 $\pm$ 13.1 & 68.3 $\pm$ 3.678 (8) &  ---  &  ---  & 42.24 $\pm$ 0.07 & 42.71 $\pm$ 0.07 \\
NGC2403 & SONG & 3.6 & 9.00 (4) & -15.3 $\pm$ 0.3 & -21.6 $\pm$ 0.2 & 1.6 $\pm$ 1.7 & 540 $\pm$ 180 (14) & 39.65 $\pm$ 0.31 & 42.88 $\pm$ 0.33 & 39.69 $\pm$ 0.25 & 42.45 $\pm$ 0.26 \\
NGC2681 & STING & 17.2 & 8.61 (7) & -22.6 $\pm$ 0.3 & -23.4 $\pm$ 0.4 & 82.8 $\pm$ 2.4 & 210 $\pm$ 40 (14) & 41.97 $\pm$ 0.11 & 42.32 $\pm$ 0.11 & 42.74 $\pm$ 0.05 & 42.86 $\pm$ 0.05 \\
NGC2782 & STING & 37.3 & 8.51 (7) & -22.6 $\pm$ 0.3 & -24.0 $\pm$ 0.3 & 35.6 $\pm$ 1.4 & 230 $\pm$ 40 (14) & 43.09 $\pm$ 0.12 & 43.49 $\pm$ 0.12 & 43.73 $\pm$ 0.05 & 43.90 $\pm$ 0.05 \\
NGC2841 & SONG & 13.7 & 9.03 (4) & -22.6 $\pm$ 0.3 & -24.7 $\pm$ 0.2 & 1.2 $\pm$ 2.6 & 5160 $\pm$ 600 (14) & 41.79 $\pm$ 0.05 & 43.02 $\pm$ 0.05 & 41.62 $\pm$ 0.07 & 42.86 $\pm$ 0.07 \\
NGC2903 & SONG & 9.4 & 8.84 (4) & -20.5 $\pm$ 0.1 & -23.2 $\pm$ 0.1 & 236.6 $\pm$ 110.7 & 2740 $\pm$ 910 (14) & 41.28 $\pm$ 0.09 & 43.07 $\pm$ 0.09 & 42.94 $\pm$ 0.08 & 43.37 $\pm$ 0.09 \\
NGC2976 & SONG & 3.6 & 8.86 (3) & -17.5 $\pm$ 0.3 & -20.4 $\pm$ 0.1 & 43.0 $\pm$ 9.5 & 610 $\pm$ 140 (14) & 41.17 $\pm$ 0.11 & 41.99 $\pm$ 0.11 & 41.70 $\pm$ 0.06 & 42.68 $\pm$ 0.06 \\
NGC3031 & SONG & 3.7 & 8.67 (7) & -22.5 $\pm$ 0.3 & -23.9 $\pm$ 0.3 & 9.0 $\pm$ 8.3 & 6138 $\pm$ 600 (12) & 41.12 $\pm$ 0.05 & 42.10 $\pm$ 0.05 & 41.77 $\pm$ 0.05 & 42.45 $\pm$ 0.05 \\
NGC3147 & NUGA & 43.1 & 8.96 (4) & -24.2 $\pm$ 0.47 & -25.6 $\pm$ 0.2 & 20.7 $\pm$ 3.1 & 1360 $\pm$ 300 (14) & 42.14 $\pm$ 0.13 & 43.72 $\pm$ 0.13 & 42.88 $\pm$ 0.06 & 43.93 $\pm$ 0.06 \\
NGC3184 & SONG & 11.5 & 9.13 (4) & -18.8 $\pm$ 0.4 & -22.9 $\pm$ 0.1 & 31.1 $\pm$ 17.1 & 1120 $\pm$ 320 (14) & 41.25 $\pm$ 0.06 & 43.13 $\pm$ 0.07 & 41.73 $\pm$ 0.09 & 42.85 $\pm$ 0.09 \\
NGC3198 & STING & 14.5 & 8.57 (7) & -19.5 $\pm$ 0.5 & -23.1 $\pm$ 0.1 & 54.3 $\pm$ 7.6 & 900 $\pm$ 100 (11) & 40.68 $\pm$ 0.15 & 43.61 $\pm$ 0.16 & 42.38 $\pm$ 0.05 & 43.04 $\pm$ 0.05 \\
NGC3344 & SONG & 6.1 & 9.20 (4) & -18.4 $\pm$ 0.5 & -21.5 $\pm$ 0.2 & <3.1   & 520 $\pm$ 130 (14) & 40.98 $\pm$ 0.06 & 42.52 $\pm$ 0.07 & 41.48 $\pm$ 0.08 & 42.32 $\pm$ 0.08 \\
NGC3351 & SONG & 9.8 & 8.69 (7) & -21.2 $\pm$ 0.3 & -23.3 $\pm$ 0.5 & 278.4 $\pm$ 53.5 & 700 $\pm$ 190 (14) & 42.01 $\pm$ 0.05 & 42.66 $\pm$ 0.05 & 42.86 $\pm$ 0.12 & 43.05 $\pm$ 0.12 \\
NGC3368 & SONG & 10.9 & 8.75 (3) & -22.5 $\pm$ 0.2 & -23.3 $\pm$ 0.2 & 631.9 $\pm$ 10.3 & 733 $\pm$ 200 (13) & 41.25 $\pm$ 0.05 & 42.65 $\pm$ 0.05 & 42.16 $\pm$ 0.06 & 42.52 $\pm$ 0.06 \\
NGC3486 & STING & 13.7 & 8.66 (7) & -21.3 $\pm$ 0.7 & -22.3 $\pm$ 0.6 & 8.9 $\pm$ 3.6 & 480 $\pm$ 240 (14) & 41.96 $\pm$ 0.18 & 43.76 $\pm$ 0.19 & 41.69 $\pm$ 0.06 & 42.84 $\pm$ 0.06 \\
NGC3521 & SONG & 12.2 & 8.91 (3) & -22.1 $\pm$ 0.7 & -24.4 $\pm$ 0.3 & 56.4 $\pm$ 56.4 & 4920 $\pm$ 1400 (14) & 40.87 $\pm$ 0.07 & 43.09 $\pm$ 0.07 & 41.80 $\pm$ 0.08 & 43.59 $\pm$ 0.09 \\
NGC3593 & STING & 5.5 & 8.67 (7) & -20.3 $\pm$ 0.1 & -21.3 $\pm$ 0.3 & 434.7 $\pm$ 19.8 & 910 $\pm$ 160 (14) & 40.79 $\pm$ 0.10 & 41.22 $\pm$ 0.10 & 42.33 $\pm$ 0.09 & 42.39 $\pm$ 0.09 \\
NGC3627 & SONG & 10.1 & 8.43 (5) & -21.6 $\pm$ 0.3 & -23.7 $\pm$ 0.2 & 327.2 $\pm$ 83.2 & 4660 $\pm$ 1540 (14) & 42.46 $\pm$ 0.10 & 44.17 $\pm$ 0.10 & 42.40 $\pm$ 0.06 & 43.42 $\pm$ 0.07 \\
NGC3726 & SONG & 15.6 & 9.15 (4) & -19.2 $\pm$ 0.4 & -22.9 $\pm$ 0.2 & 16.3 $\pm$ 8.0 & 720 $\pm$ 180 (14) & 41.66 $\pm$ 0.05 & 43.25 $\pm$ 0.05 & 41.79 $\pm$ 0.11 & 42.81 $\pm$ 0.11 \\
NGC3938 & SONG & 17.9 & 8.59 (7) & -20.2 $\pm$ 0.2 & -23.4 $\pm$ 0.1 & 19.6 $\pm$ 17.5 & 1750 $\pm$ 450 (14) & 41.21 $\pm$ 0.05 & 43.56 $\pm$ 0.05 & 41.30 $\pm$ 0.13 & 43.23 $\pm$ 0.14 \\
NGC3949 & STING & 18.5 & 8.53 (7) & -19.7 $\pm$ 0.2 & -22.5 $\pm$ 0.3 & 3.5 $\pm$ 4.2 & 220 $\pm$ 40 (14) & 41.84 $\pm$ 0.27 & 43.65 $\pm$ 0.28 & 42.49 $\pm$ 0.36 & 43.87 $\pm$ 0.38 \\
NGC3953 & SONG & 17.3 & 8.54 (7) & -22.0 $\pm$ 0.9 & -24.3 $\pm$ 0.4 & <0.2   & 1790 $\pm$ 450 (14) &  ---  &  ---  & 41.59 $\pm$ 0.06 & 43.10 $\pm$ 0.07 \\
NGC3992 & SONG & 24.9 & 8.68 (6) & -22.6 $\pm$ 0.7 & -25.2 $\pm$ 0.6 & 0.8 $\pm$ 0.2 & 232 $\pm$ 20 (12) &  ---  &  ---  & 41.60 $\pm$ 0.05 & 42.97 $\pm$ 0.05 \\
NGC4254 & STING & 14.4 & 8.81 (4) & -21.8 $\pm$ 0.5 & -23.6 $\pm$ 0.5 & 351.2 $\pm$ 46.5 & 3000 $\pm$ 850 (14) & 42.03 $\pm$ 0.10 & 43.59 $\pm$ 0.11 & 42.56 $\pm$ 0.05 & 43.61 $\pm$ 0.05 \\
NGC4258 & SONG & 7.5 & 9.19 (4) & -21.2 $\pm$ 0.5 & -24.0 $\pm$ 0.3 & 635.2 $\pm$ 87.9 & 1240 $\pm$ 230 (14) & 41.97 $\pm$ 0.05 & 42.95 $\pm$ 0.05 & 42.07 $\pm$ 0.05 & 42.79 $\pm$ 0.05 \\
NGC4273 & STING & 33.7 & 8.68 (7) & -20.7 $\pm$ 1.6 & -23.3 $\pm$ 0.3 & 22.3 $\pm$ 18.6 & 390 $\pm$ 70 (14) & 42.34 $\pm$ 0.15 & 43.47 $\pm$ 0.15 & 42.86 $\pm$ 0.05 & 43.80 $\pm$ 0.05 \\
NGC4303 & SONG & 10.6 & 9.21 (4) & -20.4 $\pm$ 0.3 & -22.9 $\pm$ 0.2 & 135.8 $\pm$ 110.4 & 2280 $\pm$ 470 (14) & 41.68 $\pm$ 0.08 & 43.17 $\pm$ 0.09 & 42.31 $\pm$ 0.10 & 43.12 $\pm$ 0.10 \\
NGC4321 & SONG & 15.8 & 8.99 (4) & -21.8 $\pm$ 0.3 & -24.2 $\pm$ 0.3 & 545.4 $\pm$ 119.7 & 3340 $\pm$ 920 (14) & 42.46 $\pm$ 0.05 & 43.31 $\pm$ 0.05 & 43.00 $\pm$ 0.06 & 43.55 $\pm$ 0.06 \\
NGC4414 & SONG & 18.3 & 8.65 (7) & -20.9 $\pm$ 0.3 & -24.1 $\pm$ 0.1 & 6.6 $\pm$ 10.0 & 2740 $\pm$ 500 (14) & 40.67 $\pm$ 0.19 & 43.01 $\pm$ 0.20 & 41.89 $\pm$ 0.29 & 43.58 $\pm$ 0.30 \\
NGC4450 & SONG & 16.5 & 9.18 (4) & -21.7 $\pm$ 0.5 & -24.0 $\pm$ 0.2 & 1.4 $\pm$ 0.8 & 450 $\pm$ 90 (14) & 41.15 $\pm$ 0.07 & 42.53 $\pm$ 0.08 & 41.69 $\pm$ 0.06 & 42.38 $\pm$ 0.06 \\
NGC4535 & SONG & 15.9 & 8.19 (2) & -19.8 $\pm$ 1.0 & -23.1 $\pm$ 0.2 & 96.1 $\pm$ 71.6 & 1570 $\pm$ 157 (14) & 41.47 $\pm$ 0.09 & 42.58 $\pm$ 0.10 & 42.62 $\pm$ 0.08 & 43.16 $\pm$ 0.08 \\
NGC4536 & STING & 14.9 & 8.69 (7) & -21.5 $\pm$ 0.3 & -22.9 $\pm$ 0.2 & 325.7 $\pm$ 56.5 & 740 $\pm$ 130 (14) & 42.09 $\pm$ 0.12 & 42.77 $\pm$ 0.12 & 43.34 $\pm$ 0.05 & 43.61 $\pm$ 0.05 \\
NGC4548 & SONG & 16.2 & 9.14 (4) & -21.6 $\pm$ 0.5 & -24.0 $\pm$ 0.4 & 5.4 $\pm$ 1.3 & 540 $\pm$ 140 (14) & 40.76 $\pm$ 0.05 & 42.52 $\pm$ 0.05 & 41.68 $\pm$ 0.06 & 42.66 $\pm$ 0.06 \\
NGC4559 & SONG & 7.9 & 8.68 (7) & -17.9 $\pm$ 0.3 & -22.2 $\pm$ 0.4 & 5.7 $\pm$ 3.1 & 284 $\pm$ 28 (12) & 41.04 $\pm$ 0.09 & 43.20 $\pm$ 0.09 & 41.26 $\pm$ 0.05 & 42.51 $\pm$ 0.05 \\
NGC4569 & SONG & 9.6 & 8.64 (7) & -20.8 $\pm$ 0.8 & -23.3 $\pm$ 0.3 & 158.6 $\pm$ 75.6 & 1500 $\pm$ 260 (14) & 41.50 $\pm$ 0.06 & 42.24 $\pm$ 0.06 & 42.37 $\pm$ 0.05 & 42.83 $\pm$ 0.05 \\
NGC4579 & SONG & 19.6 & 8.68 (7) & -23.2 $\pm$ 0.6 & -24.8 $\pm$ 0.5 & 49.4 $\pm$ 5.0 & 910 $\pm$ 200 (14) & 41.76 $\pm$ 0.05 & 42.80 $\pm$ 0.05 & 42.57 $\pm$ 0.05 & 43.02 $\pm$ 0.05 \\
NGC4605 & STING & 4.7 & 8.54 (6) & -17.9 $\pm$ 0.1 & -20.7 $\pm$ 0.1 & 6.9 $\pm$ 2.1 & 190 $\pm$ 60 (14) & 41.66 $\pm$ 0.11 & 42.39 $\pm$ 0.11 & 40.85 $\pm$ 0.05 & 41.98 $\pm$ 0.05 \\
NGC4654 & STING & 14.1 & 9.02 (4) & -18.3 $\pm$ 0.9 & -22.8 $\pm$ 0.1 & 11.9 $\pm$ 10.0 & 730 $\pm$ 150 (14) & 40.67 $\pm$ 0.31 & 42.84 $\pm$ 0.33 & 42.29 $\pm$ 0.05 & 43.99 $\pm$ 0.05 \\
NGC4699 & SONG & 23.7 & 9.05 (4) & -24.0 $\pm$ 0.5 & -25.4 $\pm$ 2.6 & 40.5 $\pm$ 11.5 &  ---   & 43.15 $\pm$ 0.05 & 43.19 $\pm$ 0.05 & 43.16 $\pm$ 0.05 & 43.24 $\pm$ 0.05 \\
NGC4725 & SONG & 13.2 & 8.64 (7) & -22.3 $\pm$ 0.3 & -24.2 $\pm$ 0.2 & 22.2 $\pm$ 4.9 & 1950 $\pm$ 700 (14) & 41.32 $\pm$ 0.05 & 42.44 $\pm$ 0.05 & 41.73 $\pm$ 0.05 & 42.71 $\pm$ 0.06 \\
NGC4736 & SONG & 5.0 & 8.69 (6) & -21.8 $\pm$ 0.2 & -23.0 $\pm$ 0.4 & 242.0 $\pm$ 62.2 & 2560 $\pm$ 690 (14) & 41.26 $\pm$ 0.06 & 42.58 $\pm$ 0.06 & 42.12 $\pm$ 0.06 & 42.71 $\pm$ 0.06 \\
NGC4826 & SONG & 5.4 & 8.63 (7) & -21.6 $\pm$ 0.5 & -23.2 $\pm$ 0.3 & 1240.3 $\pm$ 122.9 & 2170 $\pm$ 380 (14) & 41.58 $\pm$ 0.05 & 42.08 $\pm$ 0.06 & 42.39 $\pm$ 0.05 & 42.57 $\pm$ 0.05 \\
NGC5005 & SONG & 15.3 & 8.79 (4) & -22.1 $\pm$ 0.4 & -24.3 $\pm$ 0.3 & 278.5 $\pm$ 101.0 & 1260 $\pm$ 280 (14) & 40.91 $\pm$ 0.28 & 42.77 $\pm$ 0.29 & 42.60 $\pm$ 0.06 & 43.29 $\pm$ 0.06 \\
NGC5033 & SONG & 17.8 & 8.65 (7) & -23.2 $\pm$ 0.8 & -24.1 $\pm$ 0.3 & 342.4 $\pm$ 38.0 & 1640 $\pm$ 460 (14) & 42.02 $\pm$ 0.05 & 43.36 $\pm$ 0.06 & 43.15 $\pm$ 0.05 & 43.51 $\pm$ 0.05 \\
NGC5055 & SONG & 9.0 & 8.67 (7) & -21.9 $\pm$ 0.7 & -24.1 $\pm$ 0.4 & 499.9 $\pm$ 76.3 & 5670 $\pm$ 1890 (14) & 41.23 $\pm$ 0.05 & 43.00 $\pm$ 0.06 & 42.24 $\pm$ 0.06 & 43.37 $\pm$ 0.06 \\
NGC5194 & SONG & 8.0 & 8.66 (7) & -21.5 $\pm$ 0.8 & -23.9 $\pm$ 0.4 & 379.6 $\pm$ 147.5 & 9210 $\pm$ 3000 (14) & 41.88 $\pm$ 0.05 & 43.39 $\pm$ 0.06 & 42.33 $\pm$ 0.06 & 43.65 $\pm$ 0.07 \\
NGC5247 & SONG & 22.2 & 8.90 (4) & -21.0 $\pm$ 0.4 & -23.7 $\pm$ 0.2 & 65.6 $\pm$ 56.6 & 1130 $\pm$ 230 (14) & 42.10 $\pm$ 0.09 & 43.51 $\pm$ 0.10 & 42.58 $\pm$ 0.08 & 43.56 $\pm$ 0.08 \\
NGC5248 & SONG & 12.7 & 8.98 (4) & -21.5 $\pm$ 0.3 & -23.1 $\pm$ 0.3 & 484.4 $\pm$ 107.3 & 1190 $\pm$ 350 (14) & 41.77 $\pm$ 0.07 & 42.77 $\pm$ 0.07 & 42.78 $\pm$ 0.06 & 43.28 $\pm$ 0.06 \\
NGC5371 & STING & 35.3 & 8.65 (7) & -22.6 $\pm$ 0.3 & -25.4 $\pm$ 0.4 & 11.4 $\pm$ 0.4 &  ---   & 41.24 $\pm$ 0.17 & 43.66 $\pm$ 0.18 & 42.56 $\pm$ 0.19 & 44.38 $\pm$ 0.20 \\
NGC5457 & SONG & 7.2 & 8.71 (6) & -19.5 $\pm$ 0.6 & -23.7 $\pm$ 0.2 & 159.5 $\pm$ 26.0 & 2357 $\pm$ 236 (12) & 41.46 $\pm$ 0.05 & 43.89 $\pm$ 0.05 & 41.67 $\pm$ 0.06 & 43.24 $\pm$ 0.06 \\
NGC5713 & STING & 30.4 & 9.10 (4) & -23.2 $\pm$ 0.3 & -23.6 $\pm$ 0.9 & 1752.2 $\pm$ 97.5 & 680 $\pm$ 160 (14) & 43.04 $\pm$ 0.10 & 43.10 $\pm$ 0.10 & 43.97 $\pm$ 0.05 & 44.04 $\pm$ 0.05 \\
NGC6503 & STING & 5.3 & 8.97 (4) & -18.3 $\pm$ 0.2 & -21.3 $\pm$ 0.2 & 13.8 $\pm$ 6.0 & 1030 $\pm$ 340 (14) & 40.41 $\pm$ 0.16 & 42.24 $\pm$ 0.16 & 40.60 $\pm$ 0.05 & 42.07 $\pm$ 0.05 \\
NGC6946 & SONG & 6.5 & 8.81 (4) & -19.5 $\pm$ 0.7 & -23.4 $\pm$ 0.2 &
1124.0 $\pm$ 353.2 & 12370 $\pm$ 4120 (14) & --- & ---  & 42.93 $\pm$ 0.05 & 43.46 $\pm$ 0.05 \\
NGC6951 & STING & 22.6 & 8.32 (1) & -22.1 $\pm$ 0.3 & -24.2 $\pm$ 0.3 & 300.8 $\pm$ 112.7 & 1440 $\pm$ 300 (14) & 41.16 $\pm$ 0.15 & 42.30 $\pm$ 0.16 & 43.88 $\pm$ 0.05 & 44.41 $\pm$ 0.05 \\
NGC7217 & NUGA & 17.2 & 8.68 (7) & -22.5 $\pm$ 0.6 & -24.2 $\pm$ 0.5 &
2.0 $\pm$ 0.9 & 440 $\pm$ 90  (14) & 41.21 $\pm$ 0.11 & 42.73 $\pm$ 0.12 & 41.90 $\pm$ 0.05 & 42.89 $\pm$ 0.05 \\
\tablenotetext{(a)}{Metallicity adjusted to \cite{kobulnicky2004}.  Reference: (1)
 \cite{denicolo2002}; (2) \cite{engelbracht2008}; (3)
 \cite{garnett2002}; (4) \cite{moustakas2010}; (5)
 \cite{pilyugin2006}; (6) \cite{prieto2008}; (7) Metallicity
  derived from Mass-Metallicity relationship in \cite{tremonti2004}. }

\tablenotetext{(b)}{Total S$_{CO}$ Reference: (8) \cite{braine1993}; (9)
  \cite{elfhag1996}; (10) \cite{leroy2005}; (11) \cite{leroy2008}; (12)
  \cite{sheth2005}; (13) This work; (14) \cite{young1995} }
\enddata
\end{deluxetable}
\clearpage
\end{landscape}

\begin{deluxetable}{lccccccccccccc}
  \tablewidth{0pt} \tablecaption{Derived Bulge Properties}
\tablehead{\colhead{Galaxy} & \colhead{S\'ersic} &
  \colhead{log($\Sigma_{SFR}$)} & \colhead{log($\Sigma_{SFR,adj}$)} &
      \colhead{log($\Sigma_{mol}$)}& \colhead{log($\Sigma_{mol}$)} &
      \colhead{log($\Sigma_{star}$)} &\colhead{log(M/L$_H$)} &
      \colhead{Method} &
      \colhead{CO Profile} & \colhead{r$_{b=d}$}
\\
\colhead{Name} & \colhead{Index} &
  \colhead{M$_{\odot}$~yr$^{-1}$~kpc$^{-2}$} & \colhead{M$_{\odot}$~yr$^{-1}$~kpc$^{-2}$} &
      \colhead{M$_{\odot}$~pc$^{-2}$} &
      \colhead{M$_{\odot}$~pc$^{-2}$} &
      \colhead{M$_{\odot}$~kpc$^{-2}$} & \colhead{ } &
      \colhead{ } &
      \colhead{Type} & \colhead{ `` }
\\
\colhead{} & \colhead{} &
  \colhead{All Flux} & \colhead{Adjusted} &
      \colhead{X(Z)} & \colhead{X(Z,W)} & \colhead{} & \colhead{ } &
      \colhead{ } &
      \colhead{\tablenotemark{(b)}} & \colhead {\tablenotemark{(c)} }
 }
\startdata
NGC0337 &  ---  & 0.11 $\pm$ 0.34 & 0.04 $\pm$ 0.11 & 2.27 $\pm$ 0.31
& 2.25 $\pm$ 0.31 & 3.68 $\pm$ 0.10 & -0.55 & NUV-J & e & 6.2 \\
NGC0628 & 1.5 $\pm$ 0.2 & -1.80 $\pm$ 0.21 & -1.94 $\pm$ 0.23 & 1.79
$\pm$ 0.28 & 1.62 $\pm$ 0.28 & 3.35 $\pm$ 0.30 & -0.30 & NUV-J & e & 16.8 \\
NGC0772 & 3.9 $\pm$ 0.3 & -0.52 $\pm$ 0.21 & -0.62 $\pm$ 0.25 & 2.20
$\pm$ 0.37 & 1.86 $\pm$ 0.37 & 3.66 $\pm$ 0.59 & -0.29 & NUV-J & n & 13.4 \\
NGC0925 & 0.9 $\pm$ 0.2 & -0.69 $\pm$ 0.16 & -0.71 $\pm$ 0.16 & 0.64
$\pm$ 0.24 & 0.98 $\pm$ 0.24 & 2.49 $\pm$ 0.29 & -0.77 & NUV-J & e & 13.6 \\
NGC1156 &  ---  & 0.73 $\pm$ 0.28 & 0.64 $\pm$ 0.24 & 2.25 $\pm$ 0.29
& 2.05 $\pm$ 0.29 & 3.15 $\pm$ 0.10 & -0.65 & NUV-J & e & 9.6 \\
NGC1637 & 1.5 $\pm$ 0.2 & 0.61 $\pm$ 0.22 & 0.61 $\pm$ 0.22 & 3.53
$\pm$ 0.24 & 2.84 $\pm$ 0.24 & 3.62 $\pm$ 0.79 & -0.34 & B-V & b & 6.3
\\
NGC2403 & 0.8 $\pm$ 0.2 & -1.11 $\pm$ 0.62 & -1.19 $\pm$ 0.67 & 1.69
$\pm$ 0.51 & 1.93 $\pm$ 0.51 & 3.21 $\pm$ 0.37 & -0.58 & NUV-J & e & 4.0 \\
NGC2681 & 3.8 $\pm$ 0.2 & -0.65 $\pm$ 0.24 & -0.77 $\pm$ 0.28 & 2.53
$\pm$ 0.22 & 1.77 $\pm$ 0.22 & 3.68 $\pm$ 0.35 & -0.61 & SED & b & 14.7 \\
NGC2782 & 2.2 $\pm$ 0.2 & -0.11 $\pm$ 0.21 & -0.20 $\pm$ 0.37 & 2.43
$\pm$ 0.22 & 1.78 $\pm$ 0.22 & 3.23 $\pm$ 0.34 & -0.59 & SED & b &  10.2\\
NGC2841 & 3.2 $\pm$ 0.3 & -1.41 $\pm$ 0.15 & -1.73 $\pm$ 0.18 & -0.05
$\pm$ 0.93 & 0.40 $\pm$ 0.93 & 4.30 $\pm$ 0.37 & -0.08 & SED & n & 17.5\\
NGC2903 & 0.5 $\pm$ 0.1 & 0.63 $\pm$ 0.10 & 0.53 $\pm$ 0.08 & 3.84
$\pm$ 0.30 & 3.21 $\pm$ 0.30 & 3.99 $\pm$ 0.18 & -0.59 & SED & s & 8.4
\\
NGC2976 &  ---  & -0.04 $\pm$ 0.23 & -0.04 $\pm$ 0.23 & 2.21 $\pm$
0.23 & 2.05 $\pm$ 0.23 & 3.55 $\pm$ 0.10 & -0.41 & SED & e & 20.7 \\
NGC3031 & 4.1 $\pm$ 0.2 & -1.63 $\pm$ 0.15 & -2.11 $\pm$ 0.19 & -0.17
$\pm$ 0.45 & -0.12 $\pm$ 0.45 & 4.12 $\pm$ 0.35 & -0.10 & SED & n & 72.0 \\
NGC3147 & 2.0 $\pm$ 0.3 & -1.10 $\pm$ 0.17 & -1.30 $\pm$ 0.20 & 2.08
$\pm$ 0.23 & 1.56 $\pm$ 0.23 & 4.23 $\pm$ 0.48 & -0.11 & NUV-J & n & 10.2\\
NGC3184 & 1.8 $\pm$ 0.3 & -0.68 $\pm$ 0.18 & -0.76 $\pm$ 0.20 & 2.11
$\pm$ 0.32 & 2.09 $\pm$ 0.32 & 3.54 $\pm$ 0.44 & -0.17 & SED & s & 6.6 \\
NGC3198 & 1.8 $\pm$ 0.2 & -0.30 $\pm$ 0.31 & -0.40 $\pm$ 0.41 & 2.66
$\pm$ 0.23 & 2.25 $\pm$ 0.23 & 3.44 $\pm$ 0.51 & -0.37 & SED & s & 7.4 \\
NGC3344 & 2.3 $\pm$ 0.4 & -0.45 $\pm$ 0.19 & -0.53 $\pm$ 0.23 & <1.5
& <1.6 & 3.91 $\pm$ 0.53 & -0.11 & SED & n & 7.2 \\
NGC3351 & 1.4 $\pm$ 0.4 & -0.11 $\pm$ 0.17 & -0.20 $\pm$ 0.32 & 2.37
$\pm$ 0.24 & 2.09 $\pm$ 0.24 & 3.70 $\pm$ 0.34 & -0.46 & SED & s & 14.4 \\
NGC3368 & 1.6 $\pm$ 0.2 & -1.48 $\pm$ 0.12 & -1.66 $\pm$ 0.13 & 2.81
$\pm$ 0.22 & 2.15 $\pm$ 0.22 & 3.89 $\pm$ 0.40 & -0.16 & SED & n & 29.2 \\
NGC3486 & 1.8 $\pm$ 0.5 & -0.53 $\pm$ 0.09 & -1.81 $\pm$ 0.30 & 1.09
$\pm$ 0.24 & 0.86 $\pm$ 0.24 & 2.78 $\pm$ 0.56 & -0.62 & SED & e & 24.9 \\
NGC3521 & 3.7 $\pm$ 0.4 & -1.05 $\pm$ 0.17 & -1.39 $\pm$ 0.22 & 2.20
$\pm$ 0.49 & 2.09 $\pm$ 0.49 & 4.38 $\pm$ 0.70 & -0.28 & NUV-J & e &
12.1 \\
NGC3593 & 1.2 $\pm$ 0.1 & -0.54 $\pm$ 0.32 & -0.64 $\pm$ 0.38 & 2.95
$\pm$ 0.22 & 2.14 $\pm$ 0.22 & 3.54 $\pm$ 0.21 & -0.39 & SED & s 32.2 \\
NGC3627 & 1.5 $\pm$ 0.3 & 0.00 $\pm$ 0.09 & -0.07 $\pm$ 1.35 & 3.03
$\pm$ 0.25 & 2.64 $\pm$ 0.25 & 4.56 $\pm$ 0.34 & -0.08 & SED & s &
11.7 \\
NGC3726 & 1.5 $\pm$ 0.3 & -0.61 $\pm$ 0.20 & -0.68 $\pm$ 0.22 & 2.64
$\pm$ 0.31 & 2.42 $\pm$ 0.31 & 3.09 $\pm$ 0.39 & -0.71 & SED & s & 6.4 \\
NGC3938 & 1.5 $\pm$ 0.2 & -1.42 $\pm$ 0.26 & -1.55 $\pm$ 0.28 & 2.37
$\pm$ 0.45 & 2.19 $\pm$ 0.45 & 3.43 $\pm$ 0.23 & -0.42 & SED & e & 6.8 \\
NGC3949 & 1.0 $\pm$ 0.2 & 0.03 $\pm$ 0.32 & -0.05 $\pm$ -0.50 & 2.25
$\pm$ 0.57 & 2.23 $\pm$ 0.57 & 3.52 $\pm$ 0.26 & -0.54 & SED & e & 5.0  \\
NGC3953 & 1.6 $\pm$ 0.6 & -0.95 $\pm$ 0.11 & -1.52 $\pm$ 0.18 & <0.1 &
<0.5 & 4.24 $\pm$ 0.48 & -0.01 & SED & n & 13.1 \\
NGC3992 & 3.2 $\pm$ 0.6 & -1.47 $\pm$ 0.11 & -2.71 $\pm$ 0.20 & 0.50
$\pm$ 0.26 & 0.77 $\pm$ 0.26 & 3.90 $\pm$ 0.68 & -0.07 & SED & e &
14.8 \\
NGC4254 & 2.1 $\pm$ 0.3 & -0.82 $\pm$ 0.26 & -0.92 $\pm$ 0.29 & 2.42
$\pm$ 0.23 & 1.88 $\pm$ 0.23 & 3.55 $\pm$ 0.46 & -0.41 & SED & s & 19.2 \\
NGC4258 & 3.7 $\pm$ 0.4 & -0.66 $\pm$ 0.11 & -0.75 $\pm$ 0.12 & 2.89
$\pm$ 0.23 & 2.34 $\pm$ 0.23 & 3.84 $\pm$ 0.49 & -0.38 & NUV-J & s &
20.2 \\
NGC4273 & 1.2 $\pm$ 0.4 & 0.57 $\pm$ 0.20 & 0.49 $\pm$ 0.17 & 3.55
$\pm$ 0.42 & 3.14 $\pm$ 0.42 & 3.97 $\pm$ 1.38 & -0.61 & SED & s & 4.0 \\
NGC4303 & 1.1 $\pm$ 0.2 & -0.04 $\pm$ 0.15 & -0.14 $\pm$ 0.47 & 3.09
$\pm$ 0.42 & 2.75 $\pm$ 0.42 & 3.85 $\pm$ 0.32 & -0.59 & SED & s & 6.6 \\
NGC4321 & 1.7 $\pm$ 0.2 & -0.43 $\pm$ 0.03 & -0.52 $\pm$ 0.04 & 2.60
$\pm$ 0.24 & 2.20 $\pm$ 0.24 & 3.25 $\pm$ 0.31 & -0.63 & SED & s& 16.0 \\
NGC4414 & 1.2 $\pm$ 0.2 & -0.64 $\pm$ 0.59 & -0.79 $\pm$ 0.73 & 2.37
$\pm$ 0.70 & 2.32 $\pm$ 0.70 & 4.50 $\pm$ 0.32 & -0.03 & SED & n & 4.6 \\
NGC4450 & 3.5 $\pm$ 0.3 & -1.30 $\pm$ 0.16 & -1.56 $\pm$ 0.19 & 1.06
$\pm$ 0.34 & 1.24 $\pm$ 0.34 & 4.01 $\pm$ 0.49 & -0.28 & SED & n & 9.9 \\
NGC4535 & 1.3 $\pm$ 0.4 & 0.08 $\pm$ 0.12 & -0.01 $\pm$ -0.02 & 3.42
$\pm$ 0.39 & 3.02 $\pm$ 0.39 & 3.46 $\pm$ 0.89 & -0.60 & SED & s & 5.3 \\
NGC4536 & 1.5 $\pm$ 0.2 & 0.36 $\pm$ 0.03 & 0.26 $\pm$ 0.02 & 3.03
$\pm$ 0.23 & 2.49 $\pm$ 0.23 & 4.15 $\pm$ 0.35 & -0.15 & SED & s & 11.0 \\
NGC4548 & 2.9 $\pm$ 0.4 & -1.69 $\pm$ 0.16 & -1.98 $\pm$ 0.19 & 1.30
$\pm$ 0.25 & 1.33 $\pm$ 0.25 & 3.89 $\pm$ 0.48 & -0.05 & SED & e & 13.6\\
NGC4559 & 1.7 $\pm$ 0.4 & -0.86 $\pm$ 0.20 & -0.94 $\pm$ 0.22 & 1.65
$\pm$ 0.32 & 1.71 $\pm$ 0.32 & 3.05 $\pm$ 0.33 & -0.53 & SED & e & 9.8\\
NGC4569 & 1.8 $\pm$ 0.4 & 0.26 $\pm$ 0.08 & 0.03 $\pm$ 0.01 & 3.47
$\pm$ 0.27 & 2.99 $\pm$ 0.27 & 4.10 $\pm$ 0.74 & -0.63 & SED & s & 8.9 \\
NGC4579 & 3.2 $\pm$ 0.4 & -1.13 $\pm$ 0.07 & -1.31 $\pm$ 0.08 & 2.07
$\pm$ 0.23 & 1.84 $\pm$ 0.23 & 3.98 $\pm$ 0.58 & -0.26 & SED & e & 19.1\\
NGC4605 &  $\pm$  & 0.02 $\pm$ 0.42 & 0.00 $\pm$ 0.00 & 2.31 $\pm$
0.26 & 2.16 $\pm$ 0.26 & 3.67 $\pm$ 0.10 & -0.56 & SED & e & 15.1 \\
NGC4654 & 1.1 $\pm$ 0.8 & 0.81 $\pm$ 0.34 & 0.71 $\pm$ 0.30 & 3.51
$\pm$ 0.43 & 3.17 $\pm$ 0.43 & 3.98 $\pm$ 0.86 & -0.57 & SED & s & 4.0\\
NGC4699 & 3.0 $\pm$ 0.4 & -2.06 $\pm$ 0.01 & -2.14 $\pm$ 0.01 & 0.42
$\pm$ 0.26 & 0.24 $\pm$ 0.26 & 2.42 $\pm$ 0.54 & -0.41 & SED & e & 107.5 \\
NGC4725 & 3.9 $\pm$ 0.3 & -1.70 $\pm$ 0.15 & -2.08 $\pm$ 0.18 & 1.11
$\pm$ 0.24 & 1.14 $\pm$ 0.24 & 4.00 $\pm$ 0.37 & -0.10 & SED & e & 22.3\\
NGC4736 & 1.2 $\pm$ 0.1 & -0.36 $\pm$ 0.12 & -0.52 $\pm$ 0.17 & 2.42
$\pm$ 0.25 & 2.05 $\pm$ 0.25 & 4.77 $\pm$ 0.26 & -0.13 & NUV-J & e  & 15.6\\
NGC4826 & 3.9 $\pm$ 0.4 & -0.60 $\pm$ 0.09 & -0.72 $\pm$ 0.11 & 2.46
$\pm$ 0.22 & 1.98 $\pm$ 0.22 & 3.73 $\pm$ 0.52 & -0.57 & NUV-J & s & 29.7\\
NGC5005 & 1.8 $\pm$ 0.3 & -0.27 $\pm$ 0.25 & -0.30 $\pm$ 0.28 & 3.61
$\pm$ 0.27 & 3.06 $\pm$ 0.27 & 4.38 $\pm$ 0.42 & -0.38 & SED & s & 8.4 \\
NGC5033 & 1.5 $\pm$ 0.8 & -0.62 $\pm$ 0.02 & -0.73 $\pm$ 0.02 & 2.46
$\pm$ 0.23 & 2.02 $\pm$ 0.23 & 3.79 $\pm$ 0.77 & -0.37 & SED & b & 29.5\\
NGC5055 & 1.8 $\pm$ 0.5 & -0.94 $\pm$ 0.11 & -1.10 $\pm$ 0.13 & 2.40
$\pm$ 0.23 & 1.99 $\pm$ 0.23 & 3.78 $\pm$ 0.64 & -0.48 & SED & s & 22.1\\
NGC5194 & 1.6 $\pm$ 0.5 & -0.64 $\pm$ 0.10 & -0.74 $\pm$ 0.12 & 2.28
$\pm$ 0.28 & 1.92 $\pm$ 0.28 & 3.86 $\pm$ 0.72 & -0.34 & NUV-J & e & 20.0 \\
NGC5247 & 1.9 $\pm$ 0.3 & -0.77 $\pm$ 0.10 & -0.85 $\pm$ 0.11 & 2.65
$\pm$ 0.44 & 2.36 $\pm$ 0.44 & 3.14 $\pm$ 0.39 & -0.49 & NUV-J & s & 10.0\\
NGC5248 & 1.3 $\pm$ 0.2 & -0.50 $\pm$ 0.06 & -0.60 $\pm$ 0.07 & 3.17
$\pm$ 0.24 & 2.55 $\pm$ 0.24 & 3.40 $\pm$ 0.37 & -0.59 & SED & s & 17.0 \\
NGC5371 & 1.9 $\pm$ 0.2 & -1.28 $\pm$ 0.29 & -1.43 $\pm$ 0.32 & 1.83
$\pm$ 0.22 & 1.56 $\pm$ 0.22 & 3.80 $\pm$ 0.31 & -0.09 & SED & e & 10.1 \\
NGC5457 & 1.6 $\pm$ 0.5 & -0.96 $\pm$ 0.15 & -1.03 $\pm$ 0.16 & 2.66
$\pm$ 0.23 & 2.17 $\pm$ 0.23 & 3.12 $\pm$ 0.58 & -0.57 & SED & s & 15.7 \\
NGC5713 & 2.1 $\pm$ 0.2 & -0.94 $\pm$ 0.26 & -1.03 $\pm$ 0.29 & 2.39
$\pm$ 0.22 & 1.70 $\pm$ 0.22 & 2.20 $\pm$ 0.31 & -0.83 & SED & s & 41.6 \\
NGC6503 &  ---  & -0.77 $\pm$ 0.51 & -0.90 $\pm$ 0.60 & 2.75 $\pm$
0.29 & 2.51 $\pm$ 0.29 & 4.12 $\pm$ 0.10 & -0.38 & NUV-J & e & 12.0 \\
NGC6946 & 1.6 $\pm$ 0.4 & 0.84 $\pm$ 0.15 & 0.85 $\pm$ 0.15 & 3.46
$\pm$ 0.26 & 2.88 $\pm$ 0.26 & 3.96 $\pm$ 0.63 & -0.20 & NUV-J & s & 10.0 \\
NGC6951 & 1.4 $\pm$ 0.2 & 0.64 $\pm$ 0.23 & 0.54 $\pm$ 0.19 & 3.50
$\pm$ 0.28 & 2.91 $\pm$ 0.28 & 3.93 $\pm$ 0.35 & -0.37 & NUV-J & s & 8.3 \\
NGC7217 & 3.2 $\pm$ 0.5 & -1.39 $\pm$ 0.16 & -1.73 $\pm$ 0.20 & 0.97
$\pm$ 0.29 & 0.92 $\pm$ 0.29 & 4.28 $\pm$ 0.55 & -0.05 & NUV-J & n & 11.9 \\
\tablenotetext{(a)}{s -- CO profile is as steep as stellar surface
  brightness profile. \\
b --  CO profile is steeper than exponential, but not as steep as
 stars. \\
 e -- CO
  follows exponential profile.\\
 n -- CO profile drops in bulge. \\
 }
\tablenotetext{(c)}{ The bulge radius used to measure fluxes. }

\enddata
\end{deluxetable}
\clearpage

\begin{deluxetable}{lccccccccccccc}
  \tablewidth{0pt} \tablecaption{Derived Total Galaxy Properties}
\tablehead{\colhead{Galaxy} & \colhead{log(h)} &
  \colhead{log($\Sigma_{SFR}$)} & \colhead{log($\Sigma_{SFR,adj}$)} &
      \colhead{log($\Sigma_{mol}$)}& \colhead{log($\Sigma_{mol}$)} &
      \colhead{log($\Sigma_{star}$)} & \colhead{log(M/L$_H$)} & \colhead{M/L}
      & \colhead{Disk} & \colhead{Environment}
\\
\colhead{Name} & \colhead{pc} &
  \colhead{M$_{\odot}$~yr$^{-1}$~kpc$^{-2}$} & \colhead{M$_{\odot}$~yr$^{-1}$~kpc$^{-2}$} &
      \colhead{M$_{\odot}$~pc$^{-2}$} &
      \colhead{M$_{\odot}$~pc$^{-2}$} &
      \colhead{M$_{\odot}$~kpc$^{-2}$} & \colhead{ } & \colhead{Method}
      & \colhead{Type} & \colhead{\tablenotemark{(b)}}
\\
\colhead{} & \colhead{} &
  \colhead{All Flux} & \colhead{Adjusted} &
      \colhead{X(Z)} & \colhead{X(Z,W)} & \colhead{} & \colhead{log(M/L$_H$)} & \colhead{Method}
      & \colhead{\tablenotemark{(a)}} & \colhead{ }
 }
\startdata
NGC0337 & 3.18 $\pm$ 0.02 & -1.07 $\pm$ 0.34 & -1.05 $\pm$ 0.34 & 1.41 $\pm$ 0.31 & 1.71 $\pm$ 0.31 & 2.66 $\pm$ 0.12 & -0.56 & NUV-J & U & I \\
NGC0628 & 3.42 $\pm$ 0.02 & -2.30 $\pm$ 0.21 & -1.03 $\pm$ 0.21 & 0.99 $\pm$ 0.26 & 0.89 $\pm$ 0.26 & 1.88 $\pm$ 0.23 & -0.50 & NUV-J & U & N \\
NGC0772 & 3.96 $\pm$ 0.27 & -1.92 $\pm$ 0.21 & -1.18 $\pm$ 0.21 & 1.39 $\pm$ 0.24 & 1.16 $\pm$ 0.24 & 2.10 $\pm$ 0.60 & -0.44 & NUV-J & U & I \\
NGC0925 & 3.48 $\pm$ 0.03 & -1.71 $\pm$ 0.16 & -1.68 $\pm$ 0.16 & 1.59 $\pm$ 0.31 & 1.41 $\pm$ 0.31 & 1.92 $\pm$ 0.22 & -0.69 & NUV-J & B & N \\
NGC1156 & 2.96 $\pm$ 0.02 & -1.22 $\pm$ 0.28 & -2.28 $\pm$ 0.28 & 1.02 $\pm$ 0.22 & 1.18 $\pm$ 0.22 & 1.78 $\pm$ 0.11 & -0.63 & NUV-J & B & N \\
NGC1637 & 3.14 $\pm$ 0.03 & -1.37 $\pm$ 0.22 & -1.14 $\pm$ 0.22 & 1.21 $\pm$ 0.22 & 1.04 $\pm$ 0.22 & 2.43 $\pm$ 0.71 & -0.34 & B-V & B & I \\
NGC2403 & 3.19 $\pm$ 0.03 & -1.94 $\pm$ 0.62 & -1.52 $\pm$ 0.62 & 0.41 $\pm$ 0.26 & 0.40 $\pm$ 0.26 & 1.83 $\pm$ 0.29 & -0.64 & NUV-J & U & N \\
NGC2681 & 3.30 $\pm$ 0.05 & -1.99 $\pm$ 0.24 & -1.98 $\pm$ 0.24 & 1.50 $\pm$ 0.24 & 1.53 $\pm$ 0.24 & 2.68 $\pm$ 0.30 & -0.42 & SED & U & N \\
NGC2782 & 3.57 $\pm$ 0.04 & -1.68 $\pm$ 0.21 & -1.38 $\pm$ 0.21 & 1.52 $\pm$ 0.23 & 1.51 $\pm$ 0.23 & 2.08 $\pm$ 0.28 & -0.51 & SED & U & I \\
NGC2841 & 3.60 $\pm$ 0.02 & -2.27 $\pm$ 0.15 & -1.71 $\pm$ 0.15 & 1.53 $\pm$ 0.23 & 1.39 $\pm$ 0.23 & 2.88 $\pm$ 0.30 & -0.23 & SED & U & N \\
NGC2903 & 3.20 $\pm$ 0.01 & -0.97 $\pm$ 0.10 & -1.03 $\pm$ 0.10 & 2.80 $\pm$ 0.26 & 2.37 $\pm$ 0.26 & 3.10 $\pm$ 0.13 & -0.41 & SED & B & N \\
NGC2976 & 2.96 $\pm$ 0.02 & -2.09 $\pm$ 0.23 & -1.92 $\pm$ 0.23 & 1.00 $\pm$ 0.24 & 1.05 $\pm$ 0.24 & 2.18 $\pm$ 0.10 & -0.52 & SED & U & N \\
NGC3031 & 3.38 $\pm$ 0.03 & -2.47 $\pm$ 0.15 & -1.56 $\pm$ 0.15 & 1.14 $\pm$ 0.22 & 0.78 $\pm$ 0.22 & 3.00 $\pm$ 0.29 & -0.21 & SED & U & I \\
NGC3147 & 3.72 $\pm$ 0.03 & -1.91 $\pm$ 0.17 & -1.37 $\pm$ 0.17 & 1.99 $\pm$ 0.24 & 1.79 $\pm$ 0.24 & 2.56 $\pm$ 0.40 & -0.36 & NUV-J & U & N \\
NGC3184 & 3.48 $\pm$ 0.01 & -2.27 $\pm$ 0.18 & -2.14 $\pm$ 0.18 & 0.73 $\pm$ 0.25 & 0.78 $\pm$ 0.25 & 1.79 $\pm$ 0.36 & -0.55 & SED & B & N \\
NGC3198 & 3.58 $\pm$ 0.02 & -1.84 $\pm$ 0.31 & -2.16 $\pm$ 0.31 & 1.17 $\pm$ 0.22 & 1.04 $\pm$ 0.22 & 1.84 $\pm$ 0.42 & -0.62 & SED & B & N \\
NGC3344 & 3.07 $\pm$ 0.03 & -2.01 $\pm$ 0.19 & -2.45 $\pm$ 0.19 & 1.13 $\pm$ 0.25 & 1.09 $\pm$ 0.25 & 2.08 $\pm$ 0.45 & -0.55 & SED & B & N \\
NGC3351 & 3.37 $\pm$ 0.04 & -2.05 $\pm$ 0.17 & -2.20 $\pm$ 0.17 & 0.63 $\pm$ 0.25 & 0.76 $\pm$ 0.25 & 2.23 $\pm$ 0.31 & -0.56 & SED & B & N \\
NGC3368 & 3.43 $\pm$ 0.02 & -2.35 $\pm$ 0.12 & -1.40 $\pm$ 0.12 & 1.39 $\pm$ 0.27 & 1.17 $\pm$ 0.27 & 2.54 $\pm$ 0.84 & -0.29 & SED & B & N \\
NGC3486 & 3.54 $\pm$ 0.06 & -1.89 $\pm$ 0.09 & -1.87 $\pm$ 0.09 & 1.07 $\pm$ 0.31 & 0.96 $\pm$ 0.31 & 1.37 $\pm$ 0.60 & -0.63 & SED & B & N \\
NGC3521 & 3.47 $\pm$ 0.03 & -1.41 $\pm$ 0.17 & -1.14 $\pm$ 0.17 & 1.99 $\pm$ 0.25 & 1.77 $\pm$ 0.25 & 2.91 $\pm$ 0.62 & -0.47 & NUV-J & B & N \\
NGC3593 & 3.15 $\pm$ 0.04 & -1.70 $\pm$ 0.32 & -0.64 $\pm$ 0.32 & 2.10 $\pm$ 0.24 & 1.88 $\pm$ 0.24 & 2.55 $\pm$ 0.19 & -0.55 & SED & U & I \\
NGC3627 & 3.51 $\pm$ 0.00 & -1.01 $\pm$ 0.09 & -2.32 $\pm$ 0.09 & 1.77 $\pm$ 0.26 & 1.54 $\pm$ 0.26 & 2.66 $\pm$ 0.26 & -0.39 & SED & B & I \\
NGC3726 & 3.57 $\pm$ 0.03 & -2.10 $\pm$ 0.20 & -1.92 $\pm$ 0.20 & 1.54 $\pm$ 0.25 & 1.38 $\pm$ 0.25 & 1.74 $\pm$ 0.32 & -0.71 & SED & B & N \\
NGC3938 & 3.48 $\pm$ 0.02 & -1.88 $\pm$ 0.26 & -1.92 $\pm$ 0.26 & 1.78 $\pm$ 0.25 & 1.56 $\pm$ 0.25 & 2.12 $\pm$ 0.17 & -0.42 & SED & U & N \\
NGC3949 & 3.18 $\pm$ 0.03 & -0.58 $\pm$ 0.32 & -1.52 $\pm$ 0.32 & 2.02 $\pm$ 0.23 & 2.00 $\pm$ 0.23 & 2.46 $\pm$ 0.21 & -0.62 & SED & U & I \\
NGC3953 & 3.71 $\pm$ 0.07 & -1.97 $\pm$ 0.11 & -1.81 $\pm$ 0.11 & 1.76 $\pm$ 0.25 & 1.51 $\pm$ 0.25 & 2.57 $\pm$ 0.40 & -0.27 & SED & B & N \\
NGC3992 & 3.97 $\pm$ 0.11 & -2.74 $\pm$ 0.11 & -2.14 $\pm$ 0.11 & 0.52 $\pm$ 0.22 & 0.48 $\pm$ 0.22 & 2.37 $\pm$ 0.62 & -0.17 & SED & B & N \\
NGC4254 & 3.39 $\pm$ 0.04 & -1.15 $\pm$ 0.26 & -2.32 $\pm$ 0.26 & 1.88 $\pm$ 0.25 & 1.79 $\pm$ 0.25 & 2.45 $\pm$ 0.42 & -0.67 & SED & U & I \\
NGC4258 & 3.51 $\pm$ 0.04 & -2.11 $\pm$ 0.11 & -1.51 $\pm$ 0.11 & 0.95 $\pm$ 0.23 & 0.87 $\pm$ 0.23 & 2.84 $\pm$ 0.41 & -0.21 & NUV-J & B & N \\
NGC4273 & 3.27 $\pm$ 0.03 & -1.11 $\pm$ 0.20 & -1.35 $\pm$ 0.20 & 2.19 $\pm$ 0.23 & 2.16 $\pm$ 0.23 & 2.33 $\pm$ 1.28 & -0.62 & SED & B & I \\
NGC4303 & 3.25 $\pm$ 0.02 & -1.65 $\pm$ 0.15 & -1.51 $\pm$ 0.15 & 1.76 $\pm$ 0.24 & 1.57 $\pm$ 0.24 & 2.37 $\pm$ 0.25 & -0.44 & SED & B & N \\
NGC4321 & 3.73 $\pm$ 0.04 & -2.24 $\pm$ 0.03 & -1.30 $\pm$ 0.03 & 1.02 $\pm$ 0.25 & 0.92 $\pm$ 0.25 & 1.88 $\pm$ 0.25 & -0.53 & SED & B & N \\
NGC4414 & 3.27 $\pm$ 0.01 & -1.30 $\pm$ 0.59 & -1.60 $\pm$ 0.59 & 2.63 $\pm$ 0.24 & 2.31 $\pm$ 0.24 & 2.87 $\pm$ 0.25 & -0.51 & SED & U & N \\
NGC4450 & 3.47 $\pm$ 0.03 & -2.59 $\pm$ 0.16 & -1.86 $\pm$ 0.16 & 1.42 $\pm$ 0.24 & 1.36 $\pm$ 0.24 & 2.72 $\pm$ 0.41 & -0.27 & SED & U & N \\
NGC4535 & 3.48 $\pm$ 0.02 & -2.18 $\pm$ 0.12 & -2.31 $\pm$ 0.12 & 1.83 $\pm$ 0.22 & 1.58 $\pm$ 0.22 & 2.13 $\pm$ 0.80 & -0.39 & SED & B & N \\
NGC4536 & 3.37 $\pm$ 0.02 & -1.28 $\pm$ 0.03 & -1.98 $\pm$ 0.03 & 1.52 $\pm$ 0.23 & 1.56 $\pm$ 0.23 & 2.36 $\pm$ 0.28 & -0.56 & SED & B & N \\
NGC4548 & 3.63 $\pm$ 0.05 & -2.91 $\pm$ 0.16 & -1.62 $\pm$ 0.16 & 1.01 $\pm$ 0.25 & 0.92 $\pm$ 0.25 & 2.25 $\pm$ 0.42 & -0.28 & SED & B & N \\
NGC4559 & 3.43 $\pm$ 0.04 & -1.75 $\pm$ 0.20 & -1.21 $\pm$ 0.20 & 0.76 $\pm$ 0.22 & 0.75 $\pm$ 0.22 & 2.09 $\pm$ 0.29 & -0.53 & SED & B & N \\
NGC4569 & 3.43 $\pm$ 0.03 & -1.86 $\pm$ 0.08 & -2.01 $\pm$ 0.08 & 2.04 $\pm$ 0.24 & 1.79 $\pm$ 0.24 & 2.78 $\pm$ 0.66 & -0.48 & SED & B & N \\
NGC4579 & 3.63 $\pm$ 0.05 & -2.35 $\pm$ 0.07 & -2.50 $\pm$ 0.07 & 1.62 $\pm$ 0.24 & 1.46 $\pm$ 0.24 & 2.76 $\pm$ 0.52 & -0.32 & SED & B & N \\
NGC4605 & 2.94 $\pm$ 0.01 & -1.48 $\pm$ 0.42 & -1.76 $\pm$ 0.42 & 1.49 $\pm$ 0.26 & 1.47 $\pm$ 0.26 & 2.60 $\pm$ 0.10 & -0.47 & SED & B & N \\
NGC4654 & 3.37 $\pm$ 0.01 & -0.93 $\pm$ 0.34 & -2.13 $\pm$ 0.34 & 1.89 $\pm$ 0.24 & 1.74 $\pm$ 0.24 & 2.38 $\pm$ 0.77 & -0.51 & SED & B & N \\
NGC4699 & 3.88 $\pm$ 0.48 & -2.66 $\pm$ 0.01 & -1.77 $\pm$ 0.01 &  ---  &  ---  & 2.41 $\pm$ 0.93 & -0.32 & SED & B & N \\
NGC4725 & 3.67 $\pm$ 0.02 & -2.74 $\pm$ 0.15 & -1.78 $\pm$ 0.15 & 1.05 $\pm$ 0.27 & 1.00 $\pm$ 0.27 & 2.48 $\pm$ 0.30 & -0.31 & SED & B & I \\
NGC4736 & 2.81 $\pm$ 0.06 & -1.26 $\pm$ 0.12 & -2.35 $\pm$ 0.12 & 1.86 $\pm$ 0.25 & 1.73 $\pm$ 0.25 & 3.46 $\pm$ 0.24 & -0.25 & NUV-J & U & N \\
NGC4826 & 3.25 $\pm$ 0.04 & -2.16 $\pm$ 0.09 & -2.06 $\pm$ 0.09 & 0.99 $\pm$ 0.23 & 0.97 $\pm$ 0.23 & 2.73 $\pm$ 0.45 & -0.43 & NUV-J & U & I \\
NGC5005 & 3.37 $\pm$ 0.03 & -1.64 $\pm$ 0.25 & -2.20 $\pm$ 0.25 & 2.16 $\pm$ 0.24 & 1.94 $\pm$ 0.24 & 2.98 $\pm$ 0.36 & -0.48 & SED & B & N \\
NGC5033 & 3.73 $\pm$ 0.03 & -1.53 $\pm$ 0.02 & -1.98 $\pm$ 0.02 & 1.77 $\pm$ 0.25 & 1.67 $\pm$ 0.25 & 2.61 $\pm$ 0.69 & -0.56 & SED & U & N \\
NGC5055 & 3.48 $\pm$ 0.03 & -1.77 $\pm$ 0.11 & -3.05 $\pm$ 0.11 & 1.47 $\pm$ 0.26 & 1.29 $\pm$ 0.26 & 2.59 $\pm$ 0.57 & -0.47 & SED & U & N \\
NGC5194 & 3.48 $\pm$ 0.04 & -1.55 $\pm$ 0.10 & -2.76 $\pm$ 0.10 & 1.45 $\pm$ 0.26 & 1.23 $\pm$ 0.26 & 2.39 $\pm$ 0.64 & -0.50 & NUV-J & U & I \\
NGC5247 & 3.66 $\pm$ 0.02 & -2.10 $\pm$ 0.10 & -2.93 $\pm$ 0.10 & 1.51 $\pm$ 0.24 & 1.30 $\pm$ 0.24 & 1.75 $\pm$ 0.32 & -0.55 & NUV-J & U & N \\
NGC5248 & 3.43 $\pm$ 0.04 & -1.83 $\pm$ 0.06 & -1.75 $\pm$ 0.06 & 1.73 $\pm$ 0.26 & 1.52 $\pm$ 0.26 & 2.45 $\pm$ 0.31 & -0.28 & SED & B/O & N \\
NGC5371 & 3.99 $\pm$ 0.09 & -2.11 $\pm$ 0.29 & -2.19 $\pm$ 0.29 &  ---  &  ---  & 1.88 $\pm$ 0.25 & -0.42 & SED & B & N \\
NGC5457 & 3.67 $\pm$ 0.03 & -1.89 $\pm$ 0.15 & -2.97 $\pm$ 0.15 & 0.91 $\pm$ 0.22 & 0.53 $\pm$ 0.22 & 1.75 $\pm$ 0.50 & -0.64 & SED & B & N \\
NGC5713 & 3.61 $\pm$ 0.07 & -1.69 $\pm$ 0.26 & -2.38 $\pm$ 0.26 & 1.24 $\pm$ 0.24 & 1.30 $\pm$ 0.24 & 2.01 $\pm$ 0.43 & -0.37 & SED & B & I \\
NGC6503 & 2.89 $\pm$ 0.02 & -1.50 $\pm$ 0.51 & -2.03 $\pm$ 0.51 & 2.32 $\pm$ 0.27 & 2.08 $\pm$ 0.27 & 2.78 $\pm$ 0.12 & -0.53 & NUV-J & U & N \\
NGC6946 & 3.48 $\pm$ 0.03 & -1.93 $\pm$ 0.15 & -2.79 $\pm$ 0.15 & 1.47 $\pm$ 0.26 & 1.13 $\pm$ 0.26 & 2.26 $\pm$ 0.55 & -0.37 & NUV-J & B & N \\
NGC6951 & 3.50 $\pm$ 0.04 & -1.04 $\pm$ 0.23 & -1.91 $\pm$ 0.23 & 2.03 $\pm$ 0.24 & 1.81 $\pm$ 0.24 & 2.54 $\pm$ 0.29 & -0.37 & NUV-J & B & N \\
NGC7217 & 3.34 $\pm$ 0.06 & -2.12 $\pm$ 0.16 & -2.72 $\pm$ 0.16 & 1.53
$\pm$ 0.24 & 1.48 $\pm$ 0.24 & 2.97 $\pm$ 0.49 & -0.21 & NUV-J & U & N
\\
\tablenotetext{(a)}{B -- Barred Disk; U -- Unbarred Disk; B/O --
  Ovaled disk
 }
\tablenotetext{(b)}{I -- interacting; N -- Not interacting
 }
\enddata
\end{deluxetable}
\clearpage

\end{document}